\documentclass[11pt,a4paper]{article}
\pdfoutput=1
\usepackage{jheppub}
\usepackage{gensymb}
 \usepackage{multirow}
 \usepackage[table,xcdraw]{xcolor}
\DeclareSymbolFont{matha}{OML}{txmi}{m}{it}
\DeclareMathSymbol{\varv}{\mathord}{matha}{118}
\usepackage{amsmath}
\usepackage{amssymb}
\usepackage{graphicx}
\usepackage{subfigure}
\usepackage{pstricks}
\usepackage{bm}
\usepackage{pbox}
\usepackage{placeins}
\usepackage{booktabs}
\usepackage[T1]{fontenc}
\usepackage{footnote}
\usepackage{pdfpages}
\usepackage{hhline}
\usepackage{multirow}
\usepackage{multicol}
\usepackage{enumitem}
\usepackage[toc,page]{appendix}
\allowdisplaybreaks
\usepackage{comment}
\usepackage{float}                                             
\usepackage{mathtools}                  
\usepackage{textcomp}
\usepackage{gensymb}

\usepackage{relsize}

\usepackage[numbers]{natbib}
\usepackage{notoccite}

\usepackage{rotating}
\usepackage{tabularx}
\usepackage[labelfont=bf]{caption} 

\FloatBarrier

\usepackage[many]{tcolorbox}

\renewcommand{\thetable}{\Roman{table}}

\definecolor{MyDarkBlue}{rgb}{0.1, 0.1, 0.8} 
\definecolor{SBlue}{rgb}{0.2, 0.4, 0.7} 
\definecolor{MyLightBlue}{rgb}{0.22,0.51,0.9}
\definecolor{MyGreen}{rgb}{0.0, 0.5, 0.0}
\definecolor{BrickRed}{rgb}{0.8, 0.25, 0.33}
\RequirePackage{hyperref}
\hypersetup{colorlinks, citecolor=MyGreen,linkcolor=MyDarkBlue, urlcolor=BrickRed}

\begin{document}

\preprint{OSU-HEP-20-12}

\title{
Unified Framework for $B$-Anomalies, Muon \boldmath{$g-2$}, and Neutrino Masses
 }
 \author[a]{K.S. Babu,}
 \author[b]{P.S. Bhupal Dev,}
 \author[c]{Sudip Jana,}
 \author[a]{Anil Thapa}
 
 \affiliation[a]{Department of Physics, Oklahoma State University, Stillwater, OK 74078, USA}
 \affiliation[b]{Department of Physics and McDonnell Center for the Space Sciences, Washington University, St. Louis, MO 63130, USA}
 \affiliation[c]{Max-Planck-Institut f{\"u}r Kernphysik, Saupfercheckweg 1, 69117 Heidelberg, Germany}

\emailAdd{babu@okstate.edu, bdev@wustl.edu, sudip.jana@mpi-hd.mpg.de, thapaa@okstate.edu}
\abstract{We present a model of radiative neutrino masses which also resolves anomalies reported in $B$-meson decays, $R_{D^{(\star)}}$ and $R_{K^{(\star)}}$, as well as in muon $g-2$ measurement, $\Delta a_\mu$.  Neutrino masses arise in the model through loop diagrams involving TeV-scale leptoquark (LQ) scalars $R_2$ and $S_3$. Fits to neutrino oscillation parameters are obtained satisfying all flavor constraints which also explain the anomalies in $R_{D^{(\star)}}$, $R_{K^{(\star)}}$ and $\Delta a_\mu$ within $1\, \sigma$. An isospin-3/2 Higgs quadruplet plays a crucial role in generating neutrino masses; we point out that the doubly-charged scalar contained therein can be produced in the decays of the $S_3$ LQ, which enhances its reach to 1.1 (6.2) TeV at $\sqrt s=14$ TeV high-luminosity LHC ($\sqrt s=100$ TeV FCC-hh).  We also present flavor-dependent upper limits on the Yukawa couplings of the LQs to the first two family fermions, arising from non-resonant dilepton ($pp \rightarrow \ell^+ \ell^-$) processes mediated by $t$-channel LQ exchange, which for 1 TeV LQ mass, are found to be in the range $(0.15 - 0.36)$. These limits preclude any explanation of $R_{D^{(\star)}}$ through LQ-mediated $B$-meson decays involving $\nu_e$ or $\nu_\mu$ in the final state.  We also find that the same Yukawa couplings responsible for the chirally-enhanced contribution to $\Delta a_\mu$ give rise to new contributions to the SM Higgs decays to muon and tau pairs, with the modifications to the corresponding branching ratios being at (2--6)\% level, which could be tested at future hadron colliders, such as HL-LHC and FCC-hh.
}



\maketitle
\section{Introduction}\label{sec:Intro}
Among the many reasons to consider physics beyond the Standard Model (SM), an understanding of the origin of neutrino masses stands out, as neutrino oscillations have been firmly established~\cite{Zyla:2020zbs} which require nonzero neutrino masses, in contradiction with the SM.  While neutrino masses may be accommodated at tree-level simply by the addition of three SM-singlet right-handed neutrino  fields having large Majorana masses via the type-I seesaw mechanism~\cite{Minkowski:1977sc, Mohapatra:1979ia, Yanagida:1979as,GellMann:1980vs,Glashow:1979nm, Schechter:1980gr}, or by the addition of an $SU(2)_L$-triplet scalar (or fermion) via the type-II~\cite{Schechter:1980gr, Mohapatra:1980yp, Lazarides:1980nt} (or type-III~\cite{Foot:1988aq}) seesaw, there are other interesting scenarios where small neutrino masses arise naturally as quantum corrections~\cite{Zee:1980ai,Cheng:1980qt,Zee:1985id,Babu:1988ki}.  These models of radiative neutrino masses, which we focus on in this paper, are more likely to be accessible for direct experimental tests at colliders. (For recent reviews on radiative neutrino mass models and constraints, see Refs.~\cite{Cai:2017jrq,Babu:2019mfe}.)  Here we show that the new particles that are present in these models to induce neutrino masses can also play an important role in explaining certain persistent experimental  anomalies, viz. the anomalous magnetic moment of the muon ($\Delta a_\mu$), and the lepton-flavor-universality violating decays of the $B$ meson ($R_{D^{(\star)}}$ and $R_{K^{(\star)}}$).   

There has been a long-standing discrepancy in the measured value of the anomalous magnetic moment of the muon by the E821 experiment at Brookhaven National Laboratory~\cite{Bennett:2006fi} 
and the SM theory prediction~\cite{Aoyama:2020ynm}, resulting in a value for $\Delta a_\mu \equiv a_\mu^{\rm exp} - a_\mu^{\rm SM} = 
(27.4 \pm 7.3) \times 10^{-10}$, 
which indicates a $3.7 \ \sigma$ discrepancy.  The muon $g-2$ experiment at Fermilab~\cite{Grange:2015fou} which is currently in the data accumulation stage, in conjunction with more precise calculations of the dominant hadronic vacuum polarization contribution~\cite{Blum:2018mom, Davies:2019efs, Gerardin:2019rua,Davier:2019can, Borsanyi:2020mff, Lehner:2020crt, Crivellin:2020zul}, is expected to settle in the near future whether this discrepancy is indeed due to new physics~\cite{Jegerlehner:2009ry}.  Meanwhile, it appears to be productive to envision TeV-scale new physics that can account for the observed anomaly. We shall pursue this line of thought here in the presence of an $R_2({\bf 3},{\bf 2},7/6)$  leptoquark (LQ) scalar (in the notation of Ref.~\cite{Buchmuller:1986iq}, where the numbers in parenthesis denote $SU(3)_c \times SU(2)_L \times U(1)_Y$ quantum numbers) that also takes part in radiative neutrino mass generation.

Independently, various anomalies have been reported in the semi-leptonic rare decays of the $B$-meson by BaBar~\cite{Lees:2012xj,Lees:2013uzd}, Belle~\cite{Huschle:2015rga,Hirose:2016wfn,Abdesselam:2016cgx} and LHCb~\cite{Aaij:2017tyk,Aaij:2017uff,Aaij:2019wad,Aaij:2017vbb} experiments.  The combined average ratio of branching ratios for the charged-current decay, $R_{D^{(\star)}} ={\rm BR} (B \rightarrow D^{(\star)} \tau \nu)/ {\rm BR}(B \rightarrow D^{(\star)} \ell \nu)$  (with $\ell = e,\, \mu$)~\cite{Lees:2012xj,Lees:2013uzd,Huschle:2015rga,Hirose:2016wfn,Abdesselam:2016cgx,Aaij:2017tyk,Aaij:2017uff} differs from the SM prediction~\cite{Amhis:2019ckw} by $1.4\, (2.7)\, \sigma$. The ratio of branching ratios for the neutral-current decay $R_{K^{(\star)}} = {\rm BR}(B \rightarrow K^{(\star)} \mu^+ \mu^-)/{\rm BR}(B \rightarrow K^{(\star)} e^+ e^-)$ ~\cite{Aaij:2019wad,Aaij:2017vbb} differs from the SM predictions~\cite{Bordone:2016gaq,Capdevila:2017bsm,Altmannshofer:2017yso,Aebischer:2019mlg} by $2.6\, (2.4)\,\sigma$ in the high-momentum range, while the discrepancy is $2.2\, \sigma$ in the lower-momentum range for $R_{K^{\star}}$.  These anomalies, while taken together, appear to suggest some lepton-flavor-universality violating new physics beyond the SM.  The most popular explanation of these anomalies is in terms of scalar LQs.  While the charged-current $B$-anomaly requires the relevant LQ to have a mass around 1 TeV, the neutral-current anomaly may be explained with a LQ that is somewhat heavier.  

A single scalar LQ solution to both $b\to s\ell^+\ell^-$ and $b\to c\tau\nu$ anomalies~\cite{Bauer:2015knc,Popov:2016fzr, Cai:2017wry,Popov:2019tyc} seems to be ruled out when such models are confronted with global fits to $b \to s\mu^+ \mu^-$ observables, as well as perturbativity constraints  and direct limits from the LHC~\cite{Angelescu:2018tyl} (see also Refs.~\cite{Buttazzo:2017ixm, Kumar:2018kmr}). 
The $R_{D^{(\star)}}$ anomaly may be explained by either an $S_1(\overline{\bf 3},{\bf 1},1/3)$ or an $R_2({\bf 3},{\bf 2},7/6)$ LQ, while the $R_{K^{(\star)}}$ anomaly may be explained in terms of an $S_3(\overline{\bf 3}, {\bf 3}, 1/3)$ LQ.\footnote{The $R_2$ LQ can also explain $R_{K^{(\star)}}$~\cite{Popov:2019tyc}, but only by modifying $b\to se^+e^-$ at tree-level and thus cannot explain the other $b\to s\mu^+\mu^-$ anomalies like $P_5'$~\cite{Aebischer:2019mlg}.}  Thus, in order to explain both $R_{D^{(\star)}}$ and $R_{K^{(\star)}}$ anomalies, there are two logical options: Addition of (i) $R_2({\bf 3},{\bf 2},7/6)$  and  $S_3(\overline{\bf 3}, {\bf 3}, 1/3)$ LQs, or (ii) $S_1(\overline{\bf 3},{\bf 1},1/3)$ and $S_3(\overline{\bf 3}, {\bf 3}, 1/3)$ LQs.  Among these options, we find it more compelling to adopt (i) as there is a direct connection with neutrino masses induced radiatively in this case, since both the LQs are essential to generate neutrino mass, unlike option (ii) where only one such LQ is sufficient, along with a color-sextet diquark to ensure lepton number violation~\cite{Kohda:2012sr}. 
Therefore, we adopt here a radiative neutrino mass model involving $R_2({\bf 3},{\bf 2},7/6)$ and $S_3(\overline{\bf 3}, {\bf 3}, 1/3)$ LQs, along with an isospin-3/2 Higgs field $\Delta({\bf 1},{\bf 4},3/2)$ which is needed to induce an $R_2$--$S_3^\star$ mixing that leads to lepton number violation, a requirement to generate  Majorana neutrino masses. 

We show by explicit construction that a model with $R_2({\bf 3},{\bf 2},7/6)$ and $S_3(\overline{\bf 3}, {\bf 3}, 1/3)$ LQs plus $\Delta({\bf 1},{\bf 4},3/2)$ Higgs field \cite{Popov:2019tyc}  can simultaneously explain the $R_{D^{(\star)}}$,  $R_{K^{(\star)}}$ and $\Delta a_\mu$ anomalies, while being consistent with all low-energy flavor constraints, as well as with the LHC limits. We propose a minimal Yukawa flavor structure that achieves these, while also providing excellent fits to neutrino oscillation parameters. We have also evaluated constraints from $\sqrt s=13$ TeV LHC data on the LQ Yukawa couplings to fermions of the first two families arising from non-resonant $pp \rightarrow \ell^+_i\ell^-_j$ processes mediated by $t$-channel exchange of LQs.  These limits on the couplings are found to be in the range $(0.15-0.36)$ for a 1 TeV LQ, which would preclude any solution of  $R_{D^{(\star)}}$ with new LQ-mediated decays of the $B$ meson involving $\nu_e$ or $\nu_\mu$, an {\it a priori} logical possibility. We also show that the $\Delta^{++}$ scalar from the $\Delta({\bf 1},{\bf 4},3/2)$  multiplet, which decays to same-sign dileptons for much of the parameter space, can be probed to masses as large as 1.1 TeV at the high-luminosity (HL) phase of the $\sqrt s=14$ TeV LHC with 3000 ${\rm fb}^{-1}$ of data, as it can be produced via strong interactions in the decay of $S_3^{4/3} \rightarrow (R_2^\star)^{-2/3} + \Delta^{++}$. The mass reach in this new mode is somewhat better than in the standard Drell-Yan (DY) channel.  We also find that the same Yukawa couplings responsible for the chirally-enhanced contribution to $\Delta a_\mu$ give rise to new contributions to the SM Higgs decays to muon and tau pairs, with the modifications to the corresponding branching ratios being at a few percent level with opposite signs, which could be tested at future hadron colliders, such as HL-LHC and FCC-hh.

There have been various attempts to explain radiative neutrino masses and a subset of the anomalies in $R_{D^{(\star)}}$, $R_{K^{(\star)}}$  and $\Delta a_\mu$ using scalar LQs. For instance, Ref.~\cite{Cai:2017wry} has studied neutrino masses, $R_{D^{(\star)}}$ and $\Delta a_\mu$, whereas  Refs.~\cite{Pas:2015hca, Cheung:2016fjo, Guo:2017gxp, Hati:2018fzc, Singirala:2018mio, Datta:2019tuj} address neutrino masses and $R_{K^{(\star)}}$. Similarly, Refs.~\cite{Dorsner:2017ufx, Saad:2020ucl} explain radiative neutrino masses, $R_{D^{(\star)}}$ and $R_{K^{(\star)}}$, while Ref.~\cite{Chen:2020jvl} explains neutrino masses and lepton $g-2$. In some cases such explanations are disconnected from neutrino mass generation, in the sense that removing certain particle from the model would still result in nonzero neutrino masses~\cite{Chen:2017hir, Saad:2020ihm}.  
Our approach here is similar in spirit to Ref.~\cite{ Bigaran:2019bqv}, which address all three anomalies, viz., $R_{D^{(\star)}}$, $R_{K^{(\star)}}$  and $\Delta a_\mu$, in the context of radiative neutrino masses; but unlike Ref.~\cite{Bigaran:2019bqv} we do not introduce new vector-like fermions into the model. In the model proposed here there is a close-knit connection between the $R_{D^{(\star)}}$ and $R_{K^{(\star)}}$  anomalies, $\Delta a_\mu$ and neutrino mass. In particular, neutrino mass generation requires all particles that play a role in explaining these anomalies.  Removing any new particle from the model would render the neutrino to be massless. For other models of radiative neutrino mass using LQ scalars, see Refs.~\cite{AristizabalSierra:2007nf, Babu:2010vp, Babu:2011vb, Angel:2013hla,Cai:2014kra,Cata:2019wbu}.

The rest of the paper is organized as follows. In Section~\ref{sec:model} we present the basic features of the model, including the Yukawa Lagrangian (cf.~Section~\ref{sec:yukawa}), scalar potential (cf.~Section~\ref{sec:scalar}), radiative neutrino mass generation mechanism (cf.~Section~\ref{sec:neu}) and a desired  texture for the Yukawa coupling matrices (cf.~Section~\ref{sec:texture}) consistent with flavor constraints that can explain the flavor anomalies. In Section~\ref{sec:anomalies} we discuss how the LQ scalars present in the model explain the  $R_{D^{(\star)}}$ and  $R_{K^{(\star)}}$ flavor anomalies. In Section~\ref{sec:gm2} we show how the $R_2$ LQ explains the $\Delta a_\mu$ anomaly. In this section, we also point out the difficulty in simultaneously explaining the electron $g-2$ (cf.~Section~\ref{sec:deltae}), as well as the model predictions for related processes, namely, Higgs decay to lepton pairs (cf.~Section~\ref{sec:higgs}) and muon electric dipole moment (cf.~Section~\ref{sec:edm}).  Section~\ref{sec:Constraint} summarizes the low-energy constraints on the LQ couplings and masses. Section~\ref{sec:LHCcon} analyzes the LHC constraints on the LQs. In Section~\ref{sec:results} we present our numerical results for two benchmark fits to the neutrino oscillation data that simultaneously explain  $R_{D^{(\star)}}$, $R_{K^{(\star)}}$ and $(g-2)_\mu$ anomalies, while being consistent with all the low-energy and LHC constraints. 
Section~\ref{sec:collider} further analyzes the collider phenomenology of the model relevant for the $\Delta^{++}$ scalar, and makes testable predictions for HL-LHC and future hadron colliders.  Our conclusions are given in Section~\ref{sec:conclusion}.

\section{The Model}
\label{sec:model}

The model proposed here aims to explain the $B$-physics anomalies $R_{D^{(\star)}}$ and  $R_{K^{(\star)}}$, as well as  the muon $(g-2)$ anomaly $\Delta a_\mu$, and at the same time induce small neutrino masses as radiative corrections. To this end, we choose the gauge symmetry and the fermionic content of the model to be identical to the SM, while the scalar sector is extended to include three new states, apart from the SM Higgs doublet $H$:
\begin{align}
& R_2 \ ({\bf 3}, {\bf 2},7/6) \ = \ \begin{pmatrix}
    \omega^{5/3} & \omega^{2/3} 
  \end{pmatrix}^T  , \quad \hspace{15mm}
 S_3 \ (\bar{\bf 3},{\bf 3},1/3) \ = \    \begin{pmatrix}
    \rho^{4/3} & \rho^{1/3} &  \rho^{-2/3} \\
  \end{pmatrix}^T  , \quad   \nonumber \\
& \Delta \ ({\bf 1},{\bf 4},3/2) \ = \   \begin{pmatrix}
    \Delta^{+++} & \Delta^{++}  & \Delta^{+} & \Delta^{0} \\
  \end{pmatrix}^T   , \quad 
  H \ ({\bf 1},{\bf 2},1/2) \ = \    \begin{pmatrix}
    H^+ & H^0  \\
  \end{pmatrix}^T  .
  \label{eq:field}
\end{align}
Here the numbers within brackets represent the transformation properties under the SM gauge group $SU(3)_c \times SU(2)_L \times U(1)_Y$.  The superscripts on various fields denote their respective electric charge $Q$  defined as $Q=I_3 + Y$, with $I_3$ being the third-component of $SU(2)_L$-isospin.  The $R_2$ and $S_3$ LQs are introduced to explain $R_{D^{(\star)}}$ and $R_{K^{(\star)}}$ anomalies respectively. The $R_2$ LQ also explains $\Delta a_\mu$ through a chirally-enhanced operator it induces, which is  proportional to the top quark mass. The $SU(2)_L$-quadruplet $\Delta$ field mixes $\omega^{2/3}$ from $R_2$ with $\bar{\rho}^{2/3}$ from $S_3^\star$ (the complex conjugate of $\rho^{-2/3}$), which is needed to generate Majorana neutrino masses radiatively.  This multiplet, with its characteristic triply-charged component, was introduced to generate tree-level neutrino masses from dimension ($d$)-7 effective operators in Ref.~\cite{Babu:2009aq}; here we use it for radiative mass generation, also via $d=7$ operators.
\subsection{Yukawa Couplings} \label{sec:yukawa}
In addition to the SM Yukawa couplings of the fermions involving the Higgs-doublet field $H$, the following 
Yukawa couplings of the $R_2$ and $S_3$ LQs are allowed in the model:\footnote{The field $\Delta$ has no Yukawa couplings with  fermions at the tree-level, but couples to the leptons at one-loop level (cf.~Eq.~\eqref{eq:YukD}).}
\begin{equation}
 \mathcal{L}_{Y}\  = \ \hat{f}_{a b} (u_a^{c T} C \psi_b^i) R_2^j \epsilon_{ij} - \hat{f}_{ab}^\prime (Q_a^{iT} C e_b^c) \widetilde{R}_2^j \epsilon_{ij} + \hat{y}_{ab} (Q_a^T C \tau_\alpha \psi_b) S_{3\alpha} - \hat{y}_{ab}^\prime (Q_a^T C \tau_\alpha Q_b) S_{3 \alpha}^{\star}+ \text{H.c.}
    \label{eq:mainlag}
\end{equation}
Here we have adopted a notation where all fermion fields are left-handed. 
$Q=(u \ d)^T$ and $\psi=(\nu \ e)^T$ are the SM quark and lepton doublets respectively, $\{i,j\}$ are $SU(2)$ indices, $\{a,b\}$ are flavor indices, $C$ is the charge conjugation matrix, $\epsilon_{ij}$ is the $SU(2)$ Levi-Civita tensor, $\widetilde{R}_2 = i \tau_2 R_2^{\star}$, and $\tau_\alpha$ (with $\alpha = 1,2,3$) are the Pauli matrices in the doublet representation of $SU(2)$. 
The color contraction is unique in each term, which is not shown.  It is to be noted that $S_3$ possesses both leptoquark and diquark couplings, as shown in Eq.~(\ref{eq:mainlag}), which would lead to potentially dangerous proton decay operators. Therefore, we set the diquark coupling $\hat{y}_{ab}^\prime$ to zero in Eq.~(\ref{eq:mainlag}), so that baryon number remains unbroken.  This is achieved by assigning baryon number $B = -1/3$ to $S_3$ and $R_2^\star$, along with $B= 1/3$ for quarks and $-1/3$ for anti-quarks, and $0$ for leptons and anti-leptons.


We redefine fields to go from the flavor basis ($u,\,d,\,e)$ to the mass eigenstates ($u^0,\,d^0,\,e^0)$ for the charged fermions (and similarly for the ($u^c,\,d^c,\,e^c$) fields) via the following unitary rotations in family space: 
\begin{eqnarray}
   &&u = V_u\, u^0,  \hspace{5mm} d = V_d \, d^0, \hspace{5mm} e = V_e\, e^0, \hspace{5mm} \nu = V_e\, \nu^0 \, ,  \nonumber\\
     &&u^c = V_{u^c} \,u^{c\, 0},   \hspace{5mm}  d^c = V_{d^c}\, d^{c\, 0},  \hspace{5mm}  e^c = V_{e^c}\, e^{c \,0} \, .
     \label{basisU}
\end{eqnarray}
The Cabibbo-Kobayashi-Maskawa (CKM) quark mixing matrix $V_{\rm CKM}$ is generated in the process and is given by 
\begin{equation}
    V \ = \  V_{u}^\dagger V_d \ = \ P \, {V}_{\text{CKM}} Q \, ,
    \label{eq:vckm}
\end{equation}
where  $P$, $Q$ are diagonal phase matrices which are unphysical in the SM, but  become physical in non-SM interactions, such as the ones involving the LQs. These phases will have an effect on $CP$-violating observables, such as the muon electric dipole moment (EDM), see Section~\ref{sec:edm}. Note that the unitary rotation on the neutrino fields in Eq.~(\ref{basisU}) is the same as for left-handed lepton fields $e$, and therefore no Pontecorvo-Maki-Nakagawa-Sakata (PMNS) mixing in the charged weak-current interactions of leptons is induced at this stage.  For explaining the anomalies in $B$-decays and in muon $g-2$, there is no need to go to the mass eigenstates of the neutrinos; the distinction between the mass and flavor eigenstates will only affect neutrino oscillation phenomenology.  
For convenience, we also redefine the Yukawa couplings as follows:  
\begin{equation}
    V_{u^c}^T \hat{f} V_e \ \equiv \ f, \hspace{5mm} V_{u}^T \hat{f}^\prime V_{e^c} \ \equiv \ V^T f^\prime,  \hspace{5mm} V_{u}^T \hat{y} V_e \ \equiv \ V^T y \, .
\end{equation}
Eq.~\eqref{eq:mainlag} can now be written in terms of mass eigenstate fermions (except for neutrinos which are still flavor eigenstates) and the redefined Yukawa couplings as
\begin{align}
\mathcal{L}_{Y}&  \ =  \ u^{cT} C f \nu \omega^{2/3} - u^{cT} C f e \omega^{5/3} + u^T C (V^\star f^\prime) e^c \omega^{-5/3} + d^T C f^\prime e^c \omega^{-2/3} \nonumber \\
 & \qquad - u^T C (V^\star y) \nu \rho^{-2/3} + u^T C (V^\star y) e \frac{\rho^{1/3}}{\sqrt{2}}+ d^T C  y \nu \frac{\rho^{1/3}}{\sqrt{2}}+ d^T C  y e \rho^{4/3} + \text{H.c.}
 \label{eq:expandLag}
\end{align}
Here we have dropped the superscript $^0$ in the labeling of mass eigenstates. In the discussions that follow, the quark and lepton fields are to be identified as mass eigenstates.  Note that the Yukawa coupling matrices $f'$ and $y$, which respectively appear in the $d-e^c$ and $d-e$ couplings, also appear in the $u-e^c$  and $u-e$ couplings, along with the generalized CKM matrix $V$.  Any texture adopted for $f'$ and $y$ should therefore be consistent with flavor violation in both down-type and up-type quark sectors. The flavor indices $i$ and $j$ in $f_{ij}$ (and similarly for $f'$ and $y$) refer to the quark flavor and the lepton flavor respectively.
We shall make use of these interactions in explaining the $B$-anomalies, $\Delta a_\mu$ and radiative neutrino masses.

\subsection{Scalar Sector} \label{sec:scalar}
The most general renormalizable Higgs potential involving $H$, $R_2$, $S_3$ and $\Delta$ is given by:
\begin{align}
    V  \ = \ &  - \mu_H^2 H^\dagger H 
    + \mu_R^2 R_2^\dagger R_2 
    + \mu_{S}^2 S_3^\dagger S_3
    + \mu_\Delta^2 \Delta^\dagger \Delta + \frac{\lambda_H}{2} (H^\dagger H)^2 
    + \frac{\lambda_R}{2} (R_2^\dagger R_2)^2
     \nonumber\\
   & 
     + \frac{\widetilde{\lambda}_R}{2} (R_2^{\dagger \alpha} R_{2 \beta}) (R_2^{\dagger \beta} R_{2 \alpha})
     + \frac{\lambda_S}{2} (S_3^\dagger S_3)^2 +  \frac{\lambda_S'}{2} (S_3^\dagger T_a S_3)^2 
     + \frac{\widetilde{\lambda}_S}{2} (S_3^{\dagger^\alpha} S_{3 \beta})(S_3^{\dagger \beta} S_{3 \alpha})
     \nonumber\\
     &
     +  \frac{\widetilde{\lambda'}_S}{2} (S_3^{\dagger^\alpha} T_a S_{3 \beta})(S_3^{\dagger \beta} T_a S_{3 \alpha})
     + \frac{\lambda_\Delta}{2} (\Delta^\dagger \Delta)^2 + \frac{\lambda'_\Delta}{2} (\Delta^\dagger T'_a \Delta)^2
      +\lambda_{HR}(H^\dagger H) (R_2^\dagger R_2)
    \nonumber\\
   & 
   +\lambda'_{HR}(H^\dagger \tau_a H) (R_2^\dagger \tau_a R_2)
    + \lambda_{HS}(H^\dagger H) (S_3^\dagger S_3)  + \lambda'_{HS} (H^\dagger \tau_a H)(S_3^\dagger T_a S_3) 
    \nonumber\\
   &
    + \lambda_{H\Delta} (H^\dagger H) (\Delta^\dagger \Delta)
   + \lambda'_{H\Delta} (H^\dagger \tau_a H) (\Delta^\dagger T'_a \Delta) 
   +  \lambda_{RS}(R_2^\dagger R_2)  (S_3^\dagger S_3) 
   \nonumber  \\
    & 
    +  \lambda'_{RS}(R_2^\dagger \tau_a R_2)  (S_3^\dagger T_a S_3) 
    +  \widetilde{\lambda}_{RS}(R_2^{\dagger \alpha} R_{2 \beta})  (S_3^{\dagger \alpha} S_{3 \beta}) +  \widetilde{\lambda}'_{RS}(R_2^{\dagger \alpha} \tau_a R_{2\beta})  (S_3^{\dagger \alpha} T_a S_{3 \beta}) 
    \nonumber  \\
   & 
    + \lambda_{R\Delta} (R_2^\dagger R_2) (\Delta^\dagger \Delta) 
    + \lambda'_{R\Delta} (R_2^\dagger \tau_a R_2) (\Delta^\dagger T'_a \Delta) 
    +\lambda_{S\Delta} (S_3^\dagger S_3) (\Delta^\dagger \Delta) 
   \nonumber\\
   &
   +\lambda'_{S\Delta} (S_3^\dagger T_a S_3) (\Delta^\dagger T'_a \Delta) + \lambda_{S\Delta}^{''} (S_3^\dagger T_a T_b S_3) (\Delta^\dagger T'_a T'_b \Delta) 
    \nonumber\\
   & 
   + \Big(\mu \Delta^{\star^{ijk}} R_{2_i} S_{3_{jk}} + \lambda_{RHS^2} 
   R_2^{i*}S_{3ij}S_{3kl}H_m\epsilon^{jk}\epsilon^{lm}
         + \lambda_{\Delta H^3} \Delta^{\star^{ijk}} H_iH_jH_k + \text{H.c.}\Big) \, ~. \label{eq:pot}
\end{align}
Here $\{i,j\}$ are $SU(2)_L$ indices, $\{\alpha,\beta\}$ are $SU(3)_c$ indices, $\tau_a$ are the Pauli matrices, and $T_a$, $T'_a$ (with $a=1,2,3$) are the normalized generators of $SU(2)$ in the triplet and quadruplet representations, respectively.\footnote{This potential differs considerably from the one given in Ref. \cite{Popov:2019tyc}, which is missing many terms.} Color-singlet contractions not shown explicitly are to be assumed among two colored fields within the same bracket.  For example, the $\lambda'_{RS}$ term has the color contraction $(R_2^{\dagger \alpha} \tau_a R_{2\alpha})  (S_3^{\dagger \beta} T_a S_{3 \beta})$.  
Here $S_{3_{ij}}$ and $\Delta^{ijk}$ are the completely symmetric rank-2 and rank-3 tensors of $SU(2)$,  with their components related to those given in Eq. (\ref{eq:field}) as:
\begin{eqnarray}
   && S_{3_{11}} \ = \ \rho^{4/3} \, ,   \hspace{5mm}  S_{3_{12}} \ = \  \frac{\rho^{1/3}}{\sqrt{2}} \, , \hspace{5mm} S_{3_{22}} \ = \ \rho^{-2/3} \, , \nonumber \\
  &&   \Delta_{111} \ = \ \Delta^{+++} \, ,  \hspace{5mm} \Delta_{112} \ = \ \frac{\Delta^{++}}{\sqrt{3}}\, ,  \hspace{5mm} \Delta_{122} \ = \ \frac{\Delta^{+}}{\sqrt{3}}\, ,   \hspace{5mm} \Delta_{222} \ = \ \Delta^{0} \, .
\end{eqnarray}

The presence of the quartic coupling with coefficient $\lambda_{\Delta H^3}$ in Eq.~\eqref{eq:pot} will  induce a vacuum expectation value (VEV) for the neutral component of $\Delta$, even when $\mu^2_\Delta> 0$ is chosen.  The cubic coupling with coefficient $\mu$ would then lead to mixing of $\omega^{2/3}$ and $\bar{\rho}^{2/3}$ components of $R_2$ and $S_3^\star$ LQ fields. Such a mixing is required to realize lepton number violation and to generate neutrino masses.  We shall be interested in the choice $\mu_H^2> 0$ (which leads to electroweak symmetry breaking), and  $\mu_{R}^2, \, \mu_{S}^2 > 0$ (so that electric charge and color remain unbroken), and $\mu_\Delta^2 > 0$ -- so that $\Delta^0$ acquires only an induced VEV.  To ensure that this desired vacuum is indeed a local minimum of the potential, we now proceed to derive the masses of all scalars in the model.

\subsubsection{Scalar Masses}
We denote the VEVs of $H^0$ and $\Delta^0$ fields as
\begin{equation}
    \left\langle H^0 \right \rangle = \frac{v}{\sqrt{2}} \, , \qquad \left\langle \Delta^0 \right \rangle = \frac{v_\Delta}{\sqrt{2}} \, ,
    \label{VEV}
\end{equation}
with $(v^2 +3\, v_\Delta^2) \simeq (246.2~{\rm GeV})^2$ determined from the Fermi constant $G_F$.  
While $v$ can be taken to be real by a gauge rotation, $v_\Delta$ is complex in general. However, all the complex-valued couplings of the potential, i.e. terms in the last line of Eq.~\eqref{eq:pot}, can be made real by field redefinitions, which we adopt, and consequently minimization of the potential would make $v_\Delta$ real as well.

We obtain the following  conditions for the potential to be an extremum around the VEVs of Eq. (\ref{VEV}), assuming that $v \neq 0$:
\begin{eqnarray}
     - \mu_H^2 + \frac{1}{2} \lambda_H v^2 + \frac{v_\Delta}{4} \left( 6 \lambda_{\Delta H^3} v + 2 \lambda_{H\Delta} v_\Delta + 3 \lambda'_{H\Delta} v_\Delta  \right)  & \ = \ & 0 \, ,  \\
    \mu_\Delta^2 v_\Delta + \frac{1}{2} \lambda_{\Delta H^3} v^3 + \left(\frac{1}{2} \lambda_{H\Delta} + \frac{3}{4} \lambda'_{H\Delta}\right)v^2 v_\Delta  + \left(\frac{1}{2} \lambda_\Delta + \frac{9}{8} \lambda'_\Delta \right) v_\Delta^3 & \ = \ & 0 \, .
    \label{eq:mincond}
\end{eqnarray}
We eliminate $\mu_H^2$ and $\mu_\Delta^2$ using these two  conditions. To derive the scalar mass spectrum, we construct the mass matrices from the bilinear terms resulting from expanding the potential in Eq.~\eqref{eq:pot} around the VEVs $v$ and $v_\Delta$. 

The $2 \times 2$ mass matrix involving the mixing of the charge-$2/3$ LQs in the basis $( \omega^{2/3}, \ \bar{\rho}^{2/3})$ is found to be:
\begin{equation}
M_{2/3}^{2} \ = \ \left(\begin{array}{cc}
m_{\omega^{2/3}}^2~~~~ & \mu \frac{v_{\Delta}}{\sqrt{2}} \, , \\
\mu \frac{v_{\Delta}}{\sqrt{2}} ~~~~& m_{\rho^{2/3}}^2
\end{array}\right)~,
\end{equation}
where
\begin{align}
    m^2_{\omega^{2/3}} \ & = \ \mu_{R}^{2} + \frac{v^2}{2} (\lambda_{H R} + \lambda'_{H R} )  + \frac{v_\Delta^2}{4} (2\lambda_{R \Delta} + 3 \lambda'_{R \Delta}) \, , \label{eq:omega23} \\
    m^2_{\rho^{2/3}} \ & = \ \mu_{S}^{2} + \frac{v^2}{2} ( \lambda_{H S} + \lambda'_{H S} ) + \frac{v_\Delta^2}{8} (4 \lambda_{S \Delta} + 6 \lambda'_{S \Delta} + 9 \lambda''_{S \Delta} ) \, .
\end{align}
The mass eigenstates denoted as $X_{1,2}$ are given by
\begin{align}
   X_1 \ & = \ \cos\varphi \, \omega^{2/3} + \sin\varphi \, \bar{\rho}^{2/3} \, ,  \\
   X_2 \ & = \ -\sin\varphi \, \omega^{2/3} + \cos\varphi \, \bar{\rho}^{2/3} \,,
\end{align}
where the mixing angle $\varphi$ is defined as
\begin{equation}
  \tan 2\varphi  \ = \ \frac{\sqrt{2} v_\Delta \mu}{(m_{\omega^{2/3}}^2 - m_{\rho^{2/3}}^2)} \, .
  \label{eq:mixangle}
\end{equation}
The mass eigenvalues of the charge-2/3 LQ fields are then given as
\begin{equation}
    m_{X_1,X_2}^2 \ = \  \frac{1}{2} \Big[ m_{\omega^{2/3}}^2 + m_{\rho^{2/3}}^2 \pm \sqrt{(m_{\omega^{2/3}}^2 - m_{\rho^{2/3}}^2)^2 + 2 \mu^{2} v_\Delta^2}\Big].
    \label{eq:phymass}
\end{equation}
The masses for the remaining LQ components $(\omega^{5/3}, \ \rho^{1/3}, \ \rho^{4/3})$ are obtained as follows:
\begin{align}
   m_{\omega^{5/3}}^2 \ & = \ \mu_{R}^{2} + \frac{v^2}{2} ( \lambda_{H R} - \lambda'_{H R} )  + \frac{v_\Delta^2}{4} (2\lambda_{R \Delta} - 3 \lambda'_{R \Delta}) \, ,  \label{eq:omega53}\\
   m_{\rho^{1/3}}^2 \ & = \ \ \mu_{S}^{2} + \frac{v^2}{2} \lambda_{H S}  + \frac{v_\Delta^2}{4} (2 \lambda_{S \Delta}  + 3 \lambda''_{S\Delta} ) \, , \\
   m_{\rho^{4/3}}^2 \ & = \ \mu_{S}^{2} + \frac{v^2}{2} ( \lambda_{H S} - \lambda'_{H S} ) + \frac{v_\Delta^2}{8} (4 \lambda_{S \Delta} - 6 \lambda'_{S \Delta} + 15 \lambda''_{S \Delta} ) \, .
   \end{align}

   As for the $\Delta$ fields, the masses of the triply and doubly-charged components are given by
   \begin{align}
   m_{\Delta^{+++}}^2 \ & = \ - \frac{3 \lambda'_{H\Delta} v^2}{2} - \frac{9 \lambda'_\Delta v_\Delta^2}{4} - \frac{\lambda_{\Delta H^3} v^3}{ 2 v_\Delta} \, , \\ 
   m_{\Delta^{++}}^2 \ & = \ -  \lambda'_{H\Delta} v^2 - \frac{3 \lambda'_\Delta v_\Delta^2}{2} - \frac{\lambda_{\Delta H^3} v^3}{ 2 v_\Delta} \, .
\end{align}
The singly-charged components of $H$ and  $\Delta$ will mix, with a mass matrix given by:  
\begin{equation}
M_{+}^{2} \ = \ \frac{1}{2}\left(\lambda_{\Delta H^3} v + \lambda'_{H\Delta} v_\Delta\right) \left(\begin{array}{cc}
-3 v_\Delta & \, \, \, \,  \sqrt{3} v \,  \\
\sqrt{3} v  & \, \, \, \,  -\frac{v^2}{v_\Delta}
\end{array}\right)\, .
\end{equation}
One combination of $(H^\pm,\,\Delta^\pm)$ fields is the Goldstone boson ($G^\pm$) eaten up by the $W^\pm$ gauge boson, while the other combination ($\delta^\pm$) is a physical charged Higgs field.  These fields are
\begin{align}
    G^+ \ = \ \frac{v H^+ + \sqrt{3} v_\Delta \Delta^+}{\sqrt{v^2+ 3 v_\Delta^2} } \, , \quad \quad \quad
    \delta^+ \ = \ \frac{\sqrt{3} v_\Delta H^+ - v \Delta^+}{\sqrt{v^2+ 3 v_\Delta^2} } \, ,
\end{align}
with the mass of $\delta^+$ given by\footnote{
In the limit $v_\Delta\ll v$, the physical $\delta^+$ field is nearly identical to the original $\Delta^+$ field. So we will use the same notation for $m_{\delta^+}$ and $m_{\Delta^+}$.}
\begin{equation}
    m^2_{\delta^+} \ = \ - \frac{\lambda'_{H \Delta}\left (v^2 + 3 v_\Delta^2\right)}{2} - \frac{\lambda_{\Delta H^3} \left (v^3 + 3 v_\Delta^2 v\right)}{2 v_\Delta} \, .
\end{equation}

The neutral $CP$-even scalars do not mix with the $CP$-odd scalars, since all couplings and VEVs are real. The mass matrix for the $CP$-even states in the basis 
$({\rm Re} \ H^0,\,{\rm Re} \ \Delta^0 )$
reads as:
\begin{equation}
M_{\text{even}}^{2} \ = \ \left(\begin{array}{cc}
\lambda_H v^2 + \frac{3}{2} \lambda_{\Delta H^3} v v_\Delta & \frac{v}{2} \big[3 \lambda_{\Delta H^3} v + \left(2 \lambda_{H \Delta} + 3 \lambda'_{H \Delta}\right)v_\Delta \big] \\
\frac{v}{2} \big[3 \lambda_{\Delta H^3} v + (2 \lambda_{H \Delta} + 3 \lambda'_{H \Delta})v_\Delta \big] & -\frac{\lambda_{\Delta H^3} v^3}{2 v_\Delta} + \left( \lambda_\Delta + \frac{9}{4} \lambda'_\Delta\right) v_\Delta^2
\end{array}\right) \, . 
\end{equation}
The resulting mass eigenvalues are given by
\begin{align}
    m_{h,H}^2  \ = \  \frac{1}{2} \Bigg[ \lambda_H v^2 + (\lambda_\Delta + \frac{9}{4} \lambda'_\Delta) v_\Delta^2 - \frac{\lambda_{\Delta H^3} v (v^2- 3 v_\Delta^2)}{2 v_\Delta}  \pm \sqrt{A} \Bigg] 
\end{align}
where
\begin{align}
    A \ = \ & \bigg\{\lambda_H v^2 - \left(\lambda_\Delta + \frac{9}{4} \lambda'_\Delta\right) v_\Delta^2  + \frac{\lambda_{\Delta H^3} v (v^2+3 v_\Delta^2)}{2 v_\Delta} \bigg\}^2 \nonumber \\
    & \qquad + v^2 \left[3 \lambda_{\Delta H^3} v + (2 \lambda_{H \Delta} + 3 \lambda'_{H \Delta})v_\Delta \right]^2    \, .
\end{align}
The corresponding mass eigenstates are given by
\begin{align}
    h \ &= \  \cos\alpha \,{\rm Re}(H^{0}) + \sin\alpha \,{\rm Re}(\Delta^{0}) \, , \\
    H \ &= \  -\sin\alpha\, {\rm Re}(H^{0}) + \cos\alpha\, {\rm Re}(\Delta^{0}) \, , 
\end{align}
with
\begin{equation}
    \sin 2\alpha \ = \ \frac{ v \big[3 \lambda_{\Delta H^3} v + (2 \lambda_{H \Delta} + 3 \lambda'_{H \Delta})v_\Delta \big]}{(m_H^2 - m_h^2)} \, .
\end{equation}
The field $h$ is to be identified as the SM-like Higgs boson of mass 125 GeV.  

Similarly, the $CP$-odd scalar mass matrix, in the basis $({\rm Im} \ H^0,\,{\rm Im}\ \Delta^0)$ is given by
\begin{equation}
M_{\text{odd}}^{2} \ = \ \frac{1}{2} \lambda_{\Delta H^3} v \left(\begin{array}{cc}
-9  v_\Delta & \, \, \, \,   3  v \\
3  v  & \, \, \, \,   -\frac{  v^2}{ v_\Delta} 
\end{array}\right) \, .
\end{equation}
We identify the Goldstone mode $G^0$ eaten up by the $Z^0$ gauge boson and the physical pseudoscalar Higgs boson $A^0$ as 
\begin{align}
    G^0 \ = \ \frac{v \,{\rm Im}(H^{0}) + 3 v_\Delta {\rm Im}(\Delta^{0})}{\sqrt{v^2+ 9 v_\Delta^2} } \, , \quad \quad \quad
    A^0 \ = \ \frac{3 v_\Delta {\rm Im}(H^{0}) - v\, {\rm Im}(\Delta^{0})}{\sqrt{v^2+ 9 v_\Delta^2} } \, ,
\end{align}
with the mass of $A^0$ given by
\begin{equation}
    m_A^2 \ = \ -\frac{\lambda_{\Delta H^3} v^3}{ 2 v_\Delta} - \frac{9 \lambda_{\Delta H^3} v v_\Delta}{2} \, .
\end{equation}

The VEV $v_\Delta$ must obey the condition $v_\Delta \ll v$ from electroweak $T$-parameter constraint.  In presence of $v_\Delta$, the electroweak $\rho$ parameter deviates from unity at tree-level, with the deviation given by \cite{Babu:2009aq} 
\begin{equation}
     \delta \rho\ \simeq \  - 6 \frac{ v_\Delta^2}{v^2}~.
    \label{rho}
\end{equation}
Although there are also loop-induced contributions to $\delta \rho$, arising from the mass splittings among components of $\Delta,\,R_2,\,S_3$ fields which typically have the opposite sign compared to Eq. (\ref{rho}), we assume that there is no precise cancellation between these two types of contributions. A parameter $\rho_0$,  defined as 
\begin{equation}
\rho_0 \ = \ \frac{m_W^2}{m_Z^2 \hat{c}_Z^2 \hat{\rho}} 
\end{equation}
(where $\hat{c}_Z\equiv \cos\theta_W(m_Z)$ in the $\overline{\rm MS}$ scheme, $\theta_W$ being the weak mixing angle, and $\hat{\rho}$ includes leading radiative corrections from the SM), has a global average  $\rho_0 = 1.00038 \pm 0.00020$ \cite{Zyla:2020zbs}.  Eq. (\ref{rho}) can be compared to this global value, with $\rho_0 = 1$ in the SM, which sets a limit of $|v_\Delta| \leq 1.49$ GeV, allowing for $3\, \sigma$ variation, and ignoring loop contributions proportional to mass splitting among multiplets.

In the approximation $|v_\Delta| \ll |v|$, one can solve for $v_\Delta$ from Eq.~(\ref{eq:mincond}), to get 
\begin{align}
   v_\Delta \ \simeq \ -\frac{\lambda_{\Delta H^3} v^3}{2\mu^2_\Delta} \, . 
   \label{eq:vdelta}
\end{align}
  Substituting this into the masses of the Higgs quadruplet components, we obtain~\cite{Babu:2009aq} 
\begin{equation}
    m_{\Delta_i}^2 \ \simeq \ \mu_\Delta^2 - q_i \frac{ \lambda'_{H\Delta} v^2}{2} \, ,
    \label{eq:mdelta}
\end{equation}
where $q_i$ is the (non-negative) electric charge of the component  field $\Delta_i$ (with $i=1,2,3,4$ denoting the four components of $\Delta$ given in Eq.~\eqref{eq:field}). We note that there are two possibilities for mass ordering among these components, depending on the sign of the quartic coupling $\lambda'_{H\Delta}$, with $m_{\Delta^{+++}}$ being either the heaviest or the lightest member. Phenomenology of these scenarios has been studied extensively in Refs.~\cite{Babu:2009aq, Bambhaniya:2013yca,  Ghosh:2017jbw, Ghosh:2018drw}.

By choosing all the bare mass parameters  $\mu_X^2$ (for $X = H, R_2, S_3, \Delta$) in Eq. (\ref{eq:pot})  to be positive,  and the quartic coupling $\lambda_H$ to be positive, the desired minimum can be shown to be a local minimum, as long as the masses of $\Delta,\,R_2,\, S_3$ are well above $v \simeq 246$ GeV. To verify that this minimum is also the absolute minimum of the potential for some range of parameters, further work has to be done, which is beyond the scope of this paper.  Since none of the quartic couplings, except for $\lambda_{\Delta H^3}$, plays any crucial role for our analysis, it appears possible to achieve this condition.  Similarly, there is enough freedom to choose the quartic couplings so that the potential remains bounded from below. We shall discuss below a set of necessary conditions for the potential to be bounded, which will find application in Section~\ref{sec:higgs} in the discussion of modified rates for $h \rightarrow \ell^+ \ell^-$ in the model.

\subsubsection{Necessary Conditions for Boundedness of the Potential}
\label{sec:bounded}

While the full set of necessary and sufficient conditions  on the quartic couplings of Eq. (\ref{eq:pot})  for the Higgs potential to be bounded from below is not easily tractable, certain necessary conditions of phenomenological interest (cf.~Section~\ref{sec:higgs}) can be analyzed analytically.  We focus on the quartic couplings involving only the $H$ and $R_2$ fields, which will turn out to be of phenomenological interest.  With $SU(2)_L$ and $SU(3)_c$ rotations, these fields can be brought to the form
\begin{eqnarray}
H \ = \ \left(\begin{matrix} 0 \cr v \end{matrix}\right) , \qquad 
R_2 \ = \ \left(\begin{matrix} x &~~~ 0 & ~~~0 \cr y \sin\alpha e^{i \phi} & ~~~ y \cos\alpha & ~~~ 0  \end{matrix}  \right) \, ,  \end{eqnarray}
where in $R_2$, the color indices run horizontally.  Here $v, \,x,\,y$ can be taken to be real. The quartic terms $V^{(4)}(H,R_2)$ can be then written as
\begin{eqnarray}
V^{(4)}(H,\,R_2) \ = \ \frac{1}{2} \left( \begin{matrix} v^2 & x^2 & y^2 \end{matrix} \right) {\bf \hat{\lambda}} \left( \begin{matrix} v^2 \cr x^2 \cr y^2 \end{matrix}\right),
\end{eqnarray}
where ${\bf \hat{\lambda}}$ is defined as
\begin{eqnarray}
{\bf \hat{\lambda}} \ = \ \left( \begin{matrix} \lambda_H & \lambda_{HR} - \lambda'_{HR} & \lambda_{HR} + \lambda'_{HR} \cr
\lambda_{HR} - \lambda'_{HR} & \lambda_R + \tilde{\lambda}_R & \lambda_R + \tilde{\lambda}_R \sin^2\alpha \cr
 \lambda_{HR} + \lambda'_{HR} & \lambda_R + \tilde{\lambda}_R \sin^2\alpha & \lambda_R + \tilde{\lambda}_R
\end{matrix} \right)~.
\end{eqnarray}
The necessary and sufficient conditions for boundedness of this potential can now be derived from the co-positivity of real symmetric matrices \cite{Hadeler1983OnCM, Klimenko:1984qx, Babu:2014kca}:
\begin{eqnarray}
   &~& \lambda_H \ \geq  \ 0 \, ,  \\ 
   &~& \lambda_R+\tilde{\lambda}_R \ \geq \ 0 \, , \\
&~&    \lambda_{HR} - \lambda'_{HR} \ \geq \  -\sqrt{\lambda_H(\lambda_R+\tilde{\lambda}_R)} \, ,\label{eq:bound} \\ 
&~&\lambda_{HR} + \lambda'_{HR} \ \geq \  -\sqrt{\lambda_H(\lambda_R+\tilde{\lambda}_R)}\, , \\
  &~&  \lambda_R + \tilde{\lambda}_R \sin^2\alpha  \ \geq \ -(\lambda_R +\tilde{\lambda}_R) \, ,  \\
&~&    (\lambda_{HR}-\lambda'_{HR})\sqrt{\lambda_R+\tilde{\lambda}_R}+(\lambda_R+ \tilde{\lambda}_R \sin^2\alpha)\sqrt{\lambda_H} \nonumber \\ && \quad  + (\lambda_{HR} + \lambda'_{HR})\sqrt{\lambda_R+\tilde{\lambda}_R} 
 + (\lambda_R + \tilde{\lambda}_R) \sqrt{\lambda_H}  \ \geq \ 0 ~~~ {\rm or}~~~{\rm det}({\bf \hat{\lambda}})  \ \geq \ 0~.
\end{eqnarray}
These conditions should hold for any value of the angle $\alpha$.  

Note that from Eq. (\ref{eq:bound}) it follows that if $(\lambda_{HR}-\lambda'_{HR})$ is negative, its magnitude cannot exceed about 1.33, if we demand that none of the quartic couplings should exceed $\sqrt{4\pi}$ in magnitude from perturbativity considerations, and using the fact that $\lambda_H \simeq 0.25$ is fixed from the mass of $h$, $m_h = 125$ GeV.   This result will be used in the calculation of the modified Higgs branching ratio  $h \rightarrow \ell^+ \ell^-$ in Section~\ref{sec:higgs}.

\subsection{Radiative Neutrino Masses} \label{sec:neu}

\begin{figure}[t!]
    \centering
    \includegraphics[scale=0.43]{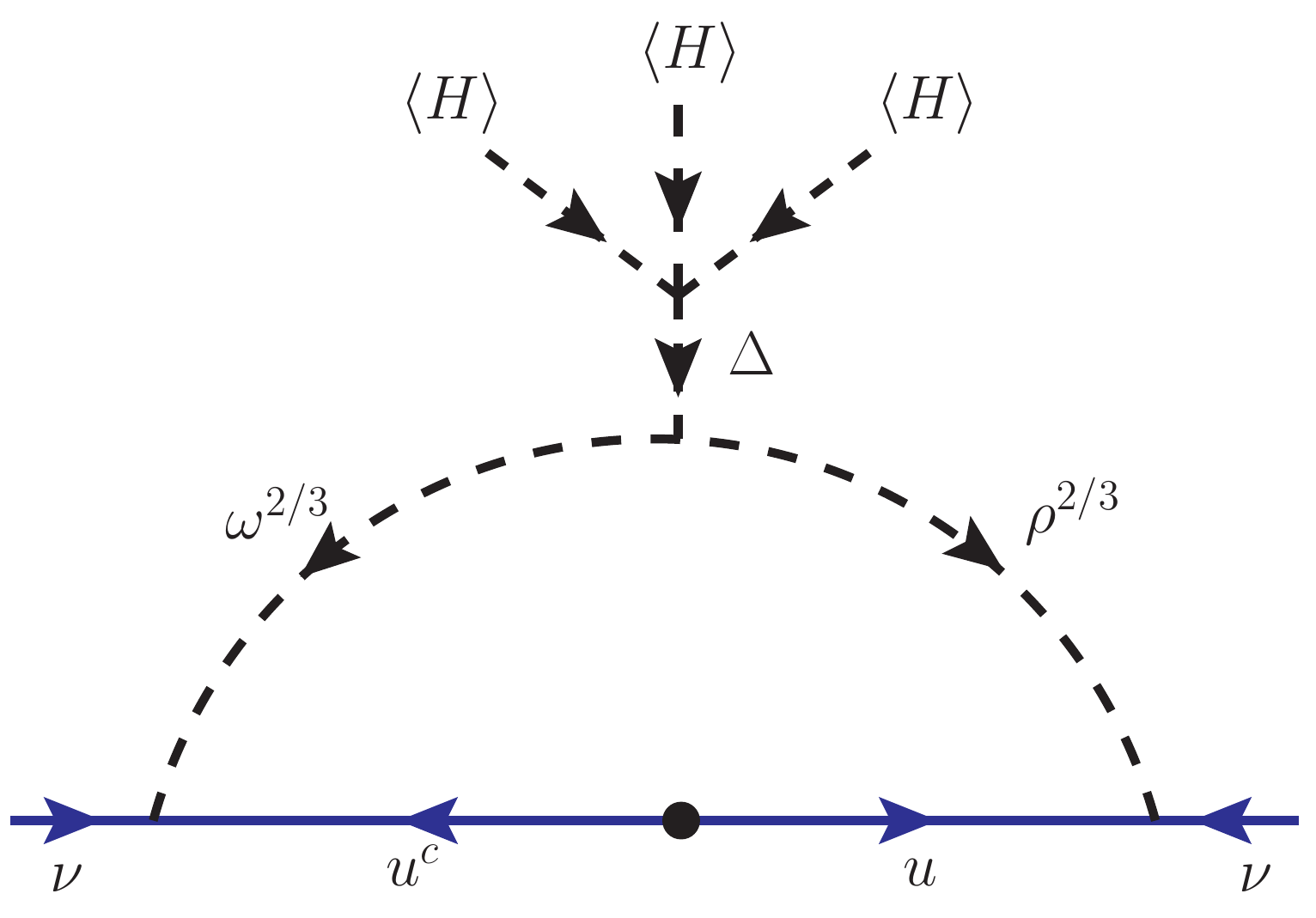} \hspace{10mm}
    \includegraphics[scale = 0.43]{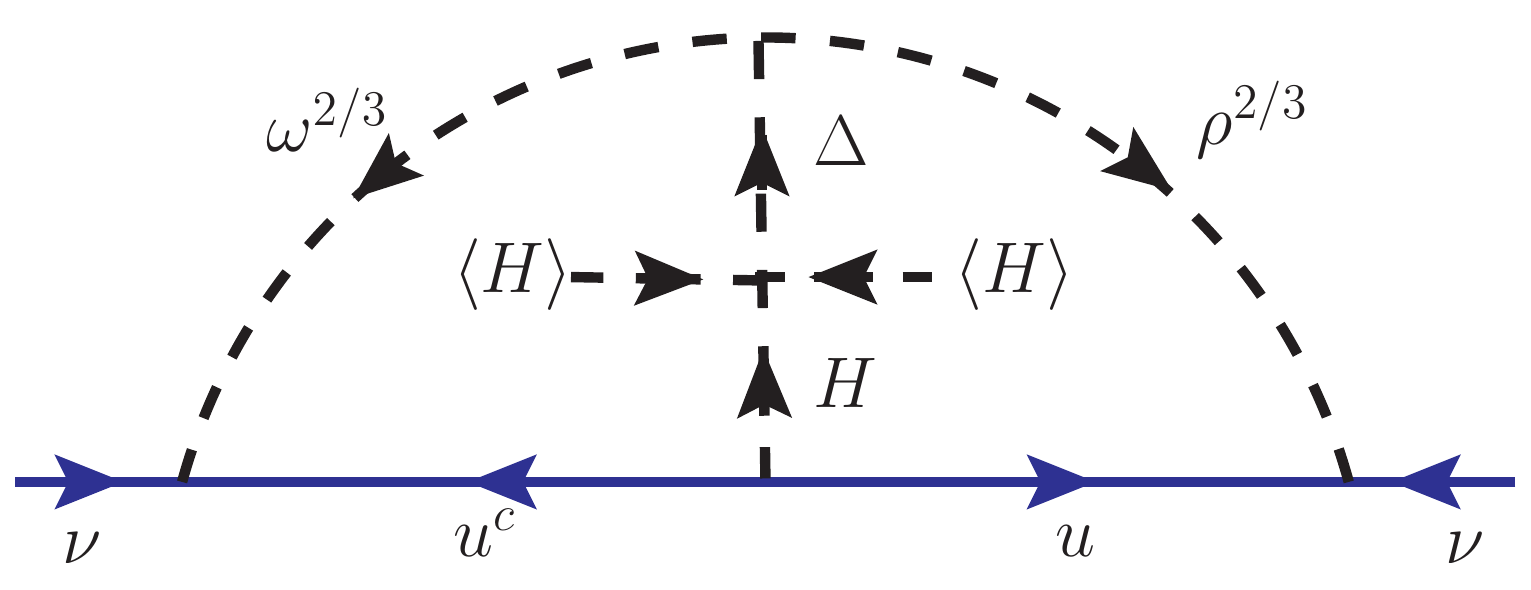}
    \caption{ Feynman diagram generating neutrino masses through the exchange of LQs in the model. The one-loop diagram shown is the leading contribution, while the two-loop diagram can be important. The dot (•) on the SM fermion line in the one-loop diagram indicates mass insertion arising from the SM Higgs doublet VEV. There is a second set of diagrams obtained by reversing the arrows on the internal particles.}
    \label{fig:numass}
\end{figure}

Neutrino masses are zero at the tree-level in the model.  However, since lepton number is not conserved, nonzero $M_\nu$ will be induced as quantum corrections.  The leading diagrams generating $M_\nu$ are shown in Fig. \ref{fig:numass}, mediated by the charge-2/3 LQs.  
The Yukawa couplings in Eq.~\eqref{eq:expandLag}, together with the $\Delta^\star R_2 S_3$ trilinear  term and the $\Delta^\star HHH$ quartic term  in the scalar potential~\eqref{eq:pot}, guarantee lepton number violation. 
These interactions result in an effective $d=9$ operator that violates lepton number by two units, given by  $\widetilde{\mathcal{O}}_1 = (\psi Q)(\psi u^c)(HH)H$~\cite{Babu:2001ex, Babu:2019mfe, deGouvea:2007qla, Cepedello:2017lyo}. Smallness of neutrino mass can be loosely understood even when the new particles have TeV scale masses, owing to a loop suppression factor and a chiral suppression affecting $M_\nu$.

The induced neutrino mass matrix arising from Fig. \ref{fig:numass} can be evaluated to be
\begin{equation}
     M_\nu \ = \ (\kappa_1 + \kappa_2) (f^T M_u V^\star y + y^T V^\dagger M_u f) \, , 
     \label{eq:numass}
\end{equation}
where $M_u = \text{diag}\{m_u, m_c, m_t\}$ is the diagonal up-quark mass matrix, and $\kappa_1$,  $\kappa_2$ are respectively the one-loop and two-loop factors given by
\begin{eqnarray}
    \kappa_1 & \ = \ & \frac{1}{16 \pi^2} \sin 2\varphi  \log{\left(\frac{m_{X_2}^2}{m_{X_1}^2} \right)} \, , 
\label{eq:kappa1} \\
  \kappa_2 & \ \approx \ & \frac{1}{(16\pi^2)^2} \frac{\lambda_{\Delta H^3} v  \mu}{M^2} \, .
  \label{eq:kappa2}
\end{eqnarray}
The leading contribution to $M_\nu$ is the one-loop term proportional to $\kappa_1$. In evaluating this loop integral we have ignored the masses of the up-type quarks in relation to the LQ masses. In Eq. (\ref{eq:kappa1}) the parameter $\varphi$ is the $\omega^{2/3}-\bar{\rho}^{2/3}$ mixing angle given in Eq. (\ref{eq:mixangle}). Since the effective operator for $M_\nu$ arising from the one-loop diagram is of the type ${\cal O}_{\rm eff}^{d=7} = \psi \psi H H H^\dagger H$, which is of $d=7$, one should also consider the lower dimensional $d=5$ operator ${\cal O}_{\rm eff}^{d=5} = \psi \psi HH$ that can be induced at the two-loop level as shown in Fig. \ref{fig:numass}.  In the approximate expression for $\kappa_2$ given in Eq. (\ref{eq:kappa2}), the relevant mass scale is that of the heaviest particle in the loop, denoted here by $M$,  
defined as  $M={\rm max}(m_{X_1}, m_{X_2}, m_{\Delta^0})$, with $m_{X_1,X_2}$ being the physical masses of the charge-2/3 LQs (cf.~Eq.~\eqref{eq:phymass}) and $m_{\Delta^0}$ being the physical masses of the quadruplet (cf.~Eq.~\eqref{eq:mdelta}). 
When $m_{X_1,X_2} \gg m_{\Delta^0}$, the ratio $\kappa_2/\kappa_1 \sim m_{\Delta^0}^2/(16 \pi^2 v^2)$, which becomes of order unity for $m_{\Delta^0} < $ 3 TeV or so.  However, as we will see later in Section~\ref{sec:results}, the $R_2$ LQ is required to have a mass not larger than about 1 TeV in order for it to explain the $R_{D^{(\star)}}$ anomaly. In this case the two-loop diagram is negligible, and therefore, we only include the one-loop contribution in the neutrino fit described in Section~\ref{sec:neutfit}, although the $\kappa_2$ term can be important in a more general setting. 
The overall factor $\kappa_1$ in Eq.~\eqref{eq:numass} is a free parameter which needs to be of $\mathcal{O}(10^{-8})$ to get the correct order of magnitude for the neutrino masses. Note that the Yukawa matrix elements $f_{ij}$ and $y_{ij}$ must have at least some entries that are of order one in order to explain the $B$-decay anomalies. $\kappa_1 \sim 10^{-8}$ can be achieved by taking either the cubic coupling $\mu$ in Eq.~\eqref{eq:pot} or the induced VEV $v_\Delta$ to be  small. Both these choices are technically natural, since if either of these parameters is set to zero, lepton number becomes a good symmetry.

We note that the same operator that leads to neutrino masses in this model also induces an effective $\Delta$-quadruplet coupling to the SM leptons. (Recall that $\Delta$ cannot couple to fermions at the tree level in the model.)  This can be seen from partner diagrams of Fig.~\ref{fig:numass}, where the $SU(2)_L$ components are chosen differently. Ignoring small $SU(2)_L$-breaking effects, these couplings would all arise from the same effective operator $(\psi \psi H^\dagger \Delta)$.  Therefore, one can write these couplings as being proportional to $M_\nu$. Explicitly, we find that the $\Delta^{++}$ coupling to leptons has the Yukawa coupling matrix given by
\begin{align}
   ( Y_{\Delta^{++} \ell \ell})_{ij} \ = \  \dfrac{\sqrt{2}}{\sqrt{3}} \dfrac{ (M_{\nu})_{ij}}{v_{\Delta}} \, ,
   \label{eq:YukD}
\end{align}
where the $1/ \sqrt{3}$ is a  Clebsch-Gordan factor for the $\Delta^{++}$ component of the quadruplet in the expansion of the $(\psi \psi H^\dagger \Delta)$ operator. Eq.~\eqref{eq:YukD} will play a crucial role in the collider phenomenology of the quadruplet, as discussed in Section~\ref{sec:collider}.

\subsection{Yukawa Textures} \label{sec:texture}
In order to minimize the number of parameters in our numerical fit to $R_D$, $R_{D^\star}$, $R_K$, $R_{K^\star}$, $(g-2)_\mu$, and the neutrino oscillation observables, while satisfying all flavor and LHC constraints, we choose the following economical textures for the Yukawa matrices $f',f$ and $y$ defined as in Eq.~\eqref{eq:expandLag} with the first (second) index corresponding to quark (lepton) flavors:
\begin{align}
f^\prime \ & = \ \left(\begin{array}{ccc}
0 & 0 & 0 \\
0 & 0 & 0 \\
0 & \textcolor{blue}{f_{32}^\prime} & \textcolor{cyan}{f_{33}^\prime}
\end{array}\right), \hspace{5mm} 
f \ = \  \left(\begin{array}{ccc}
0 & 0 & 0 \\
0 & \textcolor{cyan}{f_{22}} & \textcolor{cyan}{f_{23}} \\
0 & \textcolor{blue}{f_{32}} & \textcolor{brown}{f_{33}}
\end{array}\right) , \label{eq:ffp}  \\
y \ & = \ \left(\begin{array}{ccc}
0 & 0 & 0 \\
0 & \textcolor{violet}{y_{22}} & y_{23} \\
y_{31} & \textcolor{violet}{y_{32}} & 0
\end{array}\right) \quad ({\tt Fit-I}) \, , \hspace{2mm} {\rm or} \hspace{2mm} 
y \ = \ \left(\begin{array}{ccc}
0 & 0 & 0 \\
0 & \textcolor{violet}{y_{22}} & 0 \\
y_{31} & \textcolor{violet}{y_{32}} & y_{33}
\end{array}\right) \quad ({\tt Fit-II}) \, .
\label{eq:y} 
\end{align}
Our motivation for the above textures is as follows: Nonzero ($f^{\prime}_{32}$, $f_{32}$)  can explain the anomalous magnetic moment of the muon via chirally-enhanced top-quark loops. The couplings ($f^{\prime}_{33}$, $f_{22}, f_{23}$) are responsible for $R_{D^{(\star)}}$, while ($y_{22}$, $y_{32}$) can explain $R_{K^{(\star)}}$. Similarly, the coupling $f_{33}$ is required to suppress the lepton-flavor-violating (LFV) constraint from chirally-enhanced $\tau \to \mu \gamma$, while simultaneously explaining $(g-2)_\mu$. The remaining parameters $(y_{23~(33)},y_{31})$ in Eq.~\eqref{eq:y} are needed to satisfy the six neutrino oscillation observables ($\Delta m_{21}^2$, $\Delta m_{31}^2$, $\sin^2\theta_{13}$, $\sin^2\theta_{23}$, $\sin^2\theta_{12}$, $\delta_{\rm CP} $). For more details, see Section~\ref{sec:results}. We also note that the zeros in the coupling matrices of Eqs.~(\ref{eq:ffp})-(\ref{eq:y}) need not be exactly zero; but they need to be sufficiently small so that the flavor changing processes remain under control (cf. Section~\ref{sec:Constraint}).

\section{\texorpdfstring{$B$}{B}-physics Anomalies}
\label{sec:anomalies}
In this section, we present our strategy to
reconcile the observed tension between experiment and theory in the lepton flavor universality violating observables in the charged-current decays $B \to D^{(\star)} \ell \nu$ (with $\ell=e,\mu,\tau$) and the neutral-current decays $B \to K^{(\star)}\ell^+\ell^-$ (with $\ell=e,\mu$) by making use of the $R_2$ and $S_3$ LQs.
\subsection{Charged-current Observables}
\label{SEC-03-A}

The relevant lepton universality violating ratios $R_D$ and $R_{D^\star}$ are defined as 
\begin{align}
R_{D^{(\star)}} \ = \ \frac{{\rm BR}(B \to D^{(\star)}\tau \nu)}{{\rm BR}(B \to D^{(\star)}\ell \nu)} \qquad ({\rm with}~\ell=e,\mu)\, .
\label{eq:RD}
\end{align}
These observables  have been measured by both BaBar~\cite{Lees:2012xj,Lees:2013uzd} and  Belle~\cite{Huschle:2015rga,Hirose:2016wfn,Abdesselam:2016cgx} in the $\bar{B}^0\to D^{+(\star)}\ell^-\bar{\nu}_\ell$ decays, while   LHCb has measured only the $R_{D^{\star}}$ parameter using both $\bar{B}^0\to D^{+\star}\ell^-\bar{\nu}_\ell$~\cite{Aaij:2015yra} and $\bar{B}^0\to D^{-\star}\ell^+\nu_\ell$ decays~\cite{Aaij:2017uff}. Combining all these measurements, the average of  these ratios are found to be ~\cite{Amhis:2019ckw}:
\begin{align}
R_{D}^{\rm Exp}& \ = \ 0.340\pm 0.027\pm 0.013\,, \label{eq:RDexp}\\
R_{D^\star}^{\rm Exp}& \ = \ 0.295\pm 0.011\pm 0.008\,, \label{eq:RDstarexp}
\end{align}
which induce tensions at the levels of $1.4 \, \sigma$ and $2.5\, \sigma$ respectively with respect to the corresponding SM predictions~\cite{Fajfer:2012vx, Fajfer:2012jt, Lattice:2015rga, Na:2015kha, Bigi:2017jbd, Bernlochner:2017jka, Jaiswal:2017rve, Bernlochner:2020tfi, Jaiswal:2020wer} given by:
\begin{eqnarray}
R_{D}^{\rm SM} & \ = \ 0.299\pm 0.003\,,\\
R_{D^\star}^{\rm SM} & \ = \ 0.258\pm 0.005\,.
\end{eqnarray}
Considering the $R_D$ and $R_{D^\star}$ total correlation of $-0.38$, the combined difference with respect to the SM is about $3.08\, \sigma$.

A related observable is the ratio $R_{J/\psi}$ defined as 
\begin{eqnarray}
R_{J/\psi} \ = \ \frac{{\rm BR}(B \to J/\psi\tau \bar \nu_\tau)}{{\rm BR}(B \to J/\psi \ell \bar \nu_\ell)} \qquad ({\rm with}~\ell=e,\mu) \, ,
\end{eqnarray}
which also shows a mild discrepancy of $1.7\, \sigma$ between the experimental measurement~\cite{Aaij:2017tyk}
\begin{eqnarray}
R_{J/\psi}^{\rm Exp}  \ = \ 0.71\pm 0.17\pm 0.184\,,
\end{eqnarray}
and the corresponding SM prediction~\cite{Ivanov:2005fd, Wen-Fei:2013uea, Dutta:2017xmj, Murphy:2018sqg, Issadykov:2018myx, Watanabe:2017mip, Cohen:2018dgz,Berns:2018vpl}
\begin{eqnarray}
R_{J/\psi}^{\rm SM} \ = \ 0.289\pm 0.01\, . 
\end{eqnarray}
However, the experimental uncertainty on this measurement is rather large at the moment, and any new physics scenario that explains the $R_{D^{(\star)}}$ anomaly automatically explains the $R_{J/\psi}$ anomaly. Therefore, we will not explicitly discuss $R_{J/\psi}$ in what follows.

\begin{figure}[t!]
    \centering
    \includegraphics[scale=0.5]{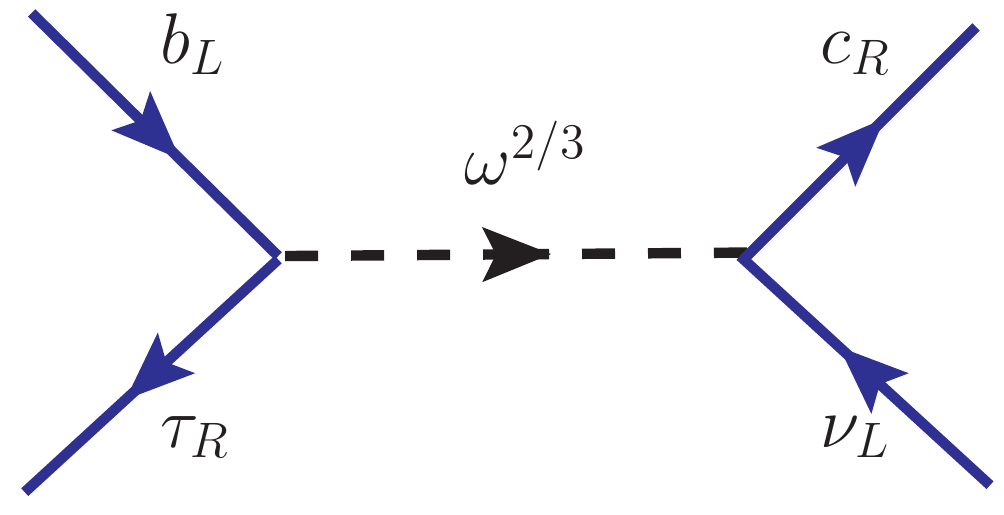} \hspace{15mm}
    \includegraphics[scale=0.5]{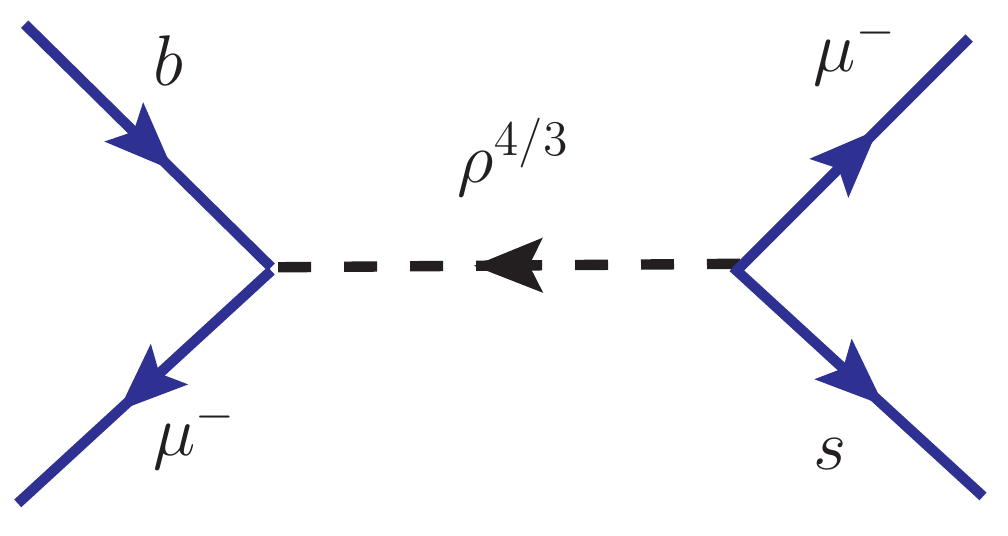}
    \caption{Feynman diagrams for the dominant LQ contributions to the $b\to c\tau^- \bar{\nu}$ (left) and $b \to s \mu^- \mu^+$ (right) transitions.}
    \label{fig:RD-RK}
\end{figure}

In order to confront our model with the experimental data for the charged-current processes, we shall consider LQ contributions to the flavor specific process $b\to c\tau^- \bar{\nu}$. Thus, only the numerator of Eq.~\eqref{eq:RD} is modified by the new LQ interactions. To this end, we consider the general low-energy effective Hamiltonian induced by SM interactions as well as the $R_2$ and $S_3$ LQs, which is given by
\begin{eqnarray}
    \mathcal{H}_{\mathrm{eff}} \ = \ &&\frac{4 G_{F}}{\sqrt{2}} V_{c b}\Big[\left(\bar{\tau}_{L} \gamma^{\mu} \nu_{\tau L}\right)\left(\bar{c}_{L} \gamma_{\mu} b_{L}\right)
    +g_V^\ell(\mu_R)\left(\bar{\tau}_{L} \gamma^{\mu} \nu_{\ell L}\right)\left(\bar{c}_{L} \gamma_{\mu} b_{L}\right)+g_{S}^\ell(\mu_R)\left(\bar{\tau}_{R} \nu_{\ell L}\right)\left(\bar{c}_{R} b_{L}\right) \nonumber \\
&& \qquad \qquad \qquad + g_{T}^\ell(\mu_R)\left(\bar{\tau}_{R} \sigma^{\mu \nu} \nu_{\ell L}\right)\left(\bar{c}_{R} \sigma_{\mu \nu} b_{L}\right)\Big]+\mathrm{H.c.} \, ,
\label{ham}
\end{eqnarray}
where the first term is the SM contribution, while the remaining terms correspond to new physics contribution, with $g^\ell_{V,S,T}$ being the Wilson coefficients defined at the appropriate renormalization scale $\mu_R$. As shown in Fig.~\ref{fig:RD-RK}, left panel, the $\omega^{2/3}$ component of the $R_2$ LQ mediates the $b \to c \tau^- \bar{\nu}_\ell$ semileptonic decay via a tree-level contribution. After integrating out the $R_2$ field, we obtain the following Wilson coefficients at the matching scale $\mu_R = m_{R_2}$: 
\begin{eqnarray}
   g_{S}^\ell\left(\mu_R=m_{R_2}\right) \ = \ 4 g_{T}^\ell \left(\mu_R=m_{R_2}\right) \ = \ \frac{ f_{2 \ell} f_{3 3}^{\prime \star}}{4 \sqrt{2} m_{R_2}^{2} G_{F} V_{c b}} \, ,
\label{eq:gSgT} 
\end{eqnarray}
where $\ell = e,\,\mu,\,\tau$ correspond to the outgoing neutrino flavors $\nu_e, \nu_\mu, \nu_\tau$ respectively. These Wilson coefficients are then run down in momentum to the $B$-meson mass scale in the leading logarithm approximation, yielding  \cite{Dorsner:2013tla}
\begin{align}
    g_S (\mu_R = m_b) & \ = \ \bigg[\frac{\alpha_s (m_b)}{ \alpha_s (m_t)} \bigg]^{-\frac{\gamma_s}{2 \beta_0^{(5)}}} \bigg[\frac{\alpha_s (m_t)}{ \alpha_s (m_{R_2})} \bigg]^{-\frac{\gamma_s}{2 \beta_0^{(6)}}} g_S(\mu_R = m_{R_2}) \, , \\
     g_T (\mu_R = m_b) & \ = \ \bigg[\frac{\alpha_s (m_b)}{ \alpha_s (m_t)} \bigg]^{-\frac{\gamma_T}{2 \beta_0^{(5)}}} \bigg[\frac{\alpha_s (m_t)}{ \alpha_s (m_{R_2})} \bigg]^{-\frac{\gamma_T}{2 \beta_0^{(6)}}} g_T(\mu_R = m_{R_2}),
\end{align}
where $\beta_0^{(n_f)} = 11 - 2 n_f/3$ is the running coefficient, with $n_f$ being the number of quark flavors effective in the relevant momentum regime \cite{Chetyrkin:1997dh, Gracey:2000am}. $\gamma_S$ and $\gamma_T$ are anomalous dimension coefficients given by $\gamma_S = -8$ and $\gamma_T = 8/3$. Thus, using $\alpha_s (m_Z) = 0.118$, which yields (using QCD running at four loops) $\alpha_s (m_b) = 0.2169$, $\alpha_s (m_t) = 0.1074$ and $\alpha_s (m_{R_2}) = 0.09$ for our benchmark value of $m_{R_2}=900~{\rm GeV}$, we obtain the following renormalization factors:\footnote{The running of $g_S$ is identical to that of the $b$-quark mass, see for e.g., Ref. \cite{Babu:2009fd}.}
\begin{align}
   g_S (\mu_R = m_b) & \ = \ 1.596 \, g_S(\mu_R = m_{R_2}) \, , \\ 
   g_T (\mu_R = m_b) & \ = \  0.855 \, g_T(\mu_R = m_{R_2}) \, .  
\end{align}
We see that the tensorial coupling $g_T$ becomes less important at $\mu_R = m_b$, with its value given by  $g_{S}(\mu_R=m_b) \approx 7.56 \, g_T (\mu_R=m_b)$~\cite{Gonzalez-Alonso:2017iyc}. We also note that we have ignored here the mixing between between the Wilson coefficients $g_S$ and $g_T$, which is an excellent approximation, as these off-diagonal terms are much smaller than the diagonal terms \cite{Gonzalez-Alonso:2017iyc}.  

The $\rho^{1/3}$ component of the $S_3$ LQ can also contribute in principle to $b \to c \tau \bar{\nu}_{\ell}$ via the Wilson coefficient $g_V^\ell$ given by 
\begin{align}
    g_{V}^\ell(\mu_R=m_{S_3}) \ = \ -\frac{y_{3\ell} (V^\star y)_{23}^\star}{8 \, m_{S_3}^2 G_F V_{cb}} \, .
    \label{eq:gVS3}
\end{align}
However, this contribution cannot accommodate $R_{D^{(\star)}}$ as the relevant Yukawa couplings are highly constrained from flavor physics. Any nonzero $y_{2\ell}$ is subject to $D^0-\bar{D}^0$ mixing and must be small (cf.~Section~\ref{sec:DDmix}), while LHC limits constrain both $y_{31}$ and $y_{32}$ (cf.~Section~\ref{sec:LHCcon}). Furthermore, the product of the Yukawa couplings $y_{2\ell}$ and $y_{3\ell'}$ is strongly constrained by processes such as $B\to K\nu\bar{\nu}$. It is also worth mentioning that one can induce Wilson coefficient $g_V^\ell$ of Eq.~\eqref{eq:gVS3} proportional to $y_{3\ell} y_{33}^\star$, in conjunction with CKM mixing. However, for $\ell = 3$, this contribution has an opposite sign compared to the SM, and therefore would require the new contribution to be twice as large as the SM one, bringing it to the non-perturbative regime.  For $\ell = 1$ or 2, there is no interference with the SM term, which would again require large non-perturbative values from the $S_3$ contribution.  
Thus we shall ignore these $S_3$-induced contributions to $R_{D^{(\star)}}$.  In Section~\ref{subsec:FitRD}, we have shown two best fit values of the Yukawa coupling matrices. 
For these choices of Yukawa couplings, shown in Eqs.~\eqref{eq:FitI} and \eqref{eq:FitII}, we get negligible contribution to $g_V^\ell=-5 \times 10^{-5}$ for Fit I and $g_V^\ell = 6 \times 10^{-6}$ for Fit II from the $S_3$ LQ, whereas the allowed $1\, \sigma$ range to explain $R_{D^{(\star)}}$ is $[0.072, 0.11]$. Therefore, we will only focus on the $R_2$ contribution to $R_{D^{(\star)}}$ induced through the Wilson coefficients $g_S^\ell$ and $g_T^\ell$.   
$R_{D}$ and $R_{D^\star}$ induced through the Wilson coefficients $g_s^\ell$ and $g_T^\ell$ at  $\mu_R = m_b$ with $\nu_\tau$ in the final state are approximately given by~\cite{Blanke:2018yud}
\begin{align}
R_D & \ \simeq \ R_D^{\rm SM} \left( 1 + 1.54 \, {\rm Re} [g_S^{\tau }] + 1.09 \, |g_S^\tau|^2 + 1.04 \, {\rm Re}[g_T^{\tau }] + 0.75 \, |g_T^\tau|^2\right) \, , \label{eq:approx1}\\
R_{D^\star} & \ \simeq \ R_{D^\star}^{\rm SM} \left( 1 - 0.13 \, {\rm Re} [g_S^{\tau }] + 0.05 \, |g_S^\tau|^2 - 5.0 \, {\rm Re}[g_T^{\tau }] + 16.27 \, |g_T^\tau|^2\right) \, ,
\label{eq:approx2}
\end{align}
where the numerical coefficients arise from the relevant form factors. These expressions are applicable for  $\nu_{e,\mu}$ final states as well, but by setting the Re$[g_S^\tau]$ and Re$[g_T^\tau]$ terms in Eqs.~(\ref{eq:approx1}) and \eqref{eq:approx2} to zero.  
This is because the new physics and the SM contributions interfere only when $\nu_\ell = \nu_\tau$. 

\begin{figure}[t!]
$$\includegraphics[width=7.3cm, height=5.9cm]{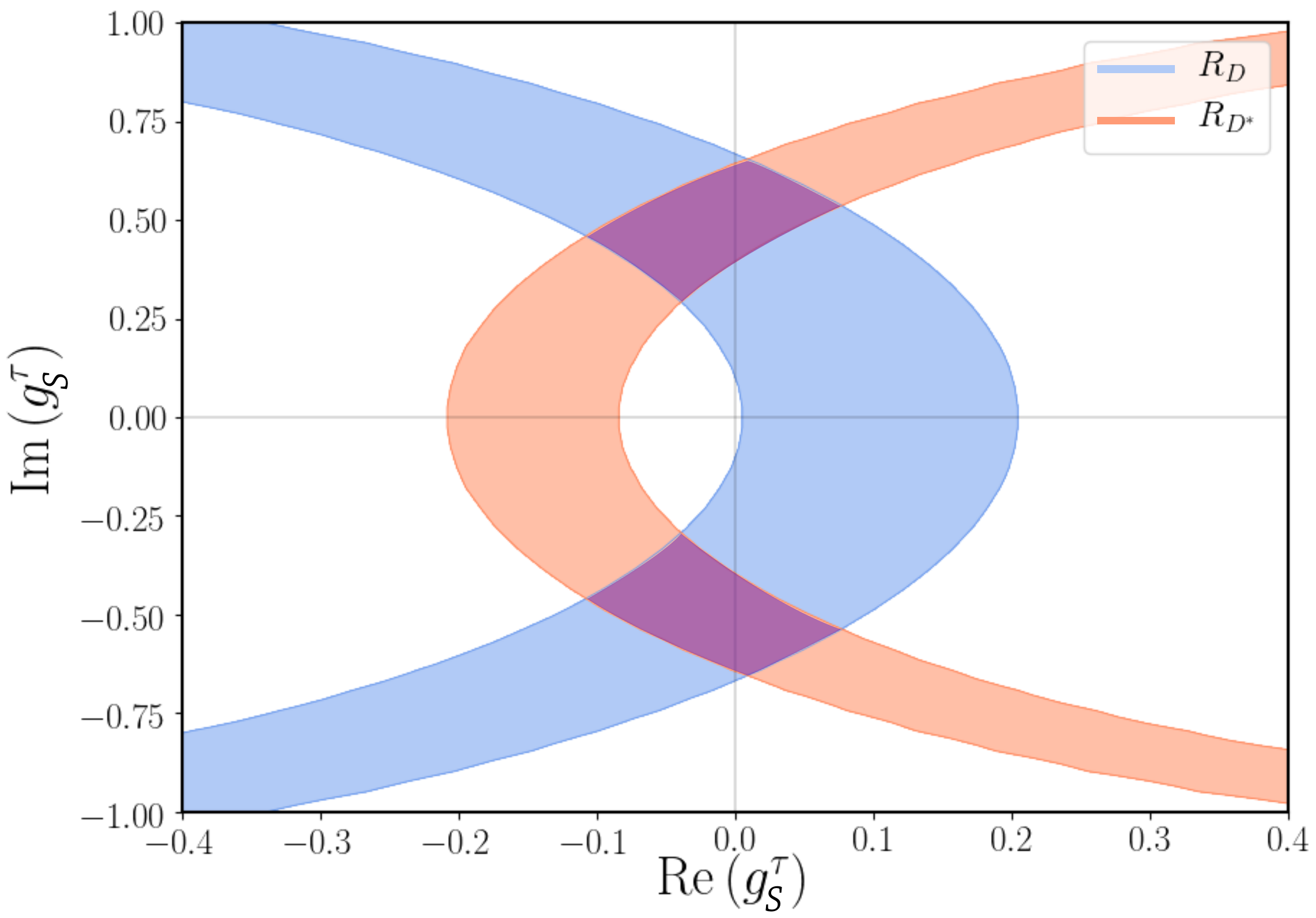} \hspace{5mm} \includegraphics[width=7.3cm, height=5.9cm]{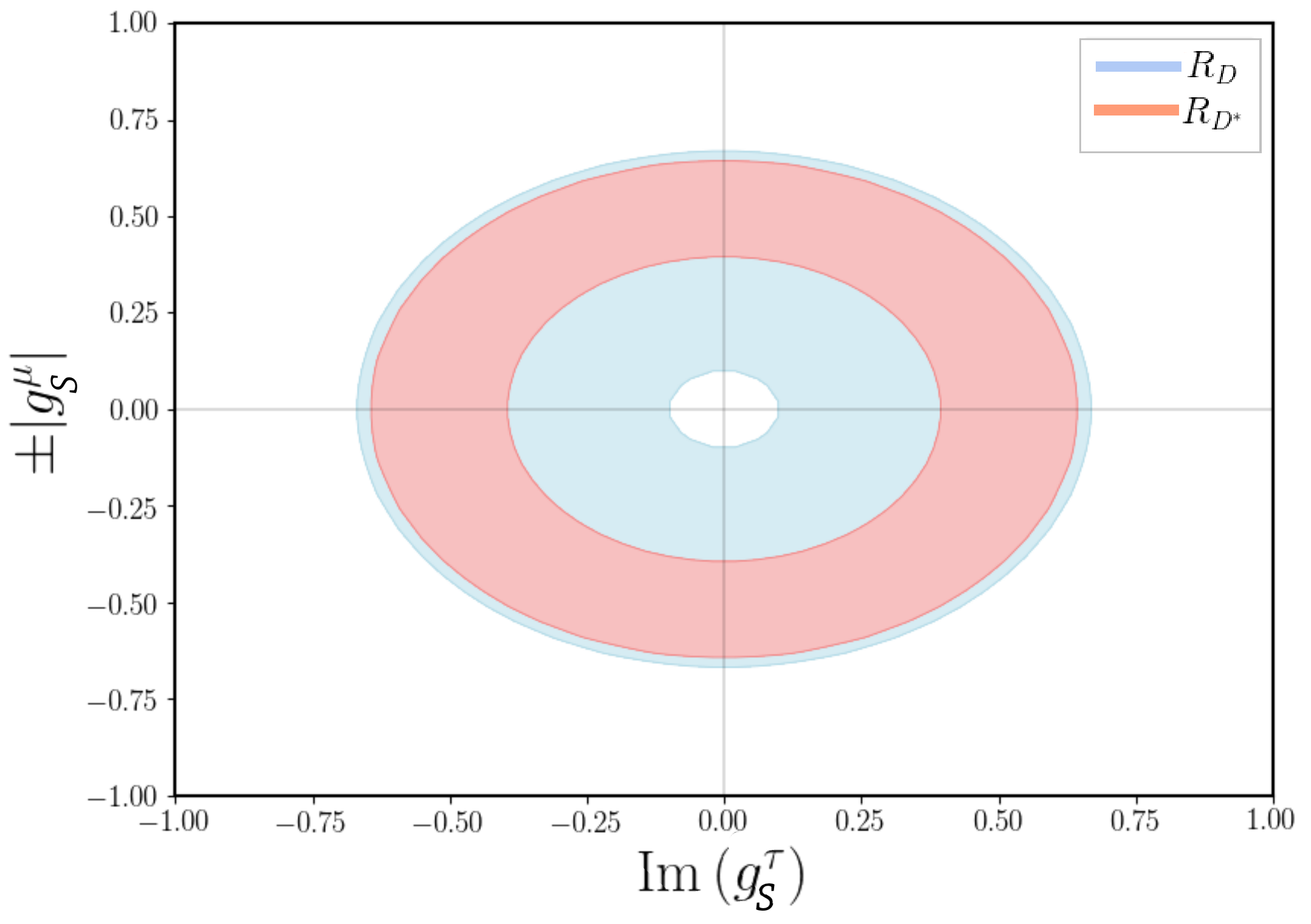} 
$$  
   \caption{{\it Left:} The $1\, \sigma$ allowed ranges for $R_D$ and $R_{D^\star}$ in the complex plane of $g_S^\tau$  with $g_S^{e,\mu} = 0$. The purple shaded regions correspond to the allowed region that explains both $R_D$ and $R_{D^\star}$.   {\it Right:} The $1\, \sigma$ allowed ranges for $R_D$ and $R_{D^\star}$ in the plane of $(g_S^\tau ,g_S^\mu)$ (with $g_S^e = 0$). The same result is obtained by replacing $g_S^\mu$ with $g_S^e$.}
    \label{fig:RDtau}
\end{figure}

The required values for the Wilson coefficient to get a simultaneous fit for both $R_D$ and $R_{D^\star}$ is depicted in Fig.~\ref{fig:RDtau}. We make use of {\tt Flavio} package~\cite{Straub:2018kue} that has NNLO QCD and NLO electroweak corrections coded in it, in generating Fig.~\ref{fig:RDtau}. The left panel  shows the $1\, \sigma$ allowed range of $R_D$ (light blue band) and $R_D^\star$ (light coral band) in the complex plane of $g_S^\tau$ with $g_S^{e, \mu} = 0$, i.e., with $f_{23}\neq 0$ and $f_{21}=f_{22}=0$ in Eq.~\eqref{eq:gSgT}. 
The intersection between the two bands, highlighted by the purple shaded regions, represents the allowed region that satisfies both anomalies. From this plot, we find that ${\rm Im}(g_S^\tau)$ must be nonzero, as first noted in Ref.~\cite{Sakaki:2013bfa}, while ${\rm Re}(g_S^\tau)$ should be nearly zero to fit $R_{D^{(\star)}}$. From Eqs. (\ref{eq:approx1}) and \eqref{eq:approx2} it is clear that any nonzero Re[$g_s^\tau]$ would pull $R_D$ and $R_D^*$ in opposite directions, in contradiction with experimental values (cf.~Eqs.~\eqref{eq:RDexp} and \eqref{eq:RDstarexp}), which is  what forces ${\rm Re}(g_S^\tau) \simeq 0$.  
In the right panel, we set ${\rm Re}(g_S^\tau)=0$, i.e., we set $g_S^\tau$ (or, equivalently, the $f_{23}$ coupling) to be purely imaginary, and switch on the $f_{22}$ coupling as well, as is the case with our texture in Eq.~\eqref{eq:ffp}. Again, the $1\, \sigma$ allowed ranges for $R_D$ and $R_{D^\star}$ are shown by the light blue and light coral bands, respectively. The same result is obtained by replacing $f_{22}$ with $f_{21}$, i.e., by using $g_S^e$ instead of $g_S^\mu$. In our numerical fit to $R_{D^{(\star)}}$ in Section~\ref{sec:results}, we fix $m_{R_2}$ ($f_{22}$) close to its minimum (maximum) allowed value from LHC constraints (discussed in Section~\ref{sec:LHCcon}), and find a neutrino mass fit for  $f_{23}$ and $f'_{33}$ such that the $g_S^{\mu,\tau}$ values are within the allowed region for both $R_D$ and $R_{D^{(\star)}}$ shown in Fig.~\ref{fig:RDtau}.   

The same effective Hamiltonian (\ref{ham}) relevant for $R_{D^{(*)}}$ also gives rise to the exclusive decay $B_c \to \tau \nu$. Within our model, the branching ratio for this decay is given by \cite{Watanabe:2017mip, Alonso:2016oyd}:
\begin{equation}
    {\rm BR} (B_c \to \tau \nu) \ = \ 0.023 \, \left|1- 4.068 \, g_S (\mu_R = m_{B_c})\right|^2 ~.
\end{equation}
Here we have used $\tau [B_c] = (0.507 \pm 0.009)$ ps, $f_{B_c} = 0.43$ GeV, and $m_{B_c} = 6.2749$ GeV. The branching ratio ${\rm BR} (B_c \to \tau \nu)$ has not been measured experimentally. Thus, $B_c$ lifetime needs to be compared with theoretical calculations \cite{Chang:2000ac, Gershtein:1994jw, Bigi:1995fs, Beneke:1996xe, Kiselev:2000pp}. With the benchmark fits shown in Section~\ref{sec:results}, we obtain branching ratio at the level of 12 \%, which is consistent with the limit quoted in Refs.~\cite{Akeroyd:2017mhr, Alonso:2016oyd, Blanke:2018yud, Bardhan:2019ljo}. 

\subsection{Neutral-current Observables}
\label{SEC-03-B}
The relevant lepton flavor universality violation ratios $R_K$ and $R_{K^\star}$ are defined as 
\begin{align} \label{Eqn:RK}
R_{K} \ & = \ \frac{{\rm BR}(B^+ \to K^{+} \mu^+ \mu^-)}{{\rm BR}(B^+ \to K^{+} e^+ e^-)} \, , \qquad 
R_{K^{\star}} \  = \ \frac{{\rm BR}(B^0 \to  K^{\star0} \mu^+ \mu^-)}{{\rm BR}( B^0 \to  K^{\star0} e^+ e^-)} \, .
\end{align}
The latest LHCb measurement of $R_K$ in the $q^2\in [1.1,6]~{\rm GeV}^2$ region ($q^2$ is the invariant mass of the lepton pair in the decays)  is~\cite{Aaij:2019wad}\footnote{For the recent update, see Ref.~\cite{Aaij:2021vac}.}
\begin{eqnarray} \label{Eqn:RK-Exp-new}
R_K^{\rm LHCb} \ = \ 0.846^{+0.060+0.016}_{-0.054-0.014}\,  ,
\end{eqnarray}
which shows a discrepancy at the level of  $2.6\, \sigma$ from its SM prediction 
~\cite{Bobeth:2007dw, Bordone:2016gaq} 
\begin{eqnarray} \label{Eqn:RK-SM}
R_K^{\rm SM} \ = \ 1.0003\pm 0.0001\,.
\end{eqnarray}

Analogously, the LHCb Collaboration has also measured the $R_{K^{\star}}$ ratio in two bins of low-$q^2$ region~\cite{Aaij:2017vbb}:
\begin{eqnarray}
R_{K^\star}^{\rm LHCb}& \ = \ & \begin{cases}0.660^{+0.110}_{-0.070}\pm 0.024 \qquad q^2\in [0.045, 1.1]~{\rm GeV}^2 \, , \\ 
0.685^{+0.113}_{-0.069}\pm 0.047 \qquad q^2\in [1.1,6.0]~{\rm GeV}^2 \, .
\end{cases}
\end{eqnarray} 
which have respectively $2.2\, \sigma$ and $2.4\, \sigma$ deviations from their corresponding SM results~\cite{Capdevila:2017bsm}:
\begin{eqnarray}
R_{K^\star}^{\rm SM} \ = \  \begin{cases} 0.92\pm 0.02 \qquad q^2\in [0.045, 1.1]~{\rm GeV}^2 \, , \\  
1.00\pm 0.01\qquad q^2\in [1.1,6.0]~{\rm GeV}^2 \, .
\end{cases}
\end{eqnarray}
In addition to these LHCb results,  Belle has recently announced new measurements on both  $R_K$~\cite{Abdesselam:2019lab} and 
$R_{K^\star}$~\cite{Abdesselam:2019wac}, but these results have comparatively larger uncertainties than the LHCb measurements on $R_{K^\star}$.

The effective Hamiltonian describing the new  physics contribution to the neutral-current process $b \to s\mu^+\mu^-$, in presence of $S_3$ LQ, is given by 
\begin{equation}
    \mathcal{H_{\text{eff}}} \ = \ - \frac{4 G_F}{\sqrt{2}} V_{tb} V_{ts}^{\star} \frac{e^2}{(4 \pi)^2} \Big[ C_9^{\mu\mu} (\bar{s} \gamma_\mu P_L b) (\bar{\mu} \gamma^\mu \mu) + C_{10}^{\mu\mu} (\bar{s} \gamma_\mu P_L b) (\bar{\mu} \gamma^\mu \gamma^5 \mu)\Big] + \text{H.c.} \, ,
\end{equation}
with $C_9^{\mu \mu}$ and $C_{10}^{\mu \mu}$ being the Wilson coefficients. Here we have assumed that the new physics couplings to electrons are negligible.  We focus on new physics contributions in the $b \to s\mu^+\mu^-$ channel, i.e. modifying only the numerator of Eq.~\eqref{Eqn:RK}. This is motivated by the fact that an explanation of $R_{K^{(\star)}}$ by modifying  the $b\to s\mu^+\mu^-$ decay provides a better global fit to other observables, as compared to modifying the $b\to se^+e^-$ decay~\cite{Aebischer:2019mlg}. It is known that both $R_K$ and $R_{K^\star}$ can be explained by either a purely vectorial Wilson coefficient $C_9^{\mu \mu} < 0$, or a purely left-handed combination, $C_9^{\mu \mu} = - C_{10}^{\mu \mu} < 0$~\cite{Angelescu:2018tyl}, with the latter combination performing better in the global analysis due to a $\sim 2 \, \sigma$ tension in the ${\rm BR}(B_s\to \mu\mu)$ decay which remains unresolved in the $C_9^{\mu\mu}$ scenario~\cite{Aebischer:2019mlg}. In our model, the dominant contribution to $b \to s\mu^+\mu^-$ comes at tree level via the mediation of the $\rho^{4/3}$ component of the $S_3$ LQ, as shown in Fig.~\ref{fig:RD-RK}, right panel. After integrating out the $S_3$ field, one can extract the Wilson coefficient for $b \to s \mu^- \mu^+$ decay as: 
\begin{equation}
C_{9}^{\mu\mu} \ = \ -C_{10}^{\mu\mu} \ = \ \frac{\pi v^{2}}{V_{t b} V_{t s}^{\star} \alpha_{\mathrm{em}}} \frac{y_{22} y_{32}^{\star}}{m_{S_3}^{2}} \, . 
\label{eq:RKC9}
 \end{equation}
The required best fit values of the Wilson coefficients at $\mu=m_b$ are $ C_9 = -C_{10} = -0.53$, with the $1 \, \sigma$ range being  $ [-0.61, -0.45]$~\cite{Aebischer:2019mlg}. In our numerical fit, $y_{22}$ and $y_{32}$ are fixed by the neutrino mass fit (up to an overall factor), which is then used to fix $m_{S_3}$ such that the best-fit value of $C_9=-C_{10}$ is obtained from Eq.~\eqref{eq:RKC9}.  

Note that the $R_2$ LQ can also give rise to $b \to s \ell^+ \ell^-$ transition at tree-level with the corresponding Wilson coefficient given by:
\begin{equation}
C_{9}^{\mu\mu} \ = \ C_{10}^{\mu\mu} \ = \ -\frac{\pi v^{2}}{V_{t b} V_{t s}^{\star} \alpha_{\mathrm{em}}} \frac{f'_{22} f_{32}^{\prime \star}}{m_{R_2}^{2}} \, .
\label{eq:RKC92}
 \end{equation}
There is no acceptable fit to  $R_{K^{(*)}}$ with $C_9 = C_{10}$.
Thus, taking the product of couplings $f^\prime_{2\alpha}$ and $f^\prime_{3\alpha}$ to be zero (or very small), one can suppress $R_2$  contribution to $R_{K^{(\star)}}$. On the other hand, a loop-level contribution to $b \to s \ell^+ \ell^-$ transition can in principle accommodate $R_{K^{(\star)}}$, but not simultaneously with $R_{D^{(\star)}}$, due to the stringent limits from $\tau\to \mu\gamma$~\cite{Becirevic:2017jtw}.  In our numerical fit, therefore, the $R_2$ contribution will not play a role in explaining $R_{K^{(\star)}}$.

\section{Muon Anomalous Magnetic Moment and Related Processes} \label{sec:gm2}
\begin{figure}[!t]
    \centering
    \includegraphics[scale=0.52]{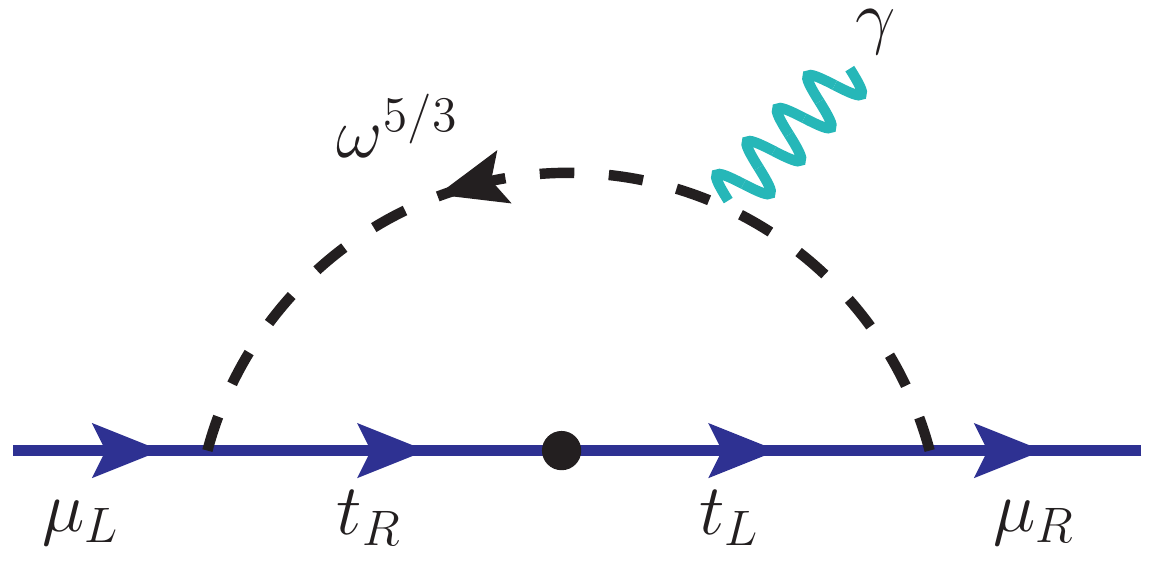} \hspace{15mm}
    \includegraphics[scale=0.51]{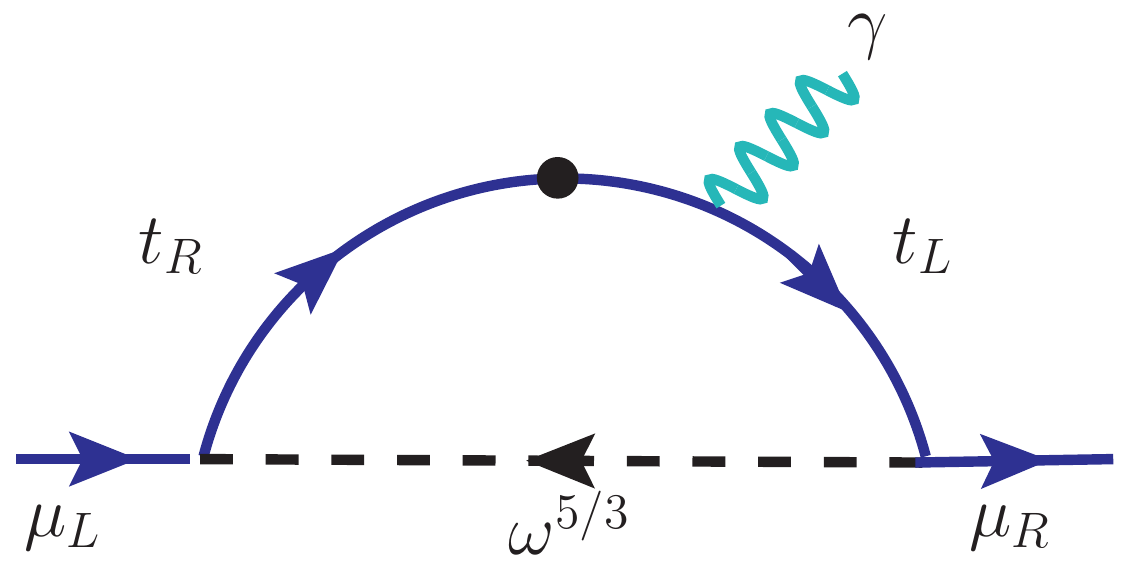} 
    \caption{Chirally-enhanced contribution from the  $R_2$ LQ to the muon anomalous magnetic moment. }
    \label{fig:g-2}
\end{figure}

Virtual corrections due to the LQ states can modify the electromagnetic interactions of charged leptons. The contribution from scalar LQ to anomalous magnetic moments has been extensively studied~\cite{Cheung:2001ip, Mandal:2019gff, Dorsner:2016wpm}. In particular, the $\omega^{5/3}$ component of the $R_2$ LQ can explain the muon (or electron)  anomalous magnetic moment, as it couples to both left-handed and right-handed fermions, see Eq.~\eqref{eq:expandLag}. 
The new contribution to the anomalous magnetic moment arising from $\omega^{5/3}$ LQ  can be written as~\cite{Cheung:2001ip, Lavoura:2003xp}:
\begin{eqnarray}
     \Delta a_{\ell}& \ = \ &-\frac{3}{16 \pi^{2}} \frac{m_{\ell}^{2}}{m_{R_2}^{2}} \sum_{q} \Big[\left(|f_{q\ell}|^{2}+ |  (V^\star f')_{q\ell}|^2\right)\left(Q_{q} F_{5}(x_q)+Q_{S} F_{2}(x_q)\right) \nonumber \\ 
&&\hspace{6mm}- \frac{m_{q}}{m_{\ell}} {\rm Re}[f_{q\ell} \,  (V^\star f^\prime)_{q\ell}^{ \star}]\left(Q_{q} F_{6}(x_q)+Q_{S} F_{3}(x_q)\right)\Big]
\label{eq:g-2}
\end{eqnarray}
where $Q_q = 2/3$ and $Q_S = 5/3$ are respectively the electric charges of the up-type quark and the LQ propagating inside the loop, as shown in Fig.~\ref{fig:g-2}.\footnote{The last term in Eq.~\eqref{eq:g-2} appears with a negative sign, as $f$ and $f^\prime$ in the Lagrangian have opposite signs, see Eq. (\ref{eq:expandLag}).}  Here $x_q = m_q^2/m_{R_2}^2$ and we have ignored terms  proportional to $m_\ell^2/m_{R_2}^2$ in the loop integral. The loop functions appearing in Eq.~\eqref{eq:g-2} are:
\begin{eqnarray}
&&F_{2}(x_q) \ = \ \frac{1}{6(1-x_q)^{4}}\left(1-6 x_q+3 x_q^{2}+2 x_q^{3}-6 x_q^{2} \ln x_q\right) \, , \label{eq:F2} \\
&&F_{3}(x_q) \ = \ \frac{1}{(1-x_q)^{3}}\left(1-x_q^{2}+2 x_q \ln x_q\right) \, ,\label{eq:F3} \\
&&F_{5}(x_q) \ = \ \frac{1}{6(1-x_q)^{4}}\left(2+3 x_q-6 x_q^{2}+x_q^{3}+6 x_q \ln x_q\right) \, ,\label{eq:F5}  \\
&&F_{6}(x_q) \ = \ \frac{1}{(1-x_q)^{3}}\left(-3+4 x_q-x_q^{2}-2 \ln x_q\right) . \label{eq:F6}
\end{eqnarray}
Note that the first term in Eq.~\eqref{eq:g-2} is the LQ contribution to the anomalous magnetic moment without chiral enhancement, whereas the second term is the chirally-enhanced one, which in our case will be proportional to the top-quark mass.

\subsection{Difficulty with Explaining \texorpdfstring{$\Delta a_e$}{Dae}} \label{sec:deltae}

A discrepancy has also been reported in the anomalous magnetic moment of the electron, denoted as
$\Delta a_{e}$, with a somewhat lower significance of $2.4 \, \sigma$~\cite{Parker:2018vye}.  The signs of $\Delta a_e$ and $\Delta a_\mu$ are opposite. We have investigated whether $\Delta a_e$ can also also explained in our framework,
but found that the model does not admit a simultaneous explanation of both anomalies, as introducing couplings of the type $f_{\alpha e}$ would lead to a chirally-enhanced contribution to the decay $\mu \to e \gamma$, which is highly constrained. One can attempt to explain both anomalies by simply avoiding chirally-enhanced  $\ell_i \to \ell_j \gamma$ decays by adopting a redefinition of $V^\star f^\prime \equiv f^\prime$ in Eq.~\eqref{eq:expandLag}. However, one introduces $V_{\rm CKM}$ in the down sector leading to strong constraints arising from processes such as $K_L \to e^{\pm} \mu^{\mp}$, $K_L \to \ell^+ \ell^-$, and $K-\bar{K}$ mixing. 

A logical option to explain $\Delta a_e$ would be to choose the Yukawa coupling $f_{21}$ to be of $\mathcal{O} (1)$, and rely on the charm-quark loop (proportional to $f_{21} f'_{21}$), while being consistent with all the flavor constraints and $R_{D^{(*)}}$. However, it turns out that the required values of the Yukawa couplings in this case have been excluded by the latest LHC dilepton constraints on LQ Yukawa couplings and masses from the non-resonant $t$-channel process $pp \to \ell^+ \ell^-$. These constraints are discussed later in Section~\ref{sec:LHCcon}, and are summarized in  ~Fig.~\ref{fig:yukbound}. Therefore, simultaneous explanation of the electron and muon anomalous magnetic moments, together with $R_{D^{(\star)}}$, is not possible in our setup. Thus, we focus on the parameter space required to explain $\Delta a_\mu$, but not $\Delta a_e$, as the former is the more persistent and significant discrepancy. In particular, we set $f_{\alpha e}=f'_{\alpha e}= 0$ in Eq.~\eqref{eq:ffp} to avoid any $\Delta a_{e}$ contribution for our numerical fits discussed in Section~\ref{sec:results}.

\subsection{Modified Higgs Decays to Lepton Pairs} \label{sec:higgs}
\begin{figure}
    \centering
    \includegraphics[scale=0.4]{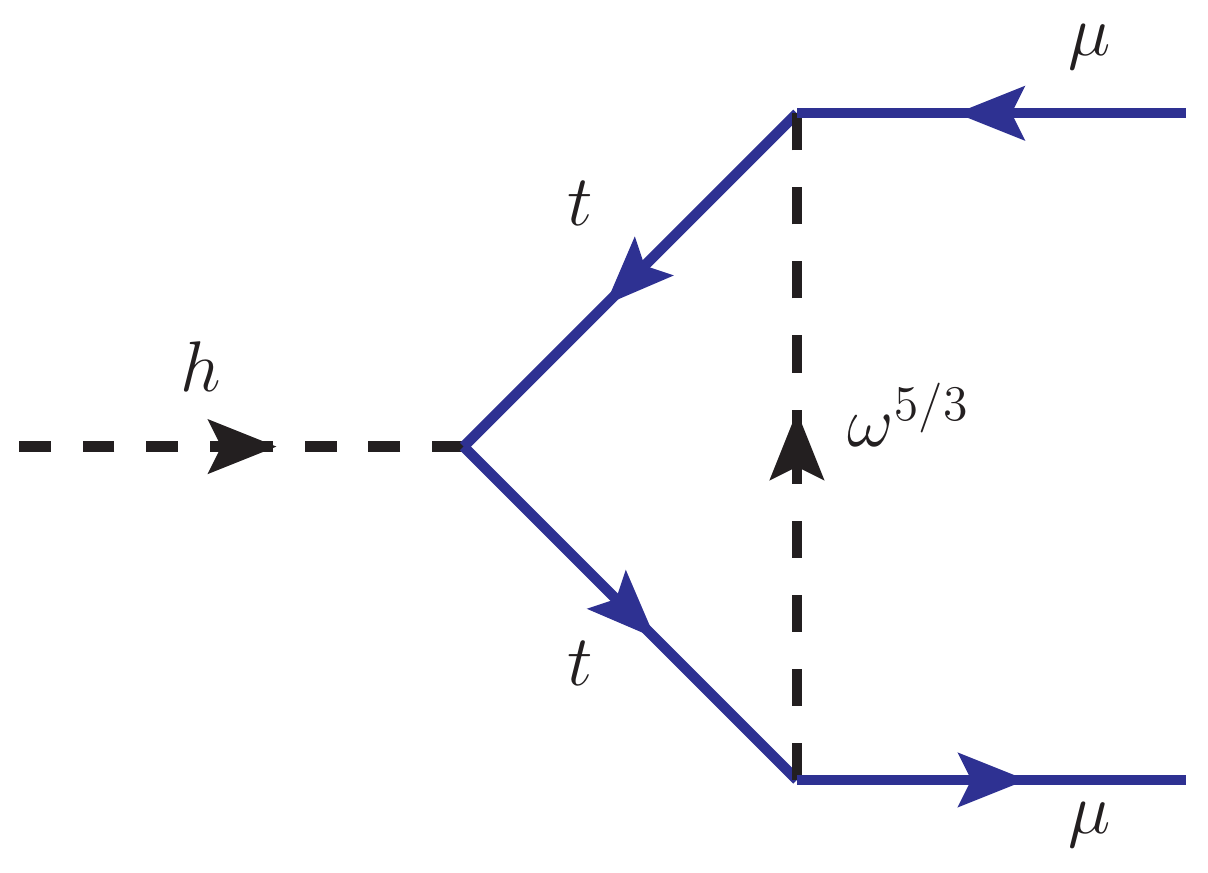} \hspace{10mm}
    \includegraphics[scale=0.4]{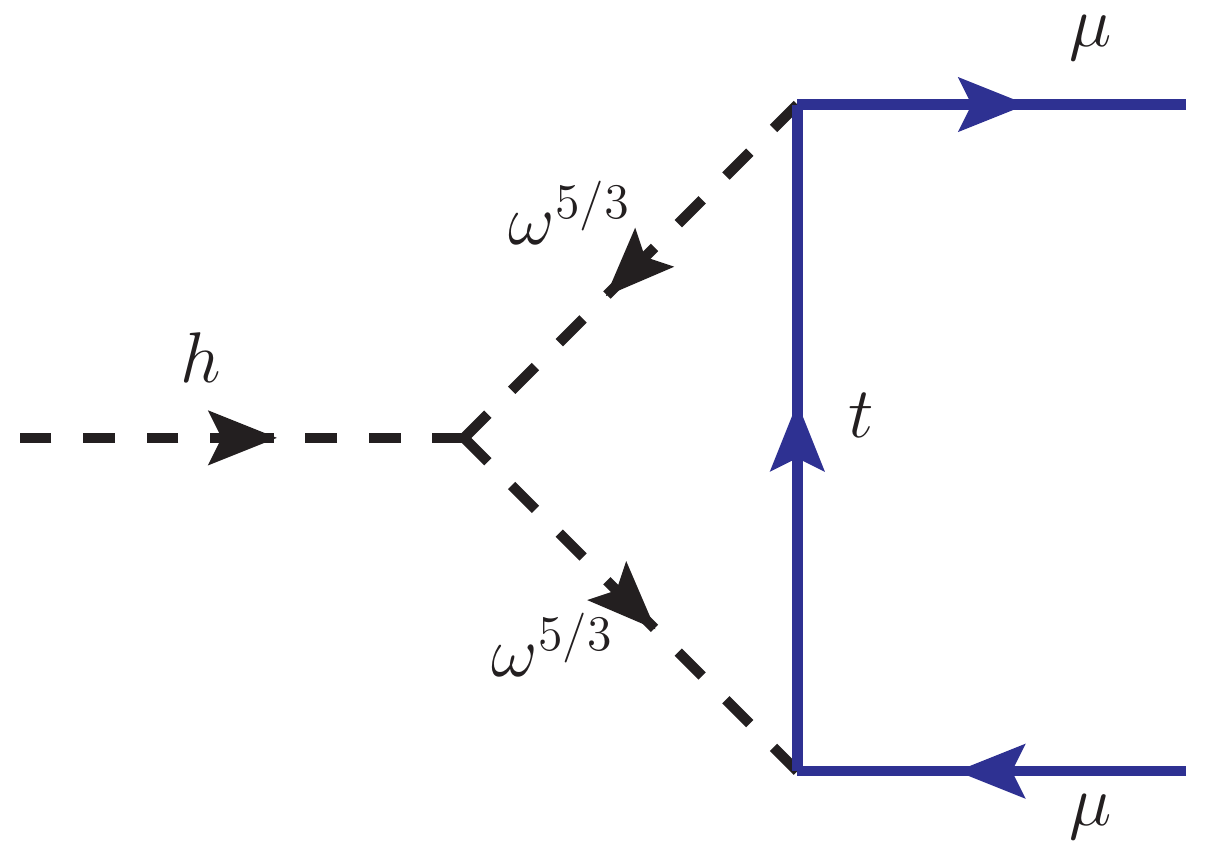}
    \caption{Feynman diagrams for the LQ contribution to $h \to \mu^+ \mu^-$ (and also $\tau^+\tau^-$) in our model.}
    \label{fig:hmumu}
\end{figure}
\begin{figure}[htb!]
    \centering
    \includegraphics[scale=0.5]{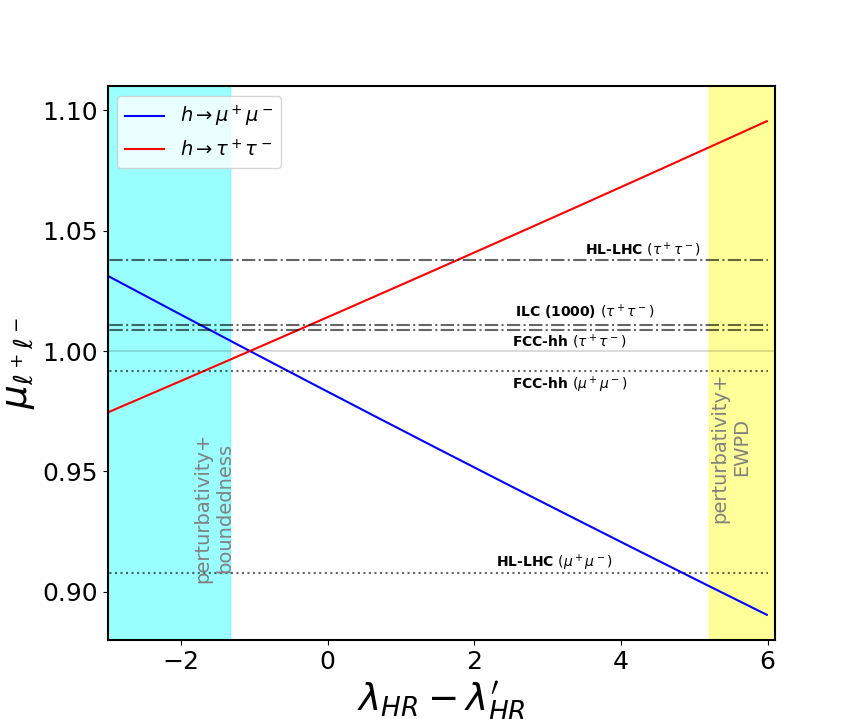} 
    \caption{Branching ratios of Higgs to dimuon (blue) and ditau (red) decays with respect to the SM predictions in our model as a function of the quartic coupling parameter $(\lambda_{HR}-\lambda'_{HR})$. The horizontal dotted (dot-dashed) lines show the sensitivities of future colliders for the $\mu^+\mu^-$ ($\tau^+\tau^-$) channel.  The shaded regions in yellow and blue are excluded by perturbativity plus  electroweak precision data, and by perturbativity plus boundedness of the potential constraints, respectively. } 
    \label{fig:hll}
\end{figure}

The same $R_2$ LQ interactions that lead to the chirally-enhanced $m_t/m_\mu$ contribution to the muon $g-2$ in Fig.~\ref{fig:g-2} will also induce a loop-level correction to the decay of the SM Higgs boson $h\to \mu^+\mu^-$.  The Feynman diagrams are shown in Fig.~\ref{fig:hmumu}.  In addition to these diagrams which modify the Yukawa couplings directly, one should also take into account correction to the muon mass arising from the $R_2$ interactions.  The relevant diagram is obtained from Fig.~\ref{fig:hmumu} by removing the Higgs boson line.  The significance of the LQ diagrams in modifying $h \rightarrow \mu^+ \mu^-$ decay has been noted recently in Ref.~\cite{Crivellin:2020tsz}.  We have carried out this calculation independently, and found full agreement with the results of Ref.~\cite{Crivellin:2020tsz}.  It is sufficient to compute the coefficient of the $d=6$ operator $(\overline{\psi}_{\mu L}\,\mu_R) H (H^\dagger H)$ which is finite, as any loop correction to the  $d=4$ operator $(\overline{\psi}_{\mu L} \,\mu_R) H$ will only renormalize the SM operator. The modification to the branching ratio BR($h
\to \mu^+ \mu^-$) is found to be
\begin{align}
   \mu_{\mu^+\mu^-} \ \equiv \ & \frac{\text{BR}(h \to \mu^+ \mu^-)}{\text{BR}(h \to \mu^+ \mu^-)_{\text{SM}}} \nonumber \\
    \ = \ & \Bigg| 1 - \frac{3}{8 \pi^2} \frac{m_t}{m_\mu} \frac{f_{32} (V^\star f^\prime)_{32}^\star}{m_{R_2}^2} \bigg\{ \frac{m_t^2}{8} \mathcal{F}\bigg( \frac{m_h^2}{m_t^2} ,  \frac{m_t^2}{m_{R_2}^2}  \bigg) + v^2 \left(\lambda_{HR} - \lambda'_{HR}\right) \bigg\} \Bigg|^2 \, .
\end{align}
The loop function $\mathcal{F}(x,y)$ can be expanded to first order in $y=m_t^2/m_{R_2}^2$ (so that  the coefficient of the $d=6$ operator is picked out), and also to the required order in $x=m_h^2/m_t^2$.  Although  $m_h^2/m_t^2 \sim 1$, the actual expansion parameter is some factor $k$ times this ratio, with $k \sim 1/10$, leading to a rapidly converging series.  The function $\mathcal{F}(x,y)$ to third order in $m_h^2/m_t^2$ is found to be
\begin{equation}
    \mathcal{F}(x,y) \ = \ - 8 + \frac{13}{3} x -\frac{1}{5} x^2 -\frac{1}{70} x^3 + 2 (x-4) \log y  \, . 
\end{equation}
For our benchmark fits (see Eqs.~\eqref{eq:FitI} and~\eqref{eq:FitII}) with $m_{R_2} = 0.9 \, \text{TeV}$, the model predictions for $\mu_{\mu^+\mu^-}$ as a function of the quartic coupling 
combination $(\lambda_{HR}-\lambda'_{HR})$ is shown in Fig.~\ref{fig:hll}. These predictions are essentially the same for the two benchmark points, so we present our results for Fit I (cf. Eq. (\ref{eq:FitI})) in Fig.~\ref{fig:hll}. 

The coupling $\lambda'_{HR}$ is responsible for the mass splitting between the $\omega^{2/3}$ and $\omega^{5/3}$ components of the $R_2$ LQ (cf.~Eqs.~\eqref{eq:omega23} and \eqref{eq:omega53})), which yields a positive contribution to the electroweak $\rho$-parameter:
\begin{align}
    \delta \rho \ \simeq \ \frac{N_cG_F}{8\sqrt 2 \pi^2}(\Delta m)^2 \, ,
\end{align}
where $N_c=3$ for color-triplets like $R_2$.
Using the current global-fit result for $\rho_0=1.00038\pm 0.00020$~\cite{Zyla:2020zbs} (with $\rho_0 = 1$ in the SM) and allowing for $3 \, \sigma$ uncertainty, we obtain an upper bound on the mass splitting $\Delta m \leq 55.9~{\rm GeV}$ (assuming that $v_\Delta$ $\leq$ few MeV, adopted in our collider physics analysis), which yields a corresponding bound on $|\lambda'_{HR}|\leq 1.66$. As discussed in Section~\ref{sec:bounded}, a necessary condition for the Higgs potential to be bounded from below (cf.~Eq.~(\ref{eq:bound})) is that for negative values of $(\lambda_{HR}-\lambda'_{HR})$, its magnitude should be below about 1.33, assuming that the magnitudes of all quartic couplings lie below $\sqrt{4\pi}$ to satisfy perturbativity.  Using the same constraint, we would then have $-1.33 \leq (\lambda_{HR}-\lambda'_{HR}) \leq 5.20$ as the preferred range, which is what we shall choose for our numerical study.

Our model prediction for $\mu_{\mu^+\mu^-}$ is shown in Fig.~\ref{fig:hll} by the solid blue line. We see that the deviation from the SM predictions in this branching is typically at the (2-6)\% level. This is fully  
consistent with the current LHC measurements: $\mu_{\mu^+\mu^-}^{\rm ATLAS}=1.2 \pm 0.6$~\cite{Aad:2020xfq} and $\mu_{\mu^+\mu^-}^{\rm CMS}=1.19^{+0.41}_{-0.39}({\rm stat.})^{+0.17}_{-0.16}({\rm syst.})$ ~\cite{CMS:2020eni}. For comparison, we quote in Table~\ref{tab:kappa} the future collider sensitivities for $\mu_{\mu^+\mu^-}$ from Ref.~\cite{deBlas:2019rxi}, and the relevant ones are also shown in Fig.~\ref{fig:hll} by the horizontal dotted lines. Thus, our predictions for the modified $h\to \mu^+\mu^-$ signal strength can be tested at the HL-LHC, HE-LHC, as well as at the FCC-hh colliders.

\begin{table}[t!]
\begin{center}
\begin{tabular}{||c|c|c||} \hline\hline
Collider & $\mu_{\mu^+\mu^-}$ & $\mu_{\tau^+\tau^-}$ \\ \hline\hline
     HL-LHC~\cite{Cepeda:2019klc} & 9.2\% & 3.8\% \\
     HE-LHC~\cite{Cepeda:2019klc} & 3.4\% & 2.2\% \\
     ILC (1000)~\cite{Fujii:2019zll} & 12.4\% & 1.1\% \\
     CLIC (3000)~\cite{Roloff:2018dqu} & 11.6\% & 1.8\% \\
    CEPC~\cite{CEPCStudyGroup:2018ghi} & 17.8\% & 2.6\% \\
    FCC-hh~\cite{Abada:2019lih} & 0.82\% & 0.88\% \\
\hline\hline
\end{tabular}
\end{center}
\caption{Expected relative precision of the Higgs signal strengths for future colliders. The numbers shown here are for the kappa-0 scenario of Ref.~\cite{deBlas:2019rxi}.}\label{tab:kappa}
\end{table}

It is also worth pointing out that the Yukawa textures needed to simultaneously explain $B$-anomalies, muon $g-2$, and neutrino mass require the $f_{33}$ entry to be nonzero, leading to a  new contribution to $h \to \tau^+ \tau^-$. This is also shown in Fig.~\ref{fig:hll} by the solid red line for our benchmark points. Our predictions for $\mu_{\tau^+\tau^-}\equiv \frac{{\rm BR}(h\to \tau^+\tau^-)}{{\rm BR}(h\to \tau^+ \tau^-)_{\rm SM}}$ are  
consistent with the current LHC measurements: $\mu_{\tau^+\tau^-}^{\rm ATLAS}=1.09^{+0.18}_{-0.17}({\rm stat.})^{+0.26}_{-0.22}({\rm syst.})^{+0.16}_{-0.11}({\rm theory~syst.})$
~\cite{Aaboud:2018pen} and $\mu_{\tau^+\tau^-}^{\rm CMS}=0.85^{+0.12}_{-0.11}$~\cite{CMS:2020dvp}.
For comparison, we quote in Table~\ref{tab:kappa} the future collider sensitivities for $\mu_{\tau^+\tau^-}$ from Ref.~\cite{deBlas:2019rxi}. Some of these are also shown in Fig.~\ref{fig:hll} by the horizontal dot-dashed lines. Thus, our predictions for the modified $h\to \tau^+\tau^-$ signal strength are potentially detectable at future colliders.

As can be seen from Fig.~\ref{fig:hll}, a characteristic feature of the model in the allowed parameter space accessible to future colliders is that while the shift in the branching ratio of $h \rightarrow \mu^+ \mu^-$ is downward compared to the SM, it is upward for the branching ratio of $h \rightarrow \tau^+ \tau^-$.  

\subsection{Muon and Neutron Electric Dipole Moments} \label{sec:edm}

LQ interactions can also lead to electric dipole moments (EDM) of the charged leptons (as well as quarks).  Existing limits from electron and muon EDMs would place strong constraints on the imaginary part of the Yukawa couplings of the $R_2$ LQ~\cite{Fuyuto:2018scm, Dekens:2018bci}. These constraints are significant only when the LQ  couples to both left- and right-handed charged leptons, as depicted in Fig.~\ref{fig:g-2}. The lepton EDM arising from these diagrams is given by~\cite{Cheung:2001ip} 
\begin{equation}
    |d_{\ell}| \ = \ \frac{3 e}{32 \pi^2} \sum_{q} \frac{m_q}{m_{R_2}^2} \big| \text{Im}[-f_{q\ell} (V^\star f')_{q\ell}^{\star}] \left(Q_{q} F_{6}(x_q)+Q_{S} F_{3}(x_q)\right) \big| \, .
    \label{eq:EDM}
\end{equation}
In particular, the constraint arising from electron couplings is quite stringent due to the ACME limit $|d_e|\leq 1.1\times 10^{-29}$ e.cm~\cite{Andreev:2018ayy}. However, since our model does not give additional contribution to $(g-2)_e$, we can simply avoid the electron EDM limit by setting the relevant couplings $f_{\alpha e}=f'_{\alpha e}=0$ in Eq.~\eqref{eq:ffp}. Furthermore, the muon EDM arising from the CKM phase, and from the phases in the  matrices $P$ and $Q$ of Eq.~\eqref{eq:vckm} when varied in their full range $[0,\,2\pi]$,  is found to be at most $3 \times 10^{-22}$ e-cm,  which is well below the current experimental limit of $|d_\mu|\leq 1.9\times 10^{-19}$ e-cm~\cite{Bennett:2008dy}, but may be potentially measurable in future experiments~\cite{Grange:2015fou, Abe:2019thb, TetsuichiKishishita:2020sxt} with  high-intensity muon sources~\cite{Janish:2020knz}.

The large Yukawa couplings necessary to explain anomalies in $b \to c \tau \nu$ decay can also lead to EDM of the tau lepton $d_\tau$, which is closely related to Im($g_s^\tau$) appearing in the $R_{D^{*}}$ calculation in Eq.~\eqref{eq:gSgT}. 
It is found to be at most $4.7 \times 10^{-21}$ e-cm when the phases in the  matrices $P$ and $Q$ of Eq.~\eqref{eq:vckm} are varied in their full range $[0,\,2\pi]$, which is below the current experimental limit of $|d_\tau| \leq 2.5 \times 10^{-17}$ e-cm \cite{Inami:2002ah}.

Similarly, the same Yukawa couplings that lead to tau EDM can also lead to charm quark EDM $d_c$ proportional to Im($g_s^\tau)$. The relevant expression is  obtained by replacing $m_q$ by $m_\tau$, $x_q$ by $x_\ell$, $Q_q$ by $Q_\ell = -1$ and $Q_S$ by $-5/3$ in Eq.~\eqref{eq:EDM}.
It is found to be at most $3.1 \times 10^{-22}$ e-cm. It is below the current experimental limit of $|d_c|\leq 1.5\times 10^{-21}$ e-cm \cite{Gisbert:2019ftm}, obtained from the limit on neutron EDM, $d_n < 3.0 \times 10^{-26}$ e-cm \cite{Afach:2015sja}. There is also a chromoelectric dipole moment of the charm quark ($\tilde{d}_c$), arising from diagrams where the photon emitted by the leptoquark is replaced by a gluon. This contribution in the model is obtained from Eq.~\eqref{eq:EDM} by keeping only the second term, and making the substitutions mentioned above.  We find that $\tilde{d}_c$ is at most $2.1 \times 10^{-23} g_s$-cm, which is below the experimental limit, $|\tilde{d}_c| < 1.0 \times 10^{-22}$ cm \cite{Gisbert:2019ftm}. Improving the neutron EDM limit by one order of magnitude can therefore directly test the leptoquark explanation of the $R_{D^{*}}$ anomaly.

\section{Low-energy Constraints}\label{sec:Constraint}

This section summarizes the most stringent low-energy flavor constraints that are relevant for our model. 

\subsection{\texorpdfstring{$\ell_\alpha \to \ell_\beta \gamma$}{llg}} \label{sec:llg}

These LFV radiative decays arising from LQ loops set some of the most stringent constraints on the couplings of the LQs to $\mu$ and $\tau$.  
As can be seen from Eq.~\eqref{eq:expandLag}, the $R_2$ LQ has both left- and right-handed couplings to charged leptons via the $f$ and $f'$ couplings; thus, it can lead to lepton decays both with and without chiral enhancement. The $S_3$ LQ on the other hand, only couples to left-handed charged leptons, so it cannot induce $\ell_\alpha \to \ell_\beta \gamma$ processes with chiral enhancement. 

The decay width for the $\ell_\alpha \to \ell_\beta \gamma$ mediated by LQ loops is given by~\cite{Lavoura:2003xp,  Benbrik:2008si, Mandal:2019gff}  
\begin{equation}
     \Gamma(\ell_\alpha \to \ell_\beta \gamma) \ = \ \frac{\alpha_{\rm em} (m_{\ell_\alpha}^2 - m_{\ell_\beta}^2)^3}{4 m_{\ell_\alpha}^3} \sum_{q} \Big(|\sigma_R^{\alpha \beta q}|^2 + |\sigma_L^{\alpha \beta q}|^2 \Big) \, .
\end{equation}
The amplitudes $\sigma_{R, L}$ arising from the exchange of $R_2$ LQ can be written as 
\begin{eqnarray}
  \sigma_R^{\alpha \beta q} & \ = \ & \frac{3}{32 \pi^2 m_{R_2}^2} \Bigg\{ \left[m_{\ell_\alpha} f_{q \alpha} f_{q \beta}^{\star} + m_{\ell_\beta} (V^\star f^\prime)_{q\beta} (V^\star f^\prime)_{q\alpha}^{\star}\right] \Big[ Q_{q} F_{5}(x_q)+Q_{S} F_{2}(x_q) \Big] \nonumber \\
    && \hspace{25mm} - m_q f_{qi} \,  (V^\star f^\prime)_{qi}^{ \star} \Big[ Q_{q} F_{6}(x_q)+Q_{S} F_{3}(x_q)\Big]  \Bigg\} \, ,
    \label{eq:sigmaR} \\
     \sigma_L^{\alpha \beta q} & \ = \ & \frac{3}{32 \pi^2 m_{R_2}^2} \Bigg\{\left[ m_{\ell_\alpha} (V^\star f^\prime)_{q \alpha} (V^\star f^\prime)_{q \beta}^{\star} + m_{\ell_\beta} f_{q\beta} f_{q\alpha}^{\star}\right] \Big[ Q_{q} F_{5}(x_q)+Q_{S} F_{2}(x_q) \Big] \nonumber \\
    && \hspace{25mm} - m_q (V^\star f^\prime)_{qi} \,  f_{qi}^{ \star} \Big[ Q_{q} F_{6}(x_q)+Q_{S} F_{3}(x_q)\Big]  \Bigg\} \, ,
    \label{eq:sigmaL}
\end{eqnarray}
with the loop functions $F_i(x_q)$ defined in Eqs.~\eqref{eq:F2}-\eqref{eq:F6}. Here we generically denote the masses of both $2/3$ and $5/3$ components of $R_2$ as $m_{R_2}$, assuming them to be degenerate.  Note that the amplitude $\sigma_L^{q}$ can be obtained from $\sigma_R^{q}$ with the substitution $f\leftrightarrow V^\star f^\prime$. The last terms in Eqs.~\eqref{eq:sigmaR} and \eqref{eq:sigmaL} which are proportional to $m_q$ are the chirally-enhanced contributions. Similarly, one can obtain the $S_3$ LQ contribution by replacing the $f$ couplings in the first term of Eq.~\eqref{eq:sigmaR} by $y$, assigning proper charges for the quark ($Q_q$) and scalar LQ ($Q_S$), and dropping the $f'$ terms in Eq.~\eqref{eq:sigmaR}. 

In the limit $m_{\ell_\beta}\to 0$, which is a very good approximation for both $\mu\to e\gamma$ and $\tau\to \ell \gamma$ (with $\ell=e,\mu$), and taking into account the $u^{cT} f e \omega^{5/3}$, $u^T (V^\star f^\prime) e^c \omega^{-5/3}$, and $d^T y e \rho^{4/3}$ terms in Eq.~\eqref{eq:expandLag}, the full expression for $\ell_\alpha\to \ell_\beta\gamma$ in our model can be written as
\begin{align}
   \Gamma \ = \ &  \frac{9 m_\alpha^5 \alpha_{\rm em}}{16 (16 \pi^2)^2} \left[\sum_{q=u,c,t}\left\{ \left| \frac{f_{q\beta}f_{q\alpha}^{\star}  }{2 m_{R_2}^2} + \left( \frac{(V^\star f^{\prime})_{q\alpha} f_{q\beta}^{\star} + f_{q\alpha}   (V^\star f^{\prime})_{q\beta}^\star }{3 m_{R_2}^2}\right) \frac{m_q}{m_\alpha} \left(1+ 4 \log x_q \right) \right|^2 \right.\right. \nonumber\\
    &+ \left. \left. \left| \frac{(V^\star f^\prime)_{q\beta} (V^\star f^\prime)_{q\alpha}^{\star} }{2 m_{R_2}^2} + \left(\frac{(V^\star f^{\prime})_{q\alpha} f_{q\beta}^{\star} + f_{q\alpha}   (V^\star f^{\prime})_{q\beta}^\star }{3 m_{R_2}^2}\right) \frac{m_q}{m_\alpha} \left(1+ 4 \log x_q \right) \right|^2 \right\} \right. \nonumber \\
    &\left. + \sum_{q'=d,s,b}\left|\frac{y_{q^\prime \beta}y_{q^\prime \alpha}^{\star}}{3 m_{S_3}^2} \right|^2 \right] \, .
    \label{eq:llgR2}
\end{align}
Here we have not included the $S_3$ contribution from the $\bar{u}_L^c e_L \rho^{1/3}$ term, because it is suppressed compared to the $d^T_L y e_L \rho^{4/3}$ contribution because of smaller electric charge, as well as due to a CKM-suppression factor and by a Clebsch factor of 2, as can be seen from Eq.~\eqref{eq:expandLag}. Similarly, the $\omega^{2/3}$ component of the $R_2$ LQ gives sub-dominant contribution proportional to $m_b^2/m_{R_2}^2$ compared to the $\omega^{5/3}$ component, owing to a GIM-like cancellation \cite{Babu:2010vp}; so we have not included it in Eq.~\eqref{eq:llgR2}. We have  displayed the constraint on the Yukawa coupling $f$ from this process in  Table~\ref{tab:llg}.

\begin{table}[!t]
    \centering
    \begin{tabular}{|c|c|c|}
    \hline\hline
        \textbf{Process} & \textbf{Experimental limit} & \textbf{Constraint} \\[3pt]  
       \hline \hline
       $\mu \to e \gamma$ & BR$ < 4.2\times 10^{-13}$~\cite{TheMEG:2016wtm}  & $|f_{q1} f_{q2}^\star| < 4.82 \times 10^{-4} \Big( \frac{m_{R_2}}{\text{TeV}}\Big)^2$ \\[3pt] \hline
      $\tau \to e \gamma$  & BR$ <3.3\times 10^{-8}$~\cite{Aubert:2009ag} & $|f_{q1} f_{q3}^\star| < 0.32 \Big( \frac{m_{R_2}}{\text{TeV}}\Big)^2$    \\[3pt] \hline
      $\tau \to \mu \gamma$  & BR$ <4.4\times 10^{-8}$~\cite{Aubert:2009ag} & $|f_{q2} f_{q3}^\star| < 0.37 \Big( \frac{m_{R_2}}{\text{TeV}}\Big)^2$  \\[3pt] \hline\hline
    \end{tabular} 
    \caption{Constraints on the Yukawa couplings as a function of LQ mass from $\ell_\alpha \to \ell_\beta \gamma$ decay. Constraints on $f^\prime$ couplings are obtained by replacing $f$ with $(V^\star f^\prime)$ for the $\omega^{5/3}$ LQ. Constraints on the $S_3$ Yukawa coupling $y$ ($V^\star y$) arising from $\bar{d}_L^c e_L \rho^{4/3}$ ($\bar{u}_L^c e_L \rho^{1/3}$)  are weaker by a factor of 3/2 (6)  in comparison to those shown here for the $f$  couplings, suppressed by smaller electric charge and Clebsch factor of 2, as can be seen from Eq.~\eqref{eq:expandLag}. }
    \label{tab:llg}
\end{table}

\subsection{\texorpdfstring{$\mu - e$}{mue} Conversion}
$\mu - e$ conversion in nuclei provides a stringent constraint on the product of the Yukawa couplings in our model. The couplings of the $S_3$ LQ, in conjunction with CKM rotation, is subject to the LFV process from coherent $\mu-e$ conversion in nuclei. 
The branching ratio for this conversion, normalized to muon capture rate, is given by.~\cite{Kuno:1999jp,Babu:2010vp, Babu:2019mfe}:
\begin{equation}
\operatorname{BR}(\mu N \rightarrow e N) \ \simeq \  \frac{\left|\vec{p}_{e}\right| E_{e} m_{\mu}^{3} \alpha_{\rm em}^{3} Z_{\text {eff }}^{4} F_{p}^{2} }{64 \pi^{2} Z \Gamma_{N}}(2 A-Z)^{2} \Bigg|\frac{(V^\star y)_{11} (y^\star V)_{12}}{2 m_{S_3}^{2}}\Bigg|^{2}
\end{equation}
where $\Gamma_N$ is the muon capture rate of the nucleus,  $\vec{p}_e$ and $E_e$ are respectively the momentum and energy of the outgoing electron, $A$, $Z$, and $Z_{\text{eff}}$ are atomic number, mass number and effective atomic number of the nucleus, whereas $F_p$ is the nuclear matrix element. The experimental limit from gold nucleus provides the most stringent bound~\cite{Bertl:2006up} of ${\rm BR} < 7.0 \times 10^{-13}$ leading to a constraint on the Yukawa coupling:
\begin{equation}
  \Big|(V^\star y)_{11} (y^\star V)_{12}\Big| \ < \ 8.58 \times 10^{-6} \left( \frac{m_{S_3}}{\text{TeV}} \right)^2 \, .
\end{equation}

\subsection{\texorpdfstring{$Z \to \tau \tau$}{Ztt} Decay} \label{sec:Ztautau}
 
Modifications of 
$Z-$boson decays to fermion pairs through one-loop radiative corrections mediated by LQs provide another important constraint on the Yukawa couplings of the LQ fields in the model. We focus our study on the leptonic $Z$ boson couplings as they are the most precisely determined by experiments~\cite{ALEPH:2005ab, Zyla:2020zbs}. 
Within our model, we require the $f_{33}^\prime$ coupling to be of $\mathcal{O}(1)$ to explain the $R_{D^{(\star)}}$ anomaly. Thus we focus on the $Z \to \tau \tau$ decay which provides a constraint of $f^\prime_{33}$. The shift in the coupling of $\tau_R$ with the $Z$ boson arising through loop corrections involving the $R_2$ LQ is given by ~\cite{Arnan:2019olv}
\begin{align}
    {\rm Re} [\delta g_{R}^{\tau \tau}] \ = \ & \frac{3 |f^\prime_{33}|^2}{16 \pi^2} \Bigg[ \frac{1}{2} x_t (1+ \log x_t) - \frac{x_z}{12} \Big\{ \log x_t \, (2 + 8/3 \sin^2 \theta_W) + (4 + 10/3 \sin^2 \theta_W) \Big\} \nonumber\\
    & \quad + \frac{x_z}{108} \Big\{ (-3 + 4 \sin^2 \theta_W) + \log x_z (18 + 12 \sin^2 \theta_W) \Big\} \Bigg] \, .
    \label{eq:zll}
\end{align}
Here we have used the definitions $x_t = \frac{m_t^2}{m_{R_2}^2}$ and $x_z = \frac{m_Z^2}{m_{R_2}^2}$, and kept terms only to linear orders in these parameters.  Using the experimental results on the effective coupling obtained by the LEP collaboration  \cite{ALEPH:2005ab}, ${\rm Re} [\delta g_{R}^{\tau \tau}] \leq 6.2 \times 10^{-4}$, we obtain the $1\, \sigma \, (2\, \sigma)$ limit on the Yukawa coupling as 
\begin{equation}
   | f^\prime_{33} | \ \leq \ 0.835 \ (1.18) 
   \label{eq:Ztautau}
\end{equation}
for the LQ mass of 900 GeV. Within the context of our model and to find a good fit to $R_{D^{(\star)}}$, we allow this coupling to be in the $2 \, \sigma$ range. A similar constraint on $f'_{32}$ can be derived, $| f^\prime_{23} | \leq 1.7$ from $Z \rightarrow \mu^+ \mu^-$ decay, which is however much weaker than the constraint that one would obtain from $\tau \rightarrow \mu \gamma$, which requires $| f^\prime_{23} f^\prime_{33} |\leq 0.3$.
%
%
%
%

\subsection{Rare \texorpdfstring{$D$}{D}-meson Decays}

Rare meson decays also put important constraints on the model parameters. The  relevant decays are $D^0 \to \mu^+ \mu^-$ and $D^+ \to \pi^+ \mu^+ \mu^-$.\footnote{In general, the decays $B\to K\nu\nu$ and $K \to \pi \nu \nu$ would provide more stringent constraint on the LQ Yukawa couplings~\cite{Buttazzo:2017ixm, DaRold:2018moy}. However, these bounds are avoided in our model by the choice of Yukawa coupling matrices}. 
For effective Lagrangian for these decays mediated by the $R_2$ and $S_3$ LQs is given by (cf.~Eq.~\eqref{eq:expandLag})
\begin{equation}
    \mathcal{L}_Y \ \supset \ u^T (V^\star f^\prime) e^c \omega^{-5/3} + u^T (V^\star y) e \frac{\rho^{1/3}}{\sqrt{2}} + \text{H.c.}
    \label{eq:LagD}
\end{equation}
There is also a contribution from the $f$ Yukawa, but it does not come with $V_{\rm CKM}$ rotation, so we do not need to consider this contribution for our choice of $f_{1\alpha}=0$, while deriving the partial decay width for the decay $D^0 \to \mu \mu$. The decay width for $D^0 \to \mu \mu$ proportional to the Yukawa couplings $f^\prime$ and $y$ is given by 
\begin{equation}
\Gamma_{D^{0} \rightarrow \mu \mu} \ = \ \frac{ |V_{u s} V_{c s}^{\star}|^{2} m_{\mu}^{2} f_{D}^{2} m_{D}}{128 \pi} 
\left(\frac{| f_{22}^\prime |^{4}}{m_{R_2}^{4}} + \frac{| y_{22} |^{4}}{4 m_{S_3}^{4}}\right) 
\left(1-\frac{4 m_{\mu}^{2}}{m_{D}^{2}}\right)^{1 / 2} \, .
\label{eq:Dmumu} 
\end{equation}
From Eq.~\eqref{eq:Dmumu}, one can obtain the constraint on $f'_{22}$ using the experimental limit ${\rm BR} (D^0 \to \mu^+ \mu^-) < 6.2 \times 10^{-9}$~\cite{Zyla:2020zbs}:
\begin{equation}
    |f_{22}^\prime| \ < \ 0.564 \,  \Big( \frac{m_{R_2}}{\text{TeV}} \Big) \, .
    \label{eq:f22}
\end{equation}

The semileptonic decay $D^+ \to \pi^+ \mu \mu$ is mediated by the same term as shown in Eq.~\eqref{eq:LagD} and we implement the calculation of Ref.~\cite{Babu:2019mfe} to obtain the following decay rate:
\begin{equation}
\Gamma_{D^{+} \rightarrow \pi^{+} \mu \mu}=\left(\frac{| f_{22}^\prime |^{4}}{m_{R_2}^{4}} + \frac{| y_{22} |^{4}}{4 m_{S_3}^{4}}\right)
\Bigg[  \frac{f_{D}}{f_{\pi}} g_{D^{\star} D \pi} |V_{u s} V_{c s}^{\star}| \Bigg]^{2} \frac{1}{64 \pi^{3} m_{D}} \mathcal{F}\, ,
\end{equation}
where the function $\mathcal{F}$ is defined as 
\begin{equation}
\mathcal{F}=\frac{m_{D^{\star}}^{2}}{12 m_{D}^{2}}\left[-2 m_{D}^{6}+9 m_{D}^{4} m_{D^{\star}}^{2}-6 m_{D}^{2} m_{D^{\star}}^{4}-6\left(m_{D^{\star}}^{2}-m_{D}^{2}\right)^{2} m_{D^{\star}}^{2} \log \left(\frac{m_{D^{\star}}^{2}-m_{D}^{2}}{m_{D^{\star}}^{2}}\right)\right] \, .
\end{equation}
The numerical value of the function $\mathcal{F} \simeq 2.98$ GeV. Using $f_{D}=212 \, \mathrm{MeV}, f_{\pi}=130 \, \mathrm{MeV}$,   $g_{D^{\star} D \pi}=0.59$
and the experimental upper limits on the corresponding branching ratio BR$(D^+ \to \pi^+ \mu \mu) < 7.3 \times 10^{-8}$, we obtain bounds on the $f^\prime$ coupling as
\begin{equation}
    |f_{22}^\prime| \ < \ 0.293 \,  \Big( \frac{m_{R_2}}{\text{TeV}} \Big) \, .
    \label{eq:f22p}
\end{equation}
Similarly, one can find the constraints on Yukawa coupling $y_{22}$, which is weaker by a factor of $\sqrt{2}$ in comparison to $f_{22}^\prime$ shown in Eqs.~\eqref{eq:f22} and \eqref{eq:f22p}, owing to a Clebsch factor.

\subsection{\texorpdfstring{$D^0-\bar{D}^0$}{D0D0} Mixing}\label{sec:DDmix}

Both $R_2$ and $S_3$ LQs can give rise to $D^0-\bar{D}^0$ mixing via box diagrams. 
Explicit calculation of the box diagram involving $R_2$ LQ gives~\cite{Crivellin:2019qnh}
\begin{equation}
    \Delta m_D \ = \ \frac{2}{3} B_1(\mu) m_D f_D^2 C_1^\prime \, ,
\end{equation}
where $f_D \simeq 212$ MeV is the $D$ meson decay constant, and $C_1^\prime$ is the Wilson coefficient given by
\begin{equation}
C_1^\prime \ (\mu=1~\text{TeV}) \ = \ \frac{1}{128 \pi^2} \frac{(f_{1\alpha} f_{2\alpha}^\star)^2}{m_{R_2}^2} \, .
\end{equation}
Here $\alpha$ is the lepton flavor that runs in the box diagrams, which is summed. The renormalized Wilson coefficients $C_1^\prime$~\cite{Buras:2000if, Ciuchini:1997bw, Golowich:2007ka} and the bag factor $B_1$~\cite{Carrasco:2014uya},  evaluated at $\mu_R=3$ GeV scale, are given by
\begin{eqnarray}
    && C_1^\prime \ (\mu_R=3~\text{GeV}) \ \approx \ 0.8 \, C_1^\prime (\mu_R=1~ \text{TeV})\, , 
    \quad B_1(\mu_R = 3\text{ GeV}) \ = \ 0.75 \, .
\end{eqnarray}
From the experimental value $ |\Delta m_D| = 0.95^{+0.41}_{-0.44} \times 10^{10}  \ {\rm s}^{-1}$ \cite{Zyla:2020zbs, Bazavov:2017lyh}, we obtain the limit 
\begin{equation}
    |f_{1\alpha} f_{2\alpha}^\star| \ < \ 0.0187 \, \Big( \frac{m_{R_2}}{\text{TeV}} \Big) \, . \label{eq:DDconst}
\end{equation}
The same constraint applies to the $f'$ coupling as well. However, in addition to the limit quoted in Eq.~\eqref{eq:DDconst}, the Yukawa $f^\prime$ is also supplemented by Cabbibo rotation, as seen from Eq.~\eqref{eq:expandLag}. Thus, for any nonzero entry in the up-sector $f'_{1\alpha}$ or charm-sector $f'_{2\alpha}$, a nonzero $D^0-\bar{D}^0$ mixing will be induced by the $(V^\star f')$ term in Eq.~\eqref{eq:expandLag}. Consequently, we get a bound on the individual couplings:
\begin{equation}
    |f_{1 \alpha}^\prime|, \, \,  |f_{2 \alpha}^\prime| \ < \ 0.305 \, \Big( \frac{m_{R_2}}{\text{TeV}} \Big)^{1/2} \, .
    \label{eq:518}
\end{equation}
Similarly, one can obtain a limit on the individual Yukawa $y$ as well, since a nonzero $y_{1 \alpha}$ (or $y_{2\alpha}$) would result in a box diagram contribution to $D^0-\bar{D}^0$ mixing, owing to the CKM mixing. This  has contributions from  $u-\nu$ term in addition to the $u-e$ term in Eq.~\eqref{eq:expandLag}. Thus for any nonzero entry in the up-sector or charm-sector in the Yukawa matrix $y$, the bound is slightly stronger than that shown in Eq.~\eqref{eq:518}:  
\begin{equation}
  |y_{1\alpha}| , \,\, |y_{2\alpha}| \ < \ 0.288 \, \Big( \frac{m_{S_3}}{\text{TeV}} \Big)^{1/2}  \, .
  \label{eq:DDy}
\end{equation}
It is worth mentioning that the Yukawa couplings $y_{3\alpha}$ and $f^\prime_{3\alpha}$ also contribute to $D$-meson mixing. However, these contributions can be safely ignored in the context of our model as they are strongly suppressed by CKM mixing angles by $V_{cb}$ and $V_{ub}$. 

\section{LHC Constraints on Leptoquarks} \label{sec:LHCcon}

At the LHC, the $R_2$ and $S_3$ LQs can be pair-produced through $gg$ and $q\bar{q}$ fusion processes, or can be singly produced in association with charged leptons via $s$- and $t$- channel quark-gluon fusion processes. The pair production of the LQs at the LHC is solely dictated by the LQ mass, irrespective of their Yukawa couplings, whereas the single production rate depends on both mass and the Yukawa coupling of the LQ. Therefore, the single-production limits are relevant only for larger Yukawa couplings $\sim {\cal O}(1)$~\cite{Babu:2019mfe, Buonocore:2020erb} to the first and second-generation quarks. 
For the benchmark points studied in Section~\ref{sec:results}, the Yukawa couplings to the first and second generation quarks are not too large ($<1$), hence the collider bounds from single-production are not so significant compared to the limits from  QCD-driven LQ pair-production. However, we will show in Section~\ref{sec:dilepton} that there are stringent limits on the Yukawa couplings of the LQ from the the dilepton processes $p p \rightarrow \ell_{i}^{+} \ell_{j}^-$. 

\subsection{Pair-production Bounds} \label{sec:pair}
\begin{figure}[t!]
         \centering
         $$
         \includegraphics[width=0.67\textwidth]{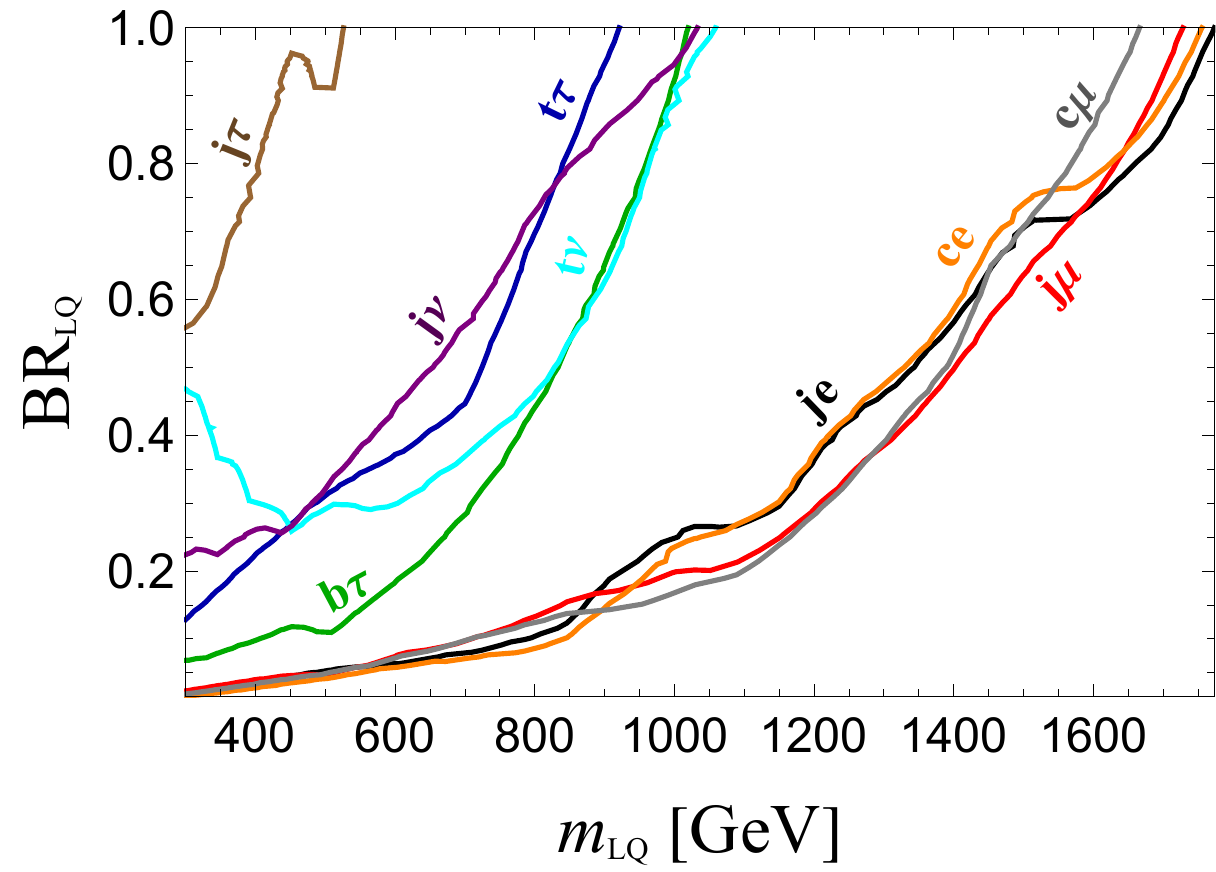}
          $$
         \caption{Summary of the updated direct limits from LQ pair-production searches at the LHC for different quark-lepton decay channels of the $R_2$ LQ. The branching ratio for a specific decay channel of the LQ as indicated in the figure is varied from 0 to 1, while the other decay channels not specified compensate for the missing branching ratios to add up to one. These limits are independent of the LQ Yukawa coupling.  }
         \label{fig:lqbound}
    \end{figure}

Once pair-produced at the LHC, each LQ will decay into a quark and a lepton, and the collider limits on these LQ masses depend on the branching ratios to different decay modes. To impose the bound on the LQ masses, we use the upper limits on the cross-sections from dedicated searches for pair production of first~\cite{CMS:2018sxp,Khachatryan:2015vaa}, second~\cite{Aaboud:2019jcc,Khachatryan:2015vaa,Sirunyan:2018kzh} and third generation~\cite{Sirunyan:2018vhk, Aaboud:2019bye,Sirunyan:2018kzh} LQs at the LHC and recast them in the context of our model, following the analysis in Ref.~\cite{Babu:2019mfe}. For this purpose, we first implement our model file in {\tt FeynRules} package~\cite{Christensen:2008py} and then analyze the signal cross sections  using {\tt MadGraph5aMC@NLO}~\cite{Alwall:2014hca}, which is then compared with the experimental upper limits on the cross section times the branching ratio, assuming that the cut efficiencies are similar in both cases. Our results for the $R_2$ LQ are shown in Fig.~\ref{fig:lqbound}, where the black, red, green, blue, cyan, purple, orange, gray, and brown solid colored lines respectively represent the current bounds from the $je$, $j\mu$, $b\tau$, $t\tau$, $t \nu$, $j\nu$, $ce$, $c\mu$, and $j\tau$ decay mode of the LQ. Here the branching ratio of each decay mode is varied from 0 to 1 individually without specifying the other decay modes, which compensate for the missing branching ratios to add up to one. As expected, the bounds on the first and second-generation LQs are much  more stringent, as compared to the third-generation case. We will use this information to our advantage while choosing our benchmark points in Section~\ref{sec:results}.

In particular, for the Yukawa ansatz of Eqs.~\eqref{eq:ffp}, the dominant decay modes of the $R_2$ LQ are:
\begin{equation}
\begin{array}{l}
\omega^{2 / 3} \ \xrightarrow[\text{}]{\text{$f$}} \ c \bar{\nu}_\mu, \ c \bar{\nu}_\tau, \ t \bar{\nu}_\mu, \ t \bar{\nu}_\tau \, , \\
\omega^{2 / 3} \ \xrightarrow[\text{}]{\text{$f'$}} \  b \tau^+, \ b\mu^+ \, ,  \\
\omega^{5 / 3} \ \xrightarrow[\text{}]{\text{$f$}} \ c \mu^+, \ c \tau^+, \ t \mu^+,  \ t \tau^+ \, , \\
\omega^{5 / 3} \ \xrightarrow[\text{}]{\text{$f'$}} \ t \tau^+, t \mu^+ \, . 
\end{array}
\label{eq:R2decay}
\end{equation}
The branching ratios for these decay modes corresponding to the fits presented in Eqs.~\eqref{eq:FitI} and~\eqref{eq:FitII} are shown in Table.~\ref{tab:branching}.  As we can see, the $\omega^{2/3}$
component of the $R_2$ LQ  dominantly  decays to $j\nu $ and $b \tau$ final states, whereas the $\omega^{5/3}$  component mostly decays to $t \tau$,  and $j\tau$ final states. Note that the mass of the $\omega^{2/3}$  component cannot be very different from that of the $\omega^{5/3}$ component due to the electroweak precision constraints, and hence,  we consider them to be almost degenerate in our analysis. Given the branching ratios in Table.~\ref{tab:branching}, the $b\bar{b} \tau^+\tau^-$ final state gives the most stringent constraint on the $R_2$ LQ mass, which is required to be larger than 859 GeV, as can be seen from Fig.~\ref{fig:lqbound}. 

 
\begin{table}[t!]
\centering
\resizebox{0.85\textwidth}{!}{%
\begin{tabular}{|c|c|c|c|c|c|c|c|c|}
\hline\hline
\multirow{3}{*}{\textbf{\begin{tabular}[]{@{}l@{}} Model Fit \end{tabular}}} &
  \multicolumn{8}{c|}{\textbf{Branching ratio}} \\ \cline{2-9} 
 &
  \multicolumn{4}{c|}{\textbf{$\omega^{2/3}$}} &
  \multicolumn{4}{c|}{\textbf{$\omega^{5/3}$}} \\ \cline{2-9} 
 & \textbf{$\nu j$} & \textbf{$b \tau$} & \textbf{$b \mu $} & \textbf{$\nu t$} & \textbf{$t \tau$} & \textbf{$\mu j$} & \textbf{$\tau j$} & \textbf{$t \mu $} \\ \hline
\multicolumn{1}{|c|}{{Fit I}} &
  $41.8 \%$ &
  $54.1 \%$ &
  $4 \%$ &
  $0.04 \%$ &
  $54.1 \%$ &
  $4 \%$ &
  $37.8 \%$ &
  $4 \%$  \\ \hline
\multicolumn{1}{|c|}{{Fit II}} &
  $41.3 \%$ &
  $54 \%$ &
  $4 \%$ &
  $0.04\%$ &
   $54.1 \%$ &
  $4 \%$ &
  $37.8 \%$ &
  $4 \%$   \\ \hline\hline
\end{tabular}%
}
\caption{Branching ratios for different decay modes of the $R_2$ LQ corresponding to the fits presented in Eqs.~\eqref{eq:FitI} and~\eqref{eq:FitII}.}
\label{tab:branching}
\end{table}

As for the $S_3$ LQ relevant for $R_{K^{(\star)}}$ anomaly, it can in principle decay to all quark and lepton flavors, due to the CKM-rotations involved in Eq.~\eqref{eq:expandLag}. However, the dominant decay modes of the $S_3$ LQ corresponding to the Yukawa ansatz in Eqs.~\eqref{eq:FitI} and~\eqref{eq:FitII} are
\begin{equation}
\begin{array}{l}
\rho^{4 / 3} \ \rightarrow \ \bar{s} \mu^+ \, , \\
\rho^{1 / 3} \ \rightarrow \ \bar{c} \mu^+, \ \bar{s} \bar{\nu} \, , \\
\rho^{-2/ 3} \ \rightarrow \ \bar{c} \bar{\nu} \, .
\end{array}
\label{eq:Sdecay1}
\end{equation}
In addition, for $m_{R_2}, m_\Delta < m_{S_3}$, the $S_3$ LQ can decay to the 
$R_2$ LQ and the quadruplet scalar $\Delta$, mediated by the  trilinear coupling $\mu$ in Eq.~\eqref{eq:pot} that is responsible for neutrino mass in our model. For our numerical analysis, we focus on the scenario with the  $R_2$ ($S_3$)
LQ mass around $\sim$ 1 TeV (2 TeV) and the quadruplet mass also around 1 TeV. In this case, the  
$S_3\to R_2+\Delta$ decay is the dominant one with $\sim 100\%$ branching ratio. In this case, the various components of $S_3$ decay as follows:
\begin{equation}
\begin{array}{l}
\rho^{4 / 3} \ \rightarrow \ \omega^{-2 / 3} \Delta^{++}\, , \ \omega^{-5 / 3} \Delta^{+++} \, , \\
\rho^{1 / 3} \ \rightarrow \ \omega^{-2 / 3} \Delta^{+} \, , \ \omega^{-5 / 3} \Delta^{++} \, , \\
\rho^{-2/ 3} \ \rightarrow \ \omega^{-5 / 3} \Delta^{+} \, , \ \omega^{-2 / 3} \Delta^{0} \, .
\end{array}
\label{eq:Sdecay}
\end{equation} 
As a consequence, limits on the $S_3$ LQ mass from the standard LHC searches are not applicable to our scenario. See Section~\ref{sec:collider} for more details on the $S_3$ decay signatures at the LHC. For this decay to occur, $S_3$ mass should exceed that of $R_2$ LQ.


\subsection{Dilepton Bounds} \label{sec:dilepton}

\begin{figure}[t!]
         \centering
         $$
         \includegraphics[width=0.67\textwidth]{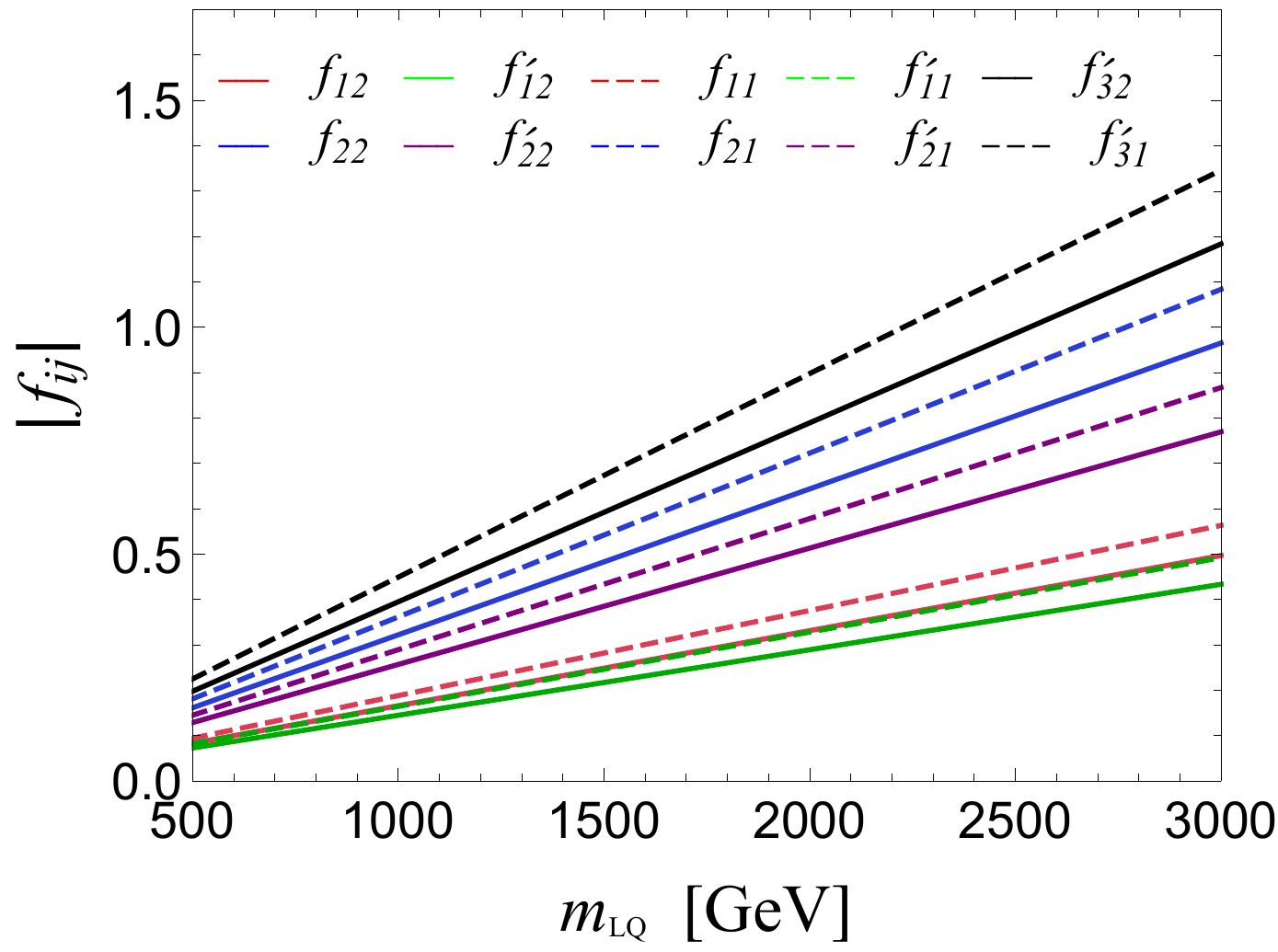}
          $$
         \caption{Summary of the new indirect constraints on the Yukawa couplings of the $R_2$ LQ as a function of its mass from a recent non-resonant dilepton search at the LHC. }
         \label{fig:yukbound}
    \end{figure}

Apart from the direct LHC limits from LQ pair-production, there also exist indirect limits from the cross section measurements on the dilepton process $p p \rightarrow \ell_{i}^{+} \ell_{j}^-$, which could get significantly modified due to a $t-$channel LQ exchange for large Yukawa couplings.  Ref.~\cite{Angelescu:2018tyl} had derived indirect limits on the LQ mass and Yukawa couplings involving the $\tau$ lepton using the previous resonant dilepton searches at the LHC. Meanwhile, a dedicated search~\cite{Aad:2020otl} for the  non-resonant signals in dielectron and dimuon final states has been performed at the $\sqrt s= 13$ TeV LHC with integrated luminosity 139 fb$^{-1}$, which is more appropriate for the $t$-channel LQ search. Therefore, we use this recent non-resonant dilepton study to derive new indirect limits on the LQ mass and Yukawa couplings. For this analysis, we first implement our model file in {\tt FeynRules} package~\cite{Christensen:2008py}, then analyze the cross section for $p p \rightarrow \ell_{i}^{+} \ell_{j}^-$ signal  using {\tt MadGraph5aMC@NLO}~\cite{Alwall:2014hca} and compare the quoted observed limits~\cite{Aad:2020otl} on the cross-section to derive the limits on the Yukawa coupling for a given LQ mass. Our results are shown in Fig.~\ref{fig:yukbound} for different Yukawa couplings $f_{i\alpha}$ and $f'_{j\alpha}$ (with $i=1,2; \ j=1,2,3; \ \alpha=1,2$) of the $R_2$ LQ. Similar bounds can also be derived for the $S_3$ LQ. There are no bounds on the $f_{31}$ and $f_{32}$ couplings quoted in Fig.~\ref{fig:yukbound}, because they involve top-quark initial states, whereas the bounds on $f'_{31}$ and $f'_{32}$ come from bottom-quark-initiated processes (cf.~Eq.~\eqref{eq:expandLag}). Similarly, we do not report any bounds on the Yukawa couplings involving $\tau$-flavor, as there is no corresponding non-resonant dilepton analysis involving taus available so far. Based on the previous analysis~\cite{Angelescu:2018tyl}, we anyway expect the tau-flavor limits to be weaker than the ones quoted here. Note that the bounds derived in Fig.~\ref{fig:yukbound} are independent of the LQ branch ratios, unlike the direct limits shown in Fig.~\ref{fig:lqbound}.  As can be seen from Fig.~\ref{fig:yukbound}, the flavor-dependent upper limits on the LQ Yukawa couplings for 1 TeV $R_2$ LQ mass to the first two family leptons and quarks are in the range $(0.15 - 0.36)$, which get slightly relaxed to $(0.15 - 0.45)$  if we include the bottom-quark. This precludes a solution of $R_{D^{(\star)}}$ with $R_2$-mediated decays of the $B$-meson involving $\nu_e$ or $\nu_\mu$ final states. Therefore, we only focus on the scenario with $\nu_\tau$ final state in our benchmark points for the explanation of $R_{D^{(\star)}}$.

\section{Numerical Fit}
\label{sec:results}
In this section, we present our numerical results for the model parameter space that explains the anomalies in $R_{D^{(\star)}}$, $R_{K^{(\star)}}$, and $\Delta a_\mu$ within their $1 \, \sigma$ measured values, while being consistent with all the low-energy and LHC constraints discussed above. It is beyond the scope of this work to explore the entire parameter space of the theory; instead we implement all the constraints and find a few benchmark points to explain the anomalies. First of all, we fix the $R_2$ LQ mass at 900 GeV to satisfy the LHC bound obtained from pair-produced $\omega^{2/3}$ decaying to $b\bar{b}\tau^+\tau^-$ (cf. Fig.~\ref{fig:lqbound} and Table~\ref{tab:branching}). 
Note that $m_{R_2}$ needs to be around 1 TeV to explain $R_{D^{(\star)}}$; making it larger would require larger $f'_{33}$ and $f_{23}$ coupling values beyond ${\cal O}(1)$. For example, with $f'_{33}={\rm Im}f_{23}=1.5$ and $f_{22}=0.45$ (to be consistent with the flavor constraints), the maximum $m_{R_2}$ we can have is 1.4 TeV. We also fix the $S_3$ LQ mass at 2 TeV for our $R_{K^{(\star)}}$ analysis, but it can be scaled up to much higher values without requiring either of the Yukawa couplings $y_{22}$ or $y_{32}$ in Eq.~\eqref{eq:RKC9} to exceed ${\cal O}(1)$ values.


\subsection{Fit to \texorpdfstring{$R_{D^{(\star)}}$}{RDs}} \label{subsec:FitRD}
\begin{figure}[t!]
    \centering
    \includegraphics[scale=0.6]{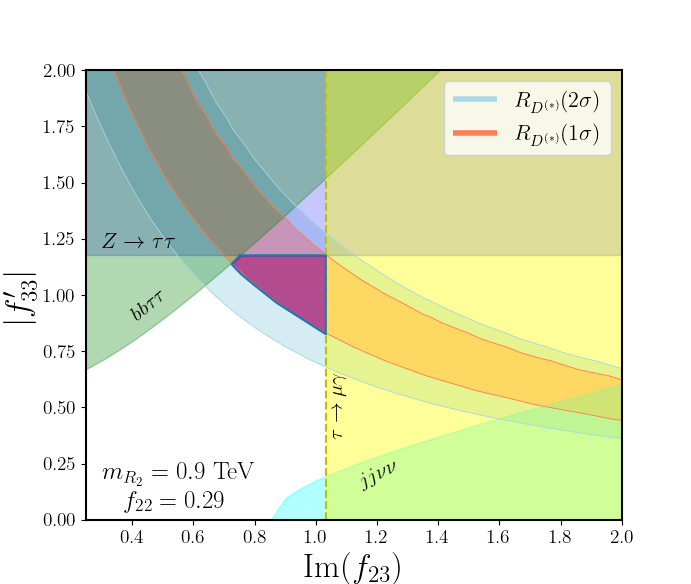}
   \caption{$1\, \sigma$ (light red) and $2\, \sigma$ (light blue) allowed range for $R_{D^{(\star)}}$ in the relevant Yukawa coupling plane, with the $R_2$ LQ mass at 900 GeV and with a fixed  $f_{22} = 0.29$.  The horizontal purple band is from the $Z \to \tau \tau$ constraint. The curved green band and cyan bands respectively represent exclusion from LQ pair production in $pp \to bb\tau\tau$ and $pp \to jj\nu\nu$ channels at LHC. The vertical yellow band corresponds to the exclusion from LFV decay $\tau \to \mu \gamma$. The dark purple shaded box represents the $1 \, \sigma$ allowed region for $R_{D^{(\star)}}$ that is consistent with all the constraints in this model.}
    \label{fig:RD_1000}
\end{figure}
In Fig.~\ref{fig:RD_1000}, we show the allowed parameter space to explain $R_{D^{(\star)}}$ at $1\, \sigma$ (orange shaded) and $2\, \sigma$ (light blue shaded) CL in the most relevant Yukawa coupling plane ${\rm Im}(f_{23})-|f^\prime_{33}|$ for a fixed $R_2$ LQ mass at 900 GeV.  We have also fixed $f_{22} = 0.29$, which is the maximum allowed value from the dilepton constraint (see Fig.~\ref{fig:yukbound}). Note that a nonzero $f_{22}$ is required by the neutrino oscillation fit for the textures we have (see Section~\ref{sec:neutfit}), and a larger $f_{22}$ helps widen the $R_{D^{(\star)}}$ region. In our numerical analysis to generate Fig.~\ref{fig:RD_1000}, we have made use of the {\tt Flavio} package \cite{Straub:2018kue}. As already noted in Section~\ref{SEC-03-A} (cf. Fig.~\ref{fig:RDtau}), the $f_{23}$ coupling needs to be complex to get a good fit to $R_{D^{(\star)}}$. Thus, while doing the minimization to get neutrino oscillation fit, we choose the $f_{23}$ coupling purely imaginary, as shown in Fig.~\ref{fig:RD_1000}. 

\begin{figure}
    \centering
    \includegraphics[height=7cm, width=8cm]{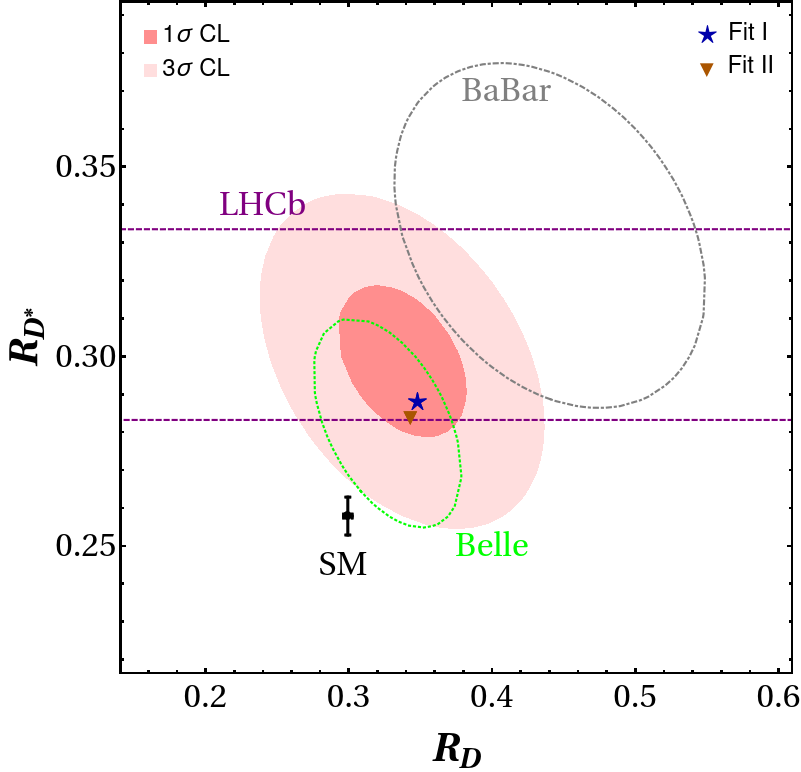}
    \caption{1$\sigma$ (dark red) and $3\sigma$ (light red) contours for experimental averages from Ref.~\cite{Altmannshofer:2020axr} for the LFUV observables $R_D$ and $R_{D^*}$ observables. Individual $1\sigma$ regions from Belle, BarBar, and LHCb are respectively shown by the dotted green, gray, and purple contours. Black error bar represent the SM prediction, whereas black and brown marker corresponds to the two model Fit I and Fit II given by Eq.~\eqref{eq:FitI} and Eq.~\eqref{eq:FitII}. }
    \label{fig:RD_RDS}
\end{figure}

The dark purple shaded area highlighted in Fig.~\ref{fig:RD_1000} represents the allowed region that is consistent with all the constraints in our model. The rest of the colored regions are excluded by various constraints discussed in the previous sections. The horizontal purple band is from $Z \to \tau \tau$ constraint (cf. Eq.~\ref{eq:Ztautau}). The green and cyan shaded regions respectively represent LHC exclusion from LQ pair-production in $b\tau$ and $j\nu$ decay modes (cf. Fig.~\ref{fig:lqbound}). The vertical yellow shaded region corresponds to the exclusion from LFV decay $\tau \to \mu \gamma$ (cf. Table~\ref{tab:llg}). In the next subsection, we will choose both $f'_{33}$ and $f_{23}$ values from within the allowed region shown in Fig.~\ref{fig:RD_1000}. Similarly, Fig.~\ref{fig:RD_RDS} shows experimental averages for $R_D$ and $R_{D^*}$ taking correlation into account between the two observables, along with benchmark fits within the model corresponding to the parameters shown in Eq.~\eqref{eq:FitI} and Eq.~\eqref{eq:FitII}. 

We note that Yukawa couplings to the third generation lepton required to explain anomalies in $R_{D^{(*)}}$ can induce $C_9^{\ell \ell}$ and $C_{10}^{\ell \ell}$ operators via penguin diagram \cite{Bobeth:2014rda, Aebischer:2019mlg}, with renormalization group equation running down to the $B$-meson mass scale. For instance, in scenarios with vector LQ, the same Yukawa couplings that explain $R_{D^{(*)}}$ induce such one-loop photon penguin diagrams \cite{Crivellin:2018yvo}. Similarly, with scalar LQs, similar log enhanced contribution can be realized \cite{Crivellin:2019dwb}. However, within our model,
although such contributions exist, the flavor structure we have adopted in Eq.~\eqref{eq:FitI} and Eq.~\eqref{eq:FitII} with $f'_{23} = 0$ ($y_{33} = 0$ or $y_{33} \ll 1$)
results in these contributions being negligible. 

\subsection{Neutrino Fit }\label{sec:neutfit} 

\begin{table}[!t]
	\centering
\begin{tabular}{|c|c|c|c|}
\hline \hline
\textbf{Oscillation} & \textbf{3\ $\sigma$ allowed range}  & \textbf{Model} & \textbf{Model}  \\
\textbf{parameters} &  \textbf{from NuFit5.0}~\cite{Esteban:2020cvm} & \textbf{{\tt Fit I}} & \textbf{{\tt Fit II}}  \\ \hline \hline
$\sin^2\theta_{12}$ & 0.269 -- 0.343 & 0.290 & 0.324   \\ \hline
$\sin^2\theta_{13}$ & 0.02032 -- 0.02410 & 0.0235 & 0.0210  \\ \hline
$\sin^2\theta_{23}$ & 0.415 -- 0.616 & 0.472 & 0.430  \\ \hline
$\Delta m_{21}^2$ $( 10^{-5} \ {\rm eV}^2)$ & 6.82 -- 8.04 & 7.39 & 7.45 \\ \hline
$\Delta m_{23}^2$ $( 10^{-3} \ {\rm eV}^2)$ & 2.435 -- 2.598 & 2.54 & 2.49  \\ \hline
$\delta$ (degree) & 107 -- 403 & 329.6 & 322.7  \\ \hline \hline
\textbf{Observable} & \textbf{$1 \, \sigma$ allowed range} &  &    \\ \hline \hline
$R_D$   &  0.310 -- 0.367~\cite{Amhis:2019ckw} & 0.348 & 0.343  \\ \hline
$R_{D\star}$   & 0.281 -- 0.308~\cite{Amhis:2019ckw} & $0.288$ & 0.284  \\ \hline
$C_9 = -C_{10}$  & $[-0.61 , -0.45]$~\cite{Aebischer:2019mlg} & $-0.52$ & $-0.51$ \\ \hline \hline
$(g-2)_\mu \, \, (10^{-10})$  & $27.4 \pm 7.3$~\cite{Blum:2018mom} & 29.7 & 34.4 \\ \hline \hline
\end{tabular}\\
\caption{Fits to the the neutrino oscillation parameters in the model with normal hierarchy, along with the  $B$-anomalies, and muon $g-2$ for two benchmark fits given in Eq.~\eqref{eq:FitI} and Eq.~\eqref{eq:FitII}. For comparison, the $3 \, \sigma$ allowed range for the oscillation parameters and the $1\, \sigma$ range for the other observables are also given. Note that correlation between $R_D$ and $R_{D^{*}}$ is not taken into account here (see Fig.~\ref{fig:RD_RDS} to see such correlation).  }
\vspace{7mm}
 \label{Tab:fit}
\end{table}

In this section, we explicitly show that the neutrino oscillation data can be explained in our model, while being consistent with the $B$-anomalies and $(g-2)_\mu$, as well as satisfying all the experimental constraints given in Sections~\ref{sec:Constraint} and \ref{sec:LHCcon}. We have performed a detailed numerical study to find the minimal texture for the Yukawa couplings to fit all the observables. We show our results for two different textures, namely, {\tt Fit I} and {\tt Fit II} as given in Eqs.~\eqref{eq:ffp} and \eqref{eq:y}. For this analysis, we fix the $R_2$ and $S_3$ LQ masses at 900 GeV and  2 TeV respectively. Furthermore, the masses of the up-type quarks entering the neutrino mass matrix (cf.~Eq.~\eqref{eq:numass}) are fixed at~\cite{Babu:2009fd, Xing:2019vks, Zyla:2020zbs} 
\begin{equation}
    m_u (\text{2 GeV}) \ = \ 2.16 \, \text{MeV}, \hspace{10mm} m_c (m_c) \ = \ 1.27 \, \text{GeV}, \hspace{10mm} m_t (m_t) \ = \ 160 \, \text{GeV}.
    \label{eq:quarkmass}
\end{equation}
We have used these input values of the running up-type quarks given in Eq.~\eqref{eq:quarkmass} and then extrapolate them to the LQ mass scale at 1 TeV in doing the numerical fit for the neutrino oscillation data. We obtain $m_u (1 \,{\rm TeV}) = 1.10$ MeV, $m_c (1 \,{\rm TeV}) = 0.532$ GeV, and $m_t (1 \,{\rm TeV}) = 150.7$ GeV \cite{Babu:2009fd, Xing:2007fb}.
The neutrino mass matrix given by Eq.~\eqref{eq:numass} is diagonalized by a unitary transformation 
\begin{equation}
    U^T_{\text{PMNS}} M_\nu U_{\text{PNMS}} \ = \ \widehat{M}_\nu \, ,
    \label{eq:nudiag}
\end{equation}
where $\widehat{M}_\nu$ is the diagonal mass matrix and $U_{\text{PMNS}}$ is the $3\times3$ PMNS lepton mixing matrix. We numerically diagonalize Eq.~\eqref{eq:nudiag} by scanning over the input parameters with two different textures as shown in Eqs.~\eqref{eq:ffp} and \eqref{eq:y}. For ease of finding the fits to oscillation data, we factor out $m_t$ into the overall factor and define $m_0 = m_t \kappa_1$, where $\kappa_1$ is given in Eq.~\eqref{eq:kappa1}. Furthermore, we perform constrained minimization in which the neutrino observables are restricted to lie within $3\, \sigma$ of their experimental measured values, for which we use the recent {\tt NuFit5.0} values (with SK atmospheric data included)~\cite{Esteban:2020cvm}. 

Our fit results for the two textures given in Eqs.~\eqref{eq:ffp} and \eqref{eq:y} are shown below:

\paragraph{Fit I:} With $m_0 =   9.9$ eV, 
\begin{equation}
f^\prime \ = \ \left(\begin{array}{ccc}
0 & 0 & 0 \\
0 & 0 & 0 \\
0 & 0.29 & -1.15 
\end{array}\right), \hspace{4mm}
f \ = \ \left(\begin{array}{ccc}
0 & 0 & 0 \\
0 & 0.29 & 0.886 i \\
0 & 0.0059 & 0.0226
\end{array}\right),  \hspace{4mm}
y \ =  \ \left(\begin{array}{ccc}
0 & 0 & 0 \\
0 & 0.124 & 0.064 \\
-0.016 & 0.028 & 0
\end{array}\right) \, . 
\label{eq:FitI}
\end{equation} 

\paragraph{Fit II:} With $m_0 =  15.1$ eV, 
\begin{equation}
f^\prime \ = \ \left(\begin{array}{ccc}
0 & 0 & 0 \\
0 & 0 & 0 \\
0 & 0.29 & -1.10 
\end{array}\right), \hspace{4mm}
f \ = \ \left(\begin{array}{ccc}
0 & 0 & 0 \\
0 & 0.29 & 0.887 i \\
0 & 0.0061 & 0.0215
\end{array}\right),  \hspace{4mm}
y \ = \ \left(\begin{array}{ccc}
0 & 0 & 0 \\
0 & 0.22 & 0 \\
0.026 & 0.0155 & -0.035
\end{array}\right) \, . 
\label{eq:FitII}
\end{equation} 


For each of these Yukawa textures, the corresponding fit results for the neutrino oscillation parameters are shown in Table \ref{Tab:fit}. It is clear that both fits are in excellent agreement with the observed experimental values. The  $f_{33}$ entry in the benchmark texture shown above is required for fine-tuning at the level of 7\% the $\tau \to \mu \gamma$ amplitude arising from top quark loop with a chiral enhancement (cf.~Section~\ref{sec:llg}). Note that the input parameter $f_{23}$ in both {\tt Fit I} and {\tt Fit IIa} is purely complex, which is required to get $R_{D^{(\star)}}$ correct (cf.~Fig.~\ref{fig:RD_1000}). Furthermore, the same coupling leads to a significant Dirac $CP$ phase, as can be seen from Table~\ref{Tab:fit}, consistent with the recent T2K result~\cite{Abe:2019vii}.

We note that the structures of $f$ and $f'$ do not change  significantly from Fit-I to Fit-II. This happens due to the various flavor violating constraints.  In this sense, the parameter space is rather limited for $f$ and $f'$. However, the structure of $y$ is different for Fits-I and II, and there is also some freedom in the overall scale of $y$, as illustrated in Eqs.~\eqref{eq:FitI} and \eqref{eq:FitII}.

We shown in Table \ref{Tab:fit} the fit results for $R_D,\ R_{D^{(\star)}}$, $R_{K^{(\star)}}$ and $(g-2)_\mu$, all of which are within $1\, \sigma$ of the experimentally allowed range.


\subsection{Non-standard Neutrino Interactions}\label{subsec:NSI}
The LQs $\omega^{2/3}$ from $R_2$ and  $\rho^{-2/3}$, $\rho^{1/3}$ from $S_3$ have couplings with neutrinos and quarks (cf.~Eq.~\eqref{eq:expandLag}). These couplings can induce charged-current NSI at tree-level~\cite{Babu:2019mfe}. Using the effective dimension-6 operators for NSI introduced in Ref.~\cite{Wolfenstein:1977ue}, the effective NSI parameters in our model are given by 
\begin{align}
    \varepsilon_{\alpha\beta} \ = \ \frac{3}{4\sqrt 2 G_F}
    \left(\frac{f^\star_{1\alpha}f_{1\beta}}{m_{\omega^{2/3}}^2}
    +\frac{(Vy^\star)_{1\alpha}(V^\star y)_{1\beta}}{m_{\rho^{-2/3}}^2}
    +\frac{y^\star_{1\alpha}y_{1\beta}}{2m_{\rho^{1/3}}^2}
    \right) \, .
    \label{eq:NSI}
\end{align}
Any non zero entry in the up-sector $f_{1\alpha}$ and $y_{1\alpha}$, relevant for generating tree-level NSI, does not affect the neutrino oscillation fit, as it is suppressed by the up-quark mass. However, Yukawa couplings to the electron and muon sector $f_{1\alpha}$ and $y_{1 \alpha}$ ($\alpha =1,2$) are highly constrained by the non-resonant dilepton searches at the LHC. The limit on $f_{11}$ and $f_{12}$ are 0.19 and 0.16, respectively, for 1 TeV LQ mass (cf.~Fig.~\ref{fig:yukbound}). Also, the limit on $y_{11}$ and $y_{12}$ are 0.16 and 0.15. Thus $\varepsilon_{11}$ and $\varepsilon_{22}$ are sub-percent level, and far beyond the reach of forthcoming neutrino experiments. Furthermore, any nonzero $y_{1\alpha}$ is in conjunction to Cabibbo rotation and induces $(V^\star y)_{2\alpha}$ leading to $D^0-\bar{D}^0$ mixing with a constraint given in Eq.~\eqref{eq:DDy}. 

As noted in Section~\ref{sec:dilepton}, the LHC limits on the LQ Yukawa couplings in the tau sector are weaker, and in principle, one can allow $\mathcal{O}(1)$ Yukawa coupling for $f_{13}$ and generate a  $\varepsilon_{33}$ which can be as large as $5.6 \%$. However, we require $f_{23}$ to be nonzero and $\mathcal{O}(1)$ to explain $R_{D^{(\star)}}$, and the constraint on the product of Yukawa couplings  $f_{13} f_{23}$ is severe due to the $D^0-\bar{D}^0$ bound, see Eq.~\eqref{eq:DDconst}. Thus the induced NSI will again be at a sub-percent level. For simplicity, we choose $f_{1\alpha}=y_{1\alpha}=0$ for all $\alpha=1,2,3$ (cf.~Eq.~\eqref{eq:y}) in both the  numerical fits discussed in Section~\ref{sec:neutfit}.

\section{Collider Implications} \label{sec:collider}

This model provides an avenue to test a unified description of $B$-anomalies,  muon anomalous magnetic moment  and neutrino masses at the LHC through a new decay channel of the $S_3$ LQ. The presence of the two scalar LQs $R_2$ and $S_3$ and the isospin-$3/2$ scalar multiplet $\Delta$ (especially its triply- and doubly-charged components) give rise to a rich phenomenology for the LHC.  In this section, we analyze the production and decay of the doubly-charged component of the scalar multiplet at the LHC and prospective smoking gun signals correlated  with the $B$-anomalies.

\subsection{Production of Doubly-charged Scalars via LQ Decay}

\begin{figure}[t!]
         \centering
         \includegraphics[scale=0.7]{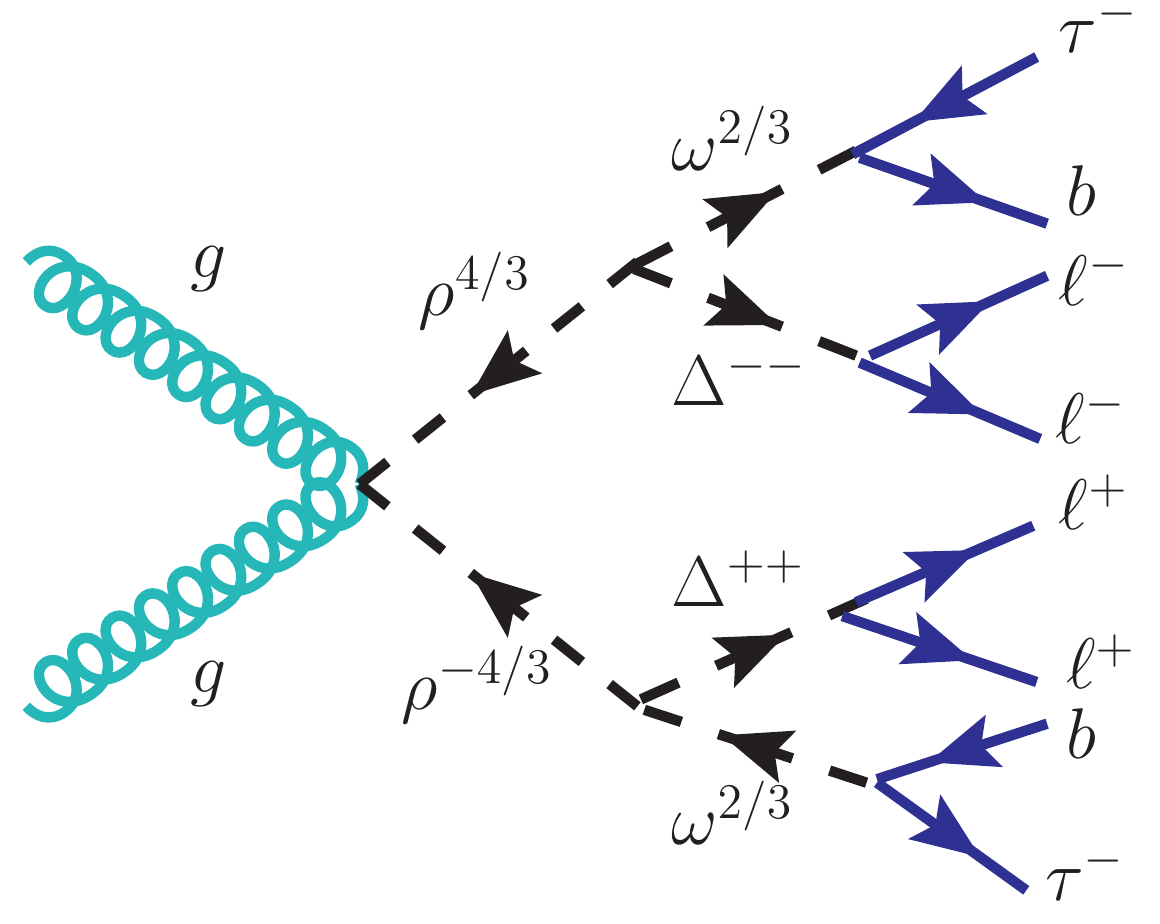}
         \caption{
         Feynman diagram for the pair-production of the $\rho^{4/3}$ component of the $S_3$ LQ ($pp \to \rho^{4/3} \rho^{-4/3}$), followed by $\rho$ decay to the $\omega^{2/3}$ component of the $R_2$ LQ and the doubly-charged component of the $\Delta$ quadruplet ($\rho^{\mp 4/3} \to \omega^{\pm 2/3} \Delta^{\mp \mp}$). The $\omega^{2/3}$ component can then decay to $b\tau$ (or $j\nu$) final state, while the doubly-charged scalar mostly decays to same-sign lepton pair (for small $v_\Delta$). This leads to the striking signal of this model: $pp\to \ell^+\ell^+\ell^-\ell^-+\tau^+\tau^-+b\bar{b} $ (where $\ell=e$ or $\mu$).} 
         \label{fig:feyn_sig}
    \end{figure}

Being part of the $SU(2)_L$-quadruplet, the charged scalars   $(\Delta^{\pm \pm \pm} , \Delta^{\pm \pm },\Delta^{\pm })$ can be pair-produced at the LHC by standard DY processes mediated by $s$-channel $Z/\gamma$ exchange. In addition, $s$-channel $W$ exchange can lead to associated production of $\Delta^{\pm \pm \pm}\Delta^{\mp \mp }$ $(\Delta^{\pm \pm}\Delta^{\mp })$. It is  important to note that being $s$-channel processes, the DY pair production cross-sections are highly suppressed for large $\Delta^{\pm \pm \pm} \, (\Delta^{\pm \pm })$ masses (similar to the doubly-charged scalar production in the type-II seesaw~\cite{Dev:2018kpa, Du:2018eaw, Babu:2016rcr}). The collider phenomenology of $SU(2)_L$-quadruplet scalars with  DY production and the same-sign dilepton (trilepton) signals from  doubly (triply)-charged scalars  has been studied  extensively in different contexts~\cite{Babu:2009aq, Bambhaniya:2013yca, Ghosh:2017jbw,Ghosh:2018drw, Bhattacharya:2016qsg, Arbelaez:2020xcg}. 

Here we  propose a unique production mechanism for the doubly-charged scalars at the LHC via the  gluon fusion process, as shown in Fig.~\ref{fig:feyn_sig}. In the  gluon-gluon fusion process, the $S_3$ LQ can be pair-produced copiously. Once produced, the various components of the $S_3$ LQ would decay dominantly to the components of the $R_2$ LQ and $\Delta$ quadruplet, if kinematically allowed (cf.~Eq.~\eqref{eq:Sdecay}). 
Here we will mainly focus on the $\rho^{\mp 4 / 3} \ \rightarrow \ \omega^{\pm 2 / 3} \Delta^{\mp \mp}$ decay channel, as $\rho^{4 / 3}$ and $\omega^{2 / 3}$ are respectively the components responsible for the  $R_{K^{(\star)}}$ and $R_{D^{(\star)}}$ anomalies in our model. Therefore, the signal shown in Fig.~\ref{fig:feyn_sig} provides a direct test of the $R_{K^{(\star)}}$ and $R_{D^{(\star)}}$ explanations at the high-energy LHC.

\begin{figure}[t!]
         \centering
         $$
         \includegraphics[width=0.65\textwidth]{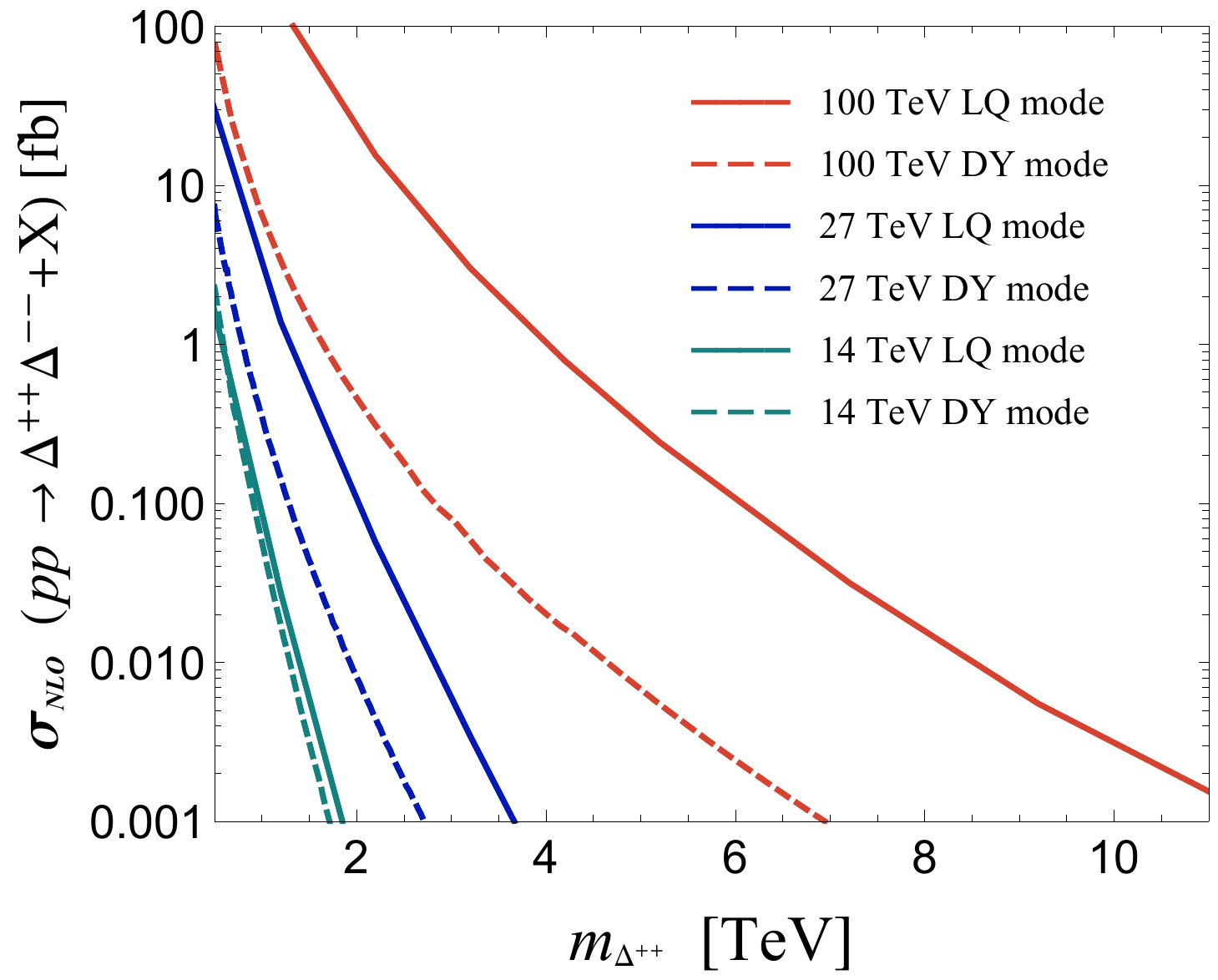} 
          $$
         \caption{Comparison of the NLO pair-production cross-sections for the doubly-charged scalars in the DY channel ($pp \to \Delta^{++} \Delta^{--}$) versus the LQ channel ($pp\to \Delta^{++}\Delta^{--}+\omega^{2/3}\omega^{-2/3}$) as a function of the doubly-charged scalar mass at $\sqrt s=$  14, 27 and 100 TeV. 
         }
         \label{fig:com}
    \end{figure}

Another reason we consider the $\Delta^{\pm\pm}$ production via $S_3$ decay is that the LQ-induced charged-scalar pair-production rate is not as highly suppressed as the DY rate for higher masses. 
In addition, there will be 
an enhancement factor for gluon luminosity compared to the quark luminosity, which becomes even more pronounced at higher center-of-mass energies. This can be seen from Fig.~\ref{fig:com}, where we compare the doubly-charged scalar pair-production cross-sections at NLO in the DY mode $pp\to \Delta^{++}\Delta^{--}$ and in the new LQ mode  $pp\to \Delta^{++}\Delta^{--}+\omega^{2/3}\omega^{-2/3}$ (in Fig.~\ref{fig:com}, $\omega^{2/3}\omega^{-2/3}$ is collectively  denoted as $X$) for center-of-mass energies $\sqrt s=$14, 27 and 100 TeV. Note that for the LQ mode, the cross section only depends on the $\rho^{4/3}$ LQ mass; however, to make a direct comparison with the DY mode, we have fixed the $\omega^{2/3}$ mass at 900 GeV (the preferred value for $R_D^{(\star)}$ explanation), and for a given $\Delta^{\pm\pm}$ mass in Fig.~\ref{fig:com}, have chosen the $\rho^{4/3}$ mass such that the $\rho^{\mp 4/3}\to \omega^{\pm 2/3}\Delta^{\mp\mp}$ decay branching ratio is $\sim 50\%$ (with the other 50\% going to $\omega^{\pm 5/3}\Delta^{\mp\mp\mp}$). From Fig.~\ref{fig:com}, we infer that the production cross-sections for the doubly-charged scalar in the LQ mode are  sizable up to the multi-TeV mass range, and  the collider reach in the inclusive mode $pp\to \Delta^{++}\Delta^{--}+X$ can be significantly enhanced, compared to the pure DY mode (see Section~\ref{sec:sens} for more details).

\subsection{Decay of Doubly-Charged Scalars}

Now we turn to the decay modes of the quadruplet scalar $\Delta $.  The doubly charged scalar $\Delta^{\pm \pm}$ can  decay to $\ell^\pm \ell^\pm$ via the leptonic coupling given by Eq.~\eqref{eq:YukD}. In addition, being a part of the $SU(2)_L$-quadruplet, the covariant derivative term leads to bosonic decay modes ($W^\pm W^\pm$) of  $\Delta^{\pm \pm}$. On the other hand, when the mass-splitting between consecutive members of the quadruplet are nonzero,  cascade decays also open up. One should note that 
depending on the quartic coupling $\lambda'_{H \Delta}$, there could be two different  hierarchies: (a) when $\lambda'_{H \Delta}>0,$ we have $m_{\Delta^{\pm \pm \pm}}<m_{\Delta^ { \pm \pm }}<m_{\Delta^\pm}<m_{\Delta^{0}}$ and (b) when $\lambda'_{H \Delta}<0,$ we have $m_{\Delta^{\pm \pm \pm}}>m_{\Delta^ { \pm \pm }}>m_{\Delta^{\pm}}<m_{\Delta^{0}}$ (cf.~Eq.~\eqref{eq:YukD}).
Therefore, due to mass-splitting, it can decay in cascades via  $\Delta^{\pm \pm \pm} X^{\mp}$ or $\Delta^{\pm} X^{\pm}$ (where $X=\pi, W^\star$) depending on whether $\Delta m > 0$ or $\Delta m < 0$. For simplicity, we consider  $\Delta^{\pm \pm \pm}$ to be the lightest  member of the $\Delta$ multiplet throughout our analysis.  The partial decay widths for different decay modes of  $\Delta^{\pm \pm}$ can be written as~\cite{Ghosh:2018drw, Ghosh:2017jbw}:
\begin{align}
& \Gamma\left(\Delta^{\pm \pm} \rightarrow \ell_{i}^{\pm} \ell_{j}^{\pm}\right)  \ = \ \frac{ m_{\Delta^{ \pm \pm}} \left(m_{\nu}\right)_{i j}^2}{6 \pi\left(1+\delta_{i j} \right)v_{\Delta}^2}\left(1-\frac{m_{i}^{2}}{m_{\Delta^{ \pm \pm}}^{2}}-\frac{m_{j}^{2}}{m_{\Delta^{ \pm \pm}}^{2}}\right)\left[\lambda\left(\frac{m_{i}^{2}}{m_{\Delta^{ \pm \pm}}^{2}}, \frac{m_{j}^{2}}{m_{\Delta^{ \pm \pm}}^{2}}\right)\right]^{1 / 2} \, ,  \label{eq:Deltadecay1} \\
& \Gamma\left(\Delta^{\pm \pm} \rightarrow W^{\pm} W^{\pm}\right)  \ = \ \frac{3g^{4} v_{\Delta}^{2} m_{\Delta^{ \pm \pm}}^{3}}{16 \pi m_{W}^{4}}\left(\frac{3 m_{W}^{4}}{m_{\Delta^{ \pm \pm}}^{4}}+ \frac{m_{W}^{2}}{m_{\Delta^{ \pm \pm}}^{2}}+\frac{1}{4}\right) \beta\left(\frac{m_{W}^{2}}{m_{\Delta^{ \pm \pm}}^{2}}\right) \, , \label{eq:Deltadecay2} \\
& \Gamma\left(\Delta^{\pm \pm} \rightarrow \Delta^{\pm \pm \pm} \pi^{\mp}\right)  \ = \ \frac{g^{4}\left|V_{u d}\right|^{2} (\Delta m)^{3} f_{\pi}^{2}}{8 \pi m_{W}^{4}} \, , \label{eq:Deltadecay3} \\
& \Gamma\left(\Delta^{\pm \pm} \rightarrow \Delta^{\pm \pm \pm} \ell^{\mp} \nu_{\ell}\right)  \ = \ \frac{g^{4} (\Delta m)^{5}}{120 \pi^{3} m_{W}^{4}} \, , \label{eq:Deltadecay4} \\
& \Gamma\left(\Delta^{\pm \pm} \rightarrow \Delta^{\pm \pm \pm} q \bar{q}^{\prime}\right)  \ = \ 3 \Gamma\left(\Delta^{\pm \pm} \rightarrow \Delta^{\pm \pm \pm } \ell^{\mp} \nu_{\ell}\right) \, , \label{eq:Deltadecay5} \\
& \Gamma\left(\Delta^{\pm \pm} \rightarrow W^{\pm} W^{\pm \star}\right)  \ = \ \frac{9 g^{6} m_{\Delta^{\pm \pm}}}{512 \pi^{3}} \frac{v_{\Delta}^{2}}{m_{W}^{2}} F\left(\frac{m_{W}^{2}}{m_{\Delta^{ \pm \pm}}^{2}}\right) \, , \label{eq:Deltadecay6}
\end{align}
 where the kinematic functions are given by~\cite{Ghosh:2018drw}
\begin{align}
& \lambda(x, y) \ = \ 1+x^{2}+y^{2}-2 x y-2 x-2 z \, , \\
& \beta(x) \ = \  \sqrt{1-4 x} \, ,  \\
& F(x) \ = \ -|1-x|\left(\frac{47}{2} x-\frac{13}{2}+\frac{1}{x}\right)+3\left(1-6 x+4 x^{2}\right)|\log \sqrt{x}| \nonumber \\
&\qquad \qquad +\frac{3\left(1-8 x+20 x^{2}\right)}{\sqrt{4 x-1}} \cos^{-1} \left(\frac{3 x-1}{2 x^{3 / 2}}\right) \, .
\end{align}
If $\Delta^{\pm\pm}$ decay to $\Delta^\pm X^\pm$ is allowed, the corresponding partial widths will be the same as in Eqs.~\eqref{eq:Deltadecay3}-\eqref{eq:Deltadecay5}. 
The different scaling factor due to the Clebsch-Gordon coefficient for the quadruplet scalar is taken into account properly for the partial decay width formulae of the doubly charged Higgs given above. For example, the leptonic decay width  given in Eq.~\eqref{eq:Deltadecay1} is suppressed by a factor of 2/3, compared to the type-II seesaw scenario~\cite{Perez:2008ha, Melfo:2011nx}. On the other hand, the bosonic and cascade decay modes are enhanced by a factor 3/2 in the quadruplet case compared to the triplet scenario~\cite{Perez:2008ha, Melfo:2011nx, Aoki:2011pz}. 
\begin{figure}[t!]
         \centering
         \includegraphics[width=0.57\textwidth]{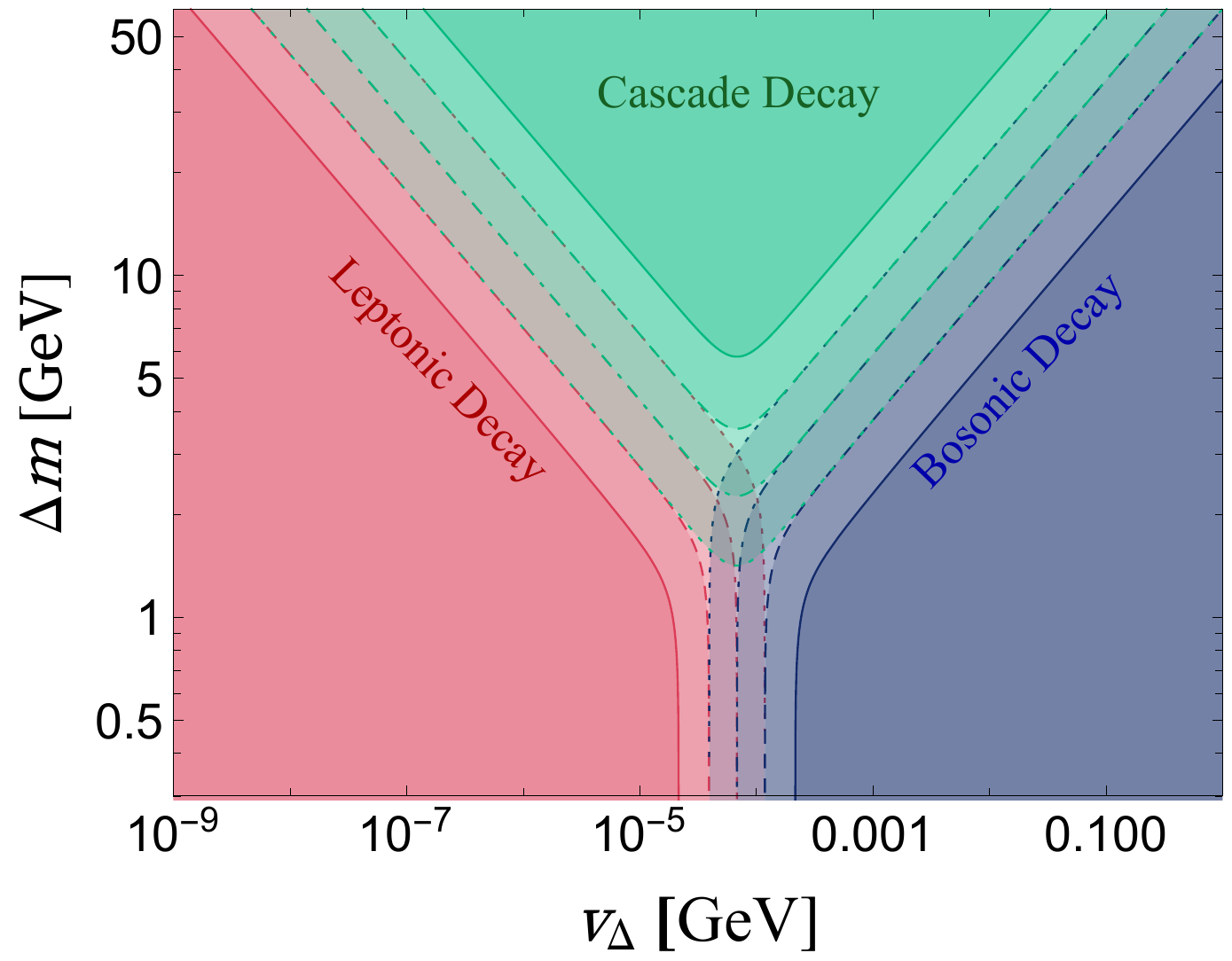}
         \caption{Generic decay phase diagram for $\Delta^{\pm \pm}$ in our model, with $m_{\Delta^{ \pm \pm}}=1 \mathrm{TeV}$.  The dotted, dot-dashed,  dashed and thick solid contours correspond to 99\%, 90\%, 50\% and $10 \%$ branching ratios respectively for the leptonic, bosonic or cascade decays, whereas $\Delta m$ is the mass splitting between the $\Delta^{++}$ and the next lightest scalar component. 
         }
         \label{fig:dch}
    \end{figure}

In Fig.~\ref{fig:dch}, we show the generic decay phase diagram for $\Delta^{\pm \pm}$ in our model, with $m_{\Delta^{ \pm \pm}}=1$ TeV.  The dotted, dot-dashed,  dashed and thick solid contours correspond to 99\%, 90\%, 50\% and $10 \%$ branching ratios into the leptonic, bosonic or cascade decay modes. The decay phase diagram clearly depicts that the branching ratio to  leptonic decay modes of $\Delta^{\pm \pm}$ decreases with $v_{\Delta}$, whereas the branching ratio to gauge boson decay mode increases with $v_{\Delta}$. The cross-over  happens at $v_{\Delta} =   10^{-4}$ GeV with $\Delta m \sim 0$, similar to the type-II seesaw case~\cite{Perez:2008ha, Melfo:2011nx}. As soon as the mass splitting is set to $\geq$ 10 GeV, cascade decays open up and start dominating depending on the exact value of $v_{\Delta}$. 
Note that the mass splitting $|\Delta m|$ between any two components of $\Delta$ cannot be larger than $\sim 50$ GeV due to stringent constraints from electroweak precision data~\cite{Ghosh:2018drw}.

\subsection{Comment on 4-body Decay of \texorpdfstring{$\Delta$}{Del}}
In addition to the two-body decays given in Eqs.~\eqref{eq:Deltadecay1}-\eqref{eq:Deltadecay6}, there will also be four-body decay modes of the doubly-charged scalar  via the virtual exchange of $R_2$  and $S_3$ LQs proportional to the $\mu$ term in Eq.~\eqref{eq:pot}: $\Delta^{\pm\pm}\to (\omega^{\pm 2/3})^\star (\rho^{\pm 4/3})^\star$, with each LQ decaying to two fermions. These decays will depend on the same parameters that lead to $\Delta^{\pm\pm} \rightarrow \ell^\pm \ell^\pm$ decays.  The phase space for these decays would appear to be comparable to the two-body decays, since the latter has a suppression of a loop factor, $1/(16\pi^2)^2$. We have evaluated these four-body decays of $\Delta^{++}$ semi-analytically following the procedure outline in Ref. \cite{Pois:1993ay}, as well as numerically.  The two methods gave very similar results.  As an example, for a benchmark values of $m_{\Delta^{++}} = 800$ GeV, $m_{R_2} = 1$ TeV, $m_{S_3} = 2$ TeV, $\mu = 246$ GeV, $v_\Delta = 10^{-4}$ GeV, and the values of the Yukawa couplings given in Fit I (cf.~Eq.~(\ref{eq:FitI})), the four-body decay width is $2.3 \times 10^{-15}$ GeV, which turns out to be much smaller than that for the dileptonic decay, which is $2 \times 10^{-9}$ GeV.  As $v_\Delta$ is increased, the four-body decay may compete with the dileptonic decay; however, in this case $\Delta^{++} \rightarrow W^+ W^+$ decay would dominate.  Consequently, the four-body decay of $\Delta^{++}$ can be safely ignored in our discussions.

\subsection{Signal Sensitivity} \label{sec:sens}

\begin{figure}[t!]
        \centering
       $$
        \includegraphics[width=0.65\textwidth]{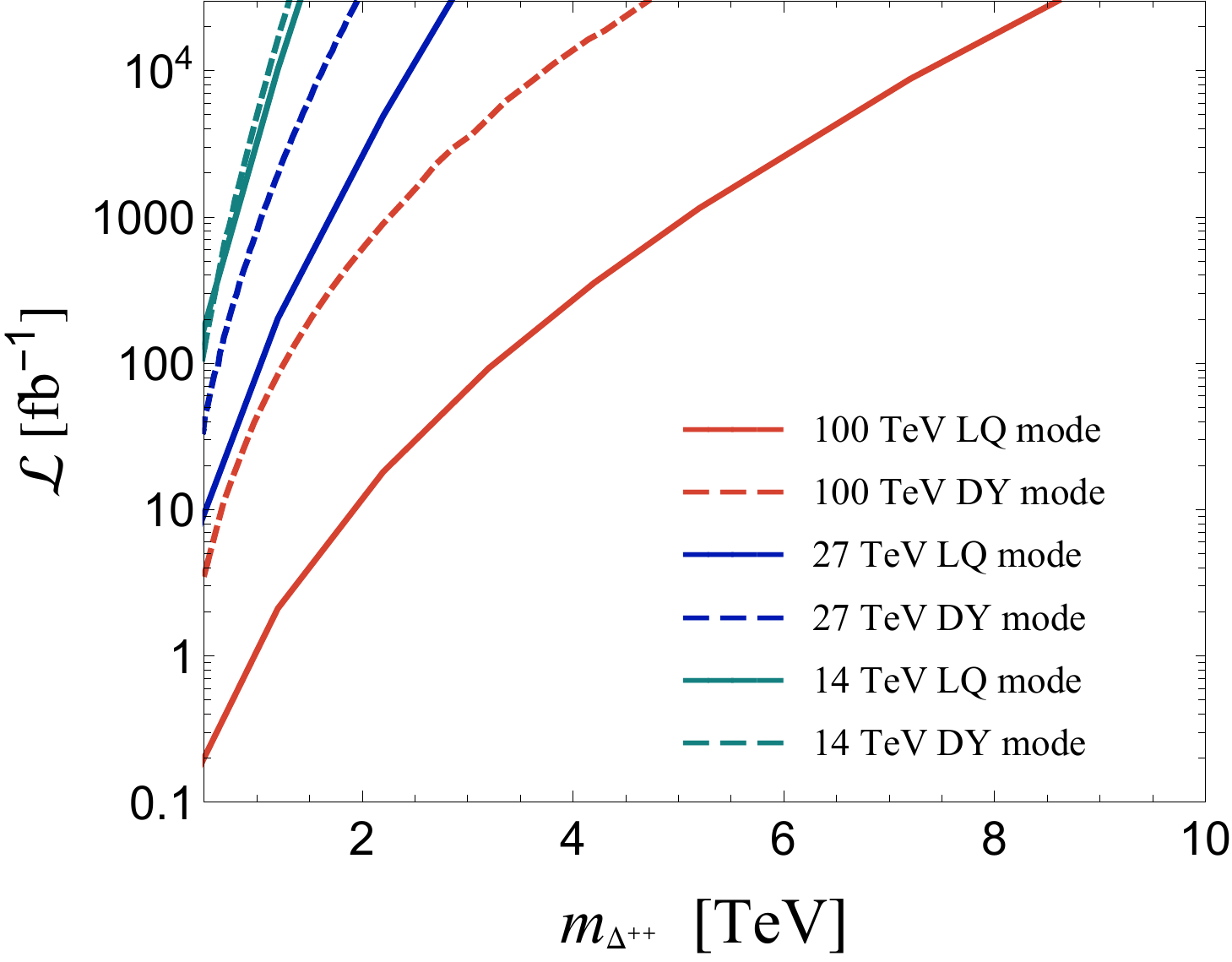} 
          $$
         \caption{Required integrated luminosities for different center-of-mass energies of the $pp$ collider to observe at least 25 events for the signal $pp \to \Delta^{++}\Delta^{--}+X\to  \mu^+\mu^+\mu^-\mu^-+X$ in the LQ and DY production modes.
         }
         \label{fig:comlum}
    \end{figure}

We focus on the small $v_\Delta$ region which gives same-sign dilepton final states from the $\Delta^{\pm\pm}$ decay, because charged leptons with large transverse momenta can be cleanly identified  with good resolution and the charge of the leptons can be identified with fairly good accuracy at hadron colliders. For the benchmark fits given in Section~\ref{sec:neutfit} with normal hierarchy, the dilepton branching ratios of the $\Delta^{\pm\pm}\to \ell_i\ell_j$ for different flavors are as follows: 
\begin{align}
    & {\rm BR}(ee) \ = \ 0 \, , \quad {\rm BR}(\mu \mu) \ = \ 0.22 \, ,  
    \quad {\rm BR}(\tau \tau) \ = \ 0.23 \, ,\nonumber \\
    & {\rm BR}(e \mu) \ = \ 0.01 \, ,  \quad 
    {\rm BR}(\mu \tau) \ = \ 0.39 \, , \quad 
    {\rm BR}(e \tau) \ = \ 0.16\, . 
\end{align}
For simplicity, we focus on the $\mu\mu$ final states and consider the signal $pp\to \Delta^{++}\Delta^{--}+X\to \mu^+\mu^+\mu^-\mu^-+X$ to derive the  sensitivity at future hadron colliders. The relevant SM background is mainly from the multi-top and 
multi-gauge boson production~\cite{Aaboud:2017qph, CMS:2017pet}. 
However, there are several discriminating characteristics of our signal: (a) the invariant mass distributions for same-sign  lepton pair from the $\Delta^{\pm\pm}$ decay would peak at a mass value much higher than the SM $Z$ boson mass; and (b) the outgoing leptons will be more energetic compared to the ones produced in the decay of SM gauge bosons,  since these leptons are produced from heavy particle $\Delta^{\pm\pm}$ decay. To derive the signal sensitivity, we first implement our model file in {\tt FeynRules} package~\cite{Christensen:2008py}, then analyze the cross section for the signal  using {\tt MadGraph5aMC@NLO}~\cite{Alwall:2014hca}, simulating the hadronization effects with {\sc Pythia8}~\cite{Sjostrand:2007gs} and detector effects with the {\tt Delphes3} package~\cite{deFavereau:2013fsa}. In order to optimize the signal efficiency over the SM background, we impose the following basic acceptance criteria:  $p_{T}^\ell>15 $ {GeV} for each lepton,  pseudorapidity $|\eta^\ell|<2.5 $ and  a veto on any opposite sign dilepton pair invariant mass being close to the $Z$ boson mass $|M(\ell^+ 
\ell^-) -m_Z| > 15 $ GeV. In addition, events are selected such that the invariant mass  for same-sign muon pair is higher than 500 GeV. After passing through all these acceptance criteria, we estimate the required luminosities to observe at least 25 events at different center-of-mass energies  ($\sqrt{s}$=14, 27, 100 $ \mathrm{TeV}$). Our results are shown in Fig.~\ref{fig:comlum}. It is clear that for a given luminosity and a given $\sqrt{s}$, the doubly-charged scalar mass reach in the LQ mode is higher than that in the DY mode. The mass reach for 3 ab$^{-1}$ integrated luminosity is summarized in Table~\ref{tab:sens} for different center-of-mass energies.

\begin{table}[t!]
\begin{center}
    \begin{tabular}{||c|c|c|c||}\hline\hline
    Production & \multicolumn{3}{c||}{$\Delta^{\pm\pm}$ mass reach for ${\cal L}=3$ ab$^{-1}$} \\ \cline{2-4}
    Channel & $\sqrt s=14$ TeV & $\sqrt s=27$ TeV & $\sqrt s=100$ TeV \\ 
    \hline\hline
    LQ-mode & 1.1 TeV & 2.0 TeV & 6.2 TeV \\
    DY-mode & 0.9 TeV & 1.3 TeV & 2.9 TeV \\ \hline\hline
    \end{tabular}
\end{center}
\caption{Comparison of the doubly-charged scalar mass reach in the LQ and DY modes (with same-sign di-muon pair final states only) for 3 ab$^{-1}$ integrated luminosity.}
\label{tab:sens}
\end{table}

Once we identify the doubly-charged scalar from the multi-lepton signal, the next step is to distinguish the underlying model. In order to identify whether the $\Delta^{\pm\pm}$'s come from the $S_3$ LQ decay, accompanied by the $\omega^{2/3}$ LQs, we can consider the decay chain given in Fig.~\ref{fig:feyn_sig}, i.e. 
\begin{align}
    pp \ \to \ \rho^{4/3} \rho^{-4/3} \ \to \ \omega^{-2/3} \Delta^{++} \omega^{2/3} \Delta^{--}  \ \to \ \ell^+\ell^+\ell^-\ell^- +\tau^+\tau^-+b \bar{b} \, .
    \label{eq:sig}
\end{align}
In this case, the right combination of the $b\tau$ invariant mass peaks at the $\omega^{2/3}$ LQ mass, if it is produced on-shell from the $\Delta$ decay. Considering the fact that the benchmark fits in our model give 54\% branching ratio of $\omega^{2/3}$ to $b\tau$ (cf.~Table~\ref{tab:branching}), and taking into account the $b$-tagging and $\tau$-identification efficiencies of $\sim 70\%$ each, we find that at least 25 signal events in the channel given by Eq.~\eqref{eq:sig} can be obtained with 3 ab$^{-1}$ luminosity for the $S_3$ LQ masses up to 1.5, 2.4 and 5.5 TeV respectively 
at $\sqrt s=14$, 27 and 100 TeV. Hence, it is possible  to independently test the unified description of $B$-anomalies, muon $g-2$ and neutrino masses in our model at future colliders. 


\section{Conclusion} \label{sec:conclusion}
 We have presented a radiative neutrino mass model involving TeV-scale scalar leptoquarks $R_2$ and $S_3$, which can simultaneously explain the $R_{D^{(\star)}}$, $R_{K^{(\star)}}$, as well as muon $g-2$ anomalies, all within $1\, \sigma$ CL, while being consistent with neutrino oscillation data, as well as all flavor and LHC constraints. The $R_2$ LQ is responsible for the $R_{D^{(\star)}}$ and $(g-2)_\mu$, while the $S_3$ LQ explains the $R_{K^{(\star)}}$ anomaly. The model also features a scalar quadruplet $\Delta$, which is required for the radiative neutrino mass generation. The same  trilinear $\Delta^\star R_2 S_3$ coupling that is responsible for neutrino mass also leads to interesting collider signatures in the $S_3$ and $\Delta$ decays that can be probed in the forthcoming run of the LHC. Similarly, the same Yukawa couplings responsible for the chirally-enhanced contribution to $\Delta a_\mu$ give rise to new contributions to the SM Higgs decays to muon and tau pairs, with the modifications to the corresponding branching ratios being at 2-6\% level, which could be tested at future hadron colliders, such as HL-LHC and FCC-hh.

\begin{acknowledgments}
 We thank Wolfgang Altmannshofer for useful discussions. The work of KB and AT are supported in part by U.S. Department of Energy Grant Number DE-SC 0016013. The work of BD is supported in part by U.S. Department of Energy under Grant No.~DE-SC0017987 and by  a Fermilab Intensity Frontier Fellowship. This work is also supported by the Neutrino Theory Network Program Grant No.~DE-AC02-07CHI11359. 
 \end{acknowledgments}

\bibliographystyle{utphys}

\bibliography{references}

\end{document}